\documentclass[showpacs,aps,prd,reprint,superscriptaddress,nofootinbib,longbibliography]{revtex4-1}
\usepackage[colorlinks=true, pdfstartview=FitV, linkcolor=magenta,citecolor=blue, urlcolor=magenta,
bookmarks=true, bookmarksnumbered=true, breaklinks]{hyperref}
\usepackage[dvipdfmx]{graphicx}
\usepackage{url}
\usepackage{amsmath,amssymb,bm,color,longtable,mathrsfs,amsfonts,slashed,ulem}

\newcommand{\Slash}[1]{{\ooalign{\hfil/\hfil\crcr$#1$}}}

\begin{document}
\begin{flushright}
\end{flushright}

\title{Heavy-quark spin polarization induced by the Kondo effect in a magnetic field
}

\author{Daiki~Suenaga}
\email[]{suenaga@rcnp.osaka-u.ac.jp}
\affiliation{Research Center for Nuclear Physics,
Osaka University, Ibaraki, 567-0048, Japan }

\author{Yasufumi Araki}
\email[]{araki.yasufumi@jaea.go.jp}
\affiliation{Advanced Science Research Center, Japan Atomic Energy Agency (JAEA), Tokai 319-1195, Japan}

\author{Kei~Suzuki}
\email[]{{k.suzuki.2010@th.phys.titech.ac.jp}}
\affiliation{Advanced Science Research Center, Japan Atomic Energy Agency (JAEA), Tokai 319-1195, Japan}

\author{Shigehiro~Yasui}
\email[]{yasuis@keio.jp}
\affiliation{Research and Education Center for Natural Sciences,\\ Keio University, Hiyoshi 4-1-1, Yokohama, Kanagawa 223-8521, Japan}
\affiliation{RIKEN iTHEMS, Wako, Saitama 351-0198, Japan}

\date{\today}

\begin{abstract}
We propose a new mechanism of the heavy-quark spin polarization (HQSP) in quark matter induced by the Kondo effect under an external magnetic field. The Kondo effect is caused by a condensate between a heavy and a light quark called the Kondo condensate leading to a mixing of the heavy- and light-quark spins. Thus, the HQSP is driven through the Kondo effect from light quarks coupling with the magnetic field in quark matter. For demonstration, we employ the Nambu-Jona-Lasinio type model under a magnetic field and investigate the HQSP within the linear response theory with vertex corrections required by the $U(1)_{\rm EM}$ electromagnetic gauge invariance.  As a result, we find that the HQSP arises significantly with the appearance of the Kondo effect. Our findings are testable in future sign-problem-free lattice simulations.
\end{abstract}

\pacs{}

\maketitle

\section{Introduction}
\label{sec:Introduction}
The Kondo effect is known as one of the important phenomena in the quantum many-body system~\cite{Kondo:1964,Hewson}. This effect arises when heavy impurities exist in matter coupling with itinerant fermions through non-Abelian interactions such as the spin exchange. The Kondo effect changes transport properties significantly; e.g., it turns the electrical resistivity from decrement to increment with the temperature lowered.

Recently, it has been discussed that the Kondo effect persists universally in high-energy physics as well, although the effect was originally discovered in condensed-matter physics. For example, in the context of quantum chromodynamics (QCD), the effect is driven when heavy quarks ($c,b$) exist as an impurity in dense quark matter formed by light quarks ($u,d$)~\cite{Yasui:2013xr,Hattori:2015hka}. This is particularly referred to as the {\it QCD Kondo effect}. In this case, the non-Abelian interaction is supplied by $SU(N_c)$ color exchange ($N_c=3$ is the number of colors). The Kondo effect is expected to arise also in the hadronic phase at lower density. In fact, it was shown that the Kondo effect arises in nuclear matter when $\Sigma_c$ and $\Sigma^*_c$  baryons or $\bar{D}$ and $\bar{D}^*$ mesons exist as impurities, where the spin and/or isospin exchange serves as the non-Abelian interaction~\cite{Yasui:2013xr,Yasui:2016hlz,Yasui:2016ngy,Yasui:2019ogk}. In addition, many works on the Kondo effect in a relativistic system have been done in the literatures~\cite{Ozaki:2015sya,Yasui:2016svc,Yasui:2016yet,Kanazawa:2016ihl,Kimura:2016zyv,Yasui:2017izi,Suzuki:2017gde,Yasui:2017bey,Kimura:2018vxj,Macias:2019vbl,Hattori:2019zig,Suenaga:2019car,Suenaga:2019jqu,Kanazawa:2020xje,Araki:2020fox,Araki:2020rok,Kimura:2020uhq,Suenaga:2020oeu,Ishikawa:2021bey}.
In this way, the interdisciplinary study bridging condensed-matter physics and QCD/hadron physics focused on the Kondo effect is attracting attention. 

In a field-theoretical treatment, the Kondo effect is caused by a condensate (a hybridization) formed by a light itinerant fermion and a heavy impurity called {\it Kondo condensate}. This condensate not only provides a gap of the itinerant fermions near the Fermi level but also causes a mixing between the itinerant fermion and the heavy impurity as induced by an $s$-$d$ interaction in the Anderson model~\cite{PhysRev.124.41}. 
In quark matter, due to the mixing, in Ref.~\cite{Suenaga:2020oeu}, the magnetically induced axial current of light quarks in quark matter, namely, the chiral separation effect (CSE)~\cite{Son:2004tq,Metlitski:2005pr,Newman:2005as}, was found to be enhanced under the Kondo effect. 
In solid states, in Ref.~\cite{Araki:2020rok}, the magnetic responses of a Dirac and nonrelativistic bands with their hybridizations were investigated, showing that the spin-orbit crossed part of the magnetic susceptibility of the Dirac (nonrelativistic) band is enhanced (suppressed) due to the presence of the hybridizations.
Those findings show that the Kondo condensate (hybridization) has a great impact on magnetic responses of the fermions.

\begin{figure}[t]
\centering
\includegraphics*[scale=0.37]{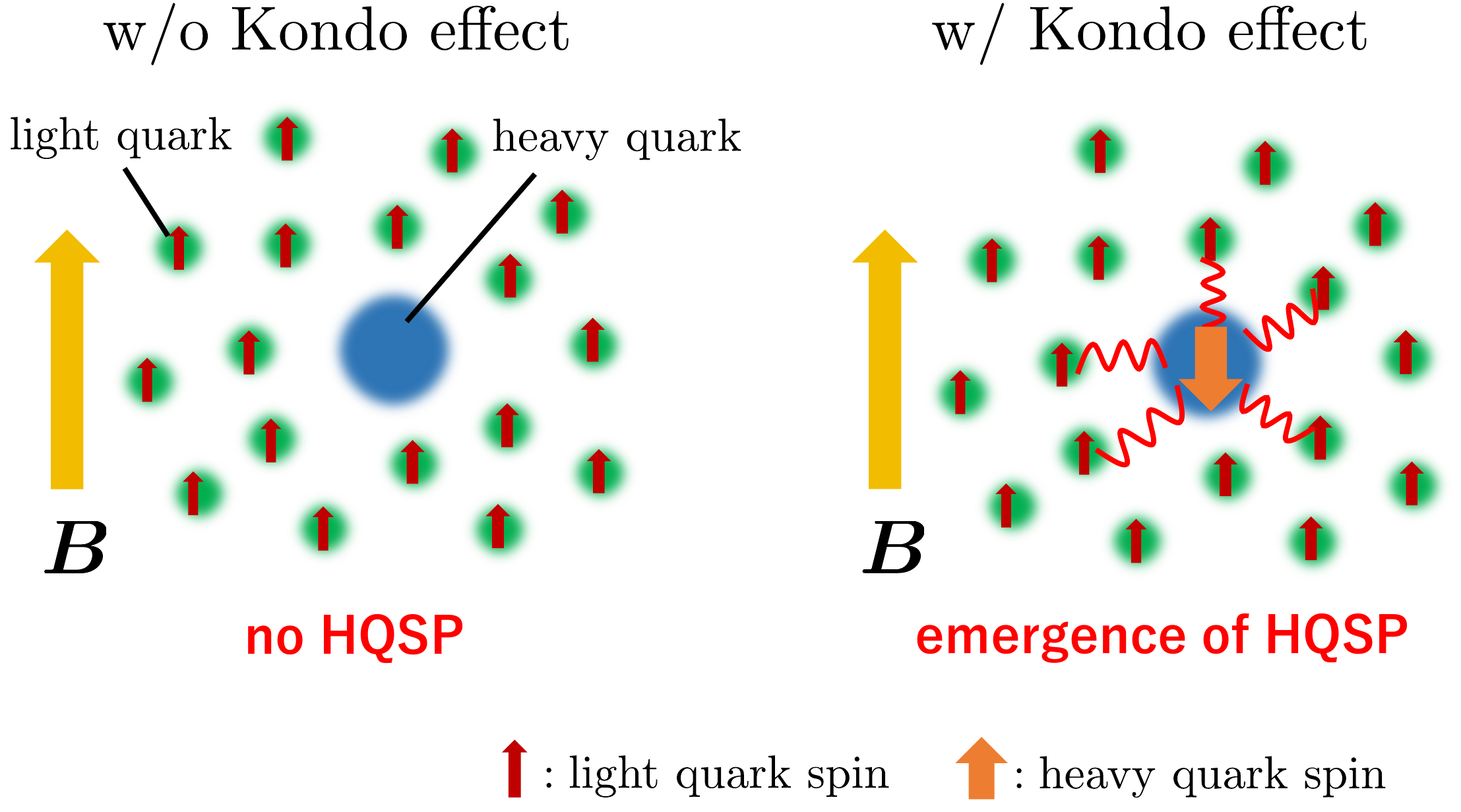}
\caption{A schematic picture of the HQSP induced by the Kondo effect under a magnetic field. The HQSP is absent without the Kondo effect in a heavy-quark mass limit $m_Q\to\infty$ (left), while it significantly emerges in the presence of the Kondo effect (right). In this figure, we have assumed that the electric charge of light quarks is positive.
}
\label{fig:HQSPFigure}
\end{figure}

In this paper, we propose a new mechanism of the heavy-quark spin polarization (HQSP) induced by the Kondo effect under a magnetic field. When the Kondo effect is absent, the HQSP does not arise significantly, since the Zeeman interaction of a heavy quark is of order $1/m_Q$ ($m_Q$ is the mass of a heavy quark) and is suppressed for sufficiently large $m_Q$. In particular, it vanishes at $m_Q\to\infty$. On the other hand, when the Kondo effect is present, the HQSP can arise without $1/m_Q$ suppression. The Kondo condensate correlates the heavy-quark spin with the light-quark spin coupling to the magnetic field in a medium. In other words, the condensate converts the magnetic response of the light-quark spin into that of heavy-quark spin. Schematically this effect is depicted in Fig.~\ref{fig:HQSPFigure}. This new mechanism of the HQSP is expected to be testable in future sign-problem-free lattice simulations.

This paper is organized as follows. In Sec.~\ref{sec:Model}, we introduce the model for demonstration and derive the Green's function of the fermions. In Sec.~\ref{sec:Calculation}, we show the analytic evaluation for the HQSP within the linear response theory with the vertex corrections. In Sec.~\ref{sec:TwoLimits}, we study the HQSP response function in the so-called dynamical and static limits in detail, toward a better understanding of the HQSP in the timelike and spacelike momentum regions. Referring to it, in Sec.~\ref{sec:Results}, we present numerical results of the HQSP in both the momentum regions. In Sec.~\ref{sec:Discussions}, we compare the HQSP induced by the Kondo effect with that by the ordinary Zeeman effect, and in Sec.~\ref{sec:Conclusions}, we conclude the present work.

\section{Model}
\label{sec:Model}
Here, we present our effective model to study the HQSP induced by the Kondo effect under a magnetic field. One of the most useful models for the demonstration is the Nambu-Jona-Lasinio (NJL) type model including a four-point interaction between a light and a heavy quark~\cite{Yasui:2016svc,Yasui:2017izi}. Namely, we start with the following Lagrangian: 
\begin{eqnarray}
{\cal L} &=& \bar{\psi}(i\Slash{D}+\mu\gamma_0)\psi +\Psi_v^\dagger iD_0\Psi_v -\lambda(\Psi^\dagger_v\Psi_v-n_Q)\nonumber\\
&& -g\Big[(\bar{\psi}\Psi_v)(\Psi^\dagger_v\psi) + (\bar{\psi}\gamma^i\Psi_v)(\Psi_v^\dagger\gamma^i\psi)\Big]\ . \label{NJLStart}
\end{eqnarray}
In this Lagrangian, $\psi$ is the light quark described by the ordinary Dirac theory, while $\Psi_v$ is the heavy one described by the heavy-quark effective theory (HQET)~\cite{Eichten:1989zv,Georgi:1990um,Neubert:1993mb,Manohar:2000dt}. Thus, $\Psi_v$ is defined by $\Psi_v = {\rm e}^{im_Q t}\frac{1+\gamma_0}{2}\Psi$, with $\Psi$ being the relativistic heavy fermion in Dirac theory, and $\mu$ is the chemical potential of the light quark included to access finite density. It should be noted that $\lambda$ is a Lagrange multiplier included to impose a condition $\Psi_v^\dagger\Psi_v=n_Q$, with $n_Q$ a space-averaged heavy-quark density. In the following analysis, we examine the HQSP for the $\lambda=0$ case as a clear demonstration. The common coupling $g$ for the scalar and vector interactions is derived by the Fierz transformation from the one-gluon exchange interaction~\cite{Ebert:1994tv}. In Eq.~(\ref{NJLStart}) the covariant derivatives are defined by
\begin{eqnarray}
D_\mu\psi &=& (\partial_\mu+ie_q A_\mu)\psi \ , \nonumber\\
D_0\Psi_v &=& (\partial_0+ie_Q A_0)\Psi_v\ ,
\end{eqnarray}
where the electromagnetic gauge field $A^\mu = (A^0,{\bm A})$ is introduced to incorporate interactions with the magnetic field. The coupling constants $e_q$ and $e_Q$ are electric charges of the light and heavy quarks, respectively. We take $e_q=+\frac{2}{3}e$ for the $u$ quark, $e_q=-\frac{1}{3}e$ for the $d$ quark, and $e_Q=+\frac{2}{3}e$ for the $c$ quark, with the elementary charge $e$.

For the Lagrangian~(\ref{NJLStart}), we introduce the Kondo condensate made of the light and heavy quarks to describe the ground state governed by the Kondo effect. For this purpose, we make use of the mean field approximation for $ \bar{\psi}\Psi_v$ and $\bar{\psi}\gamma^i\Psi_v$ by rewriting the Lagrangian~(\ref{NJLStart}) as
\begin{eqnarray}
{\cal L} &=& \bar{\psi}(i\Slash{D}+\mu\gamma_0)\psi +\Psi_v^\dagger iD_0\Psi_v \nonumber\\
&& -g\Big[\langle\bar{\psi}\Psi_v\rangle \Psi_v^\dagger\psi + \langle\bar{\psi}\gamma^i\Psi_v\rangle\Psi_v^\dagger\gamma^i\psi + ({\rm h.c.})\Big] \nonumber\\
&& +g\Big[\langle\bar{\psi}\Psi_v\rangle \langle\Psi_v^\dagger\psi\rangle + \langle\bar{\psi}\gamma^i\Psi_v\rangle\langle\Psi_v^\dagger\gamma^i\psi\rangle \Big] \ , \label{NJLStart2}
\end{eqnarray}
with $\lambda=0$ taken. In the previous works~\cite{Yasui:2016svc,Yasui:2017izi}, it was found that the reasonable ground state is realized by assuming the so-called {\it hedgehog ansatz} provided by
\begin{eqnarray}
\langle\bar{\psi}\Psi_v\rangle =\frac{1}{g}\Delta \ , \ \  \langle\bar{\psi}\gamma^i\Psi_v\rangle = \frac{1}{g}\Delta\hat{p}^i\  , \label{Ansatz}
\end{eqnarray}
with $\hat{p}^i\equiv p^i/|{\bm p}|$ defined in momentum space, where ${\bm p}$ is identical to the momentum of the heavy and light quarks.\footnote{Thus, $\hat{p}^i$ in Eq.~(\ref{Ansatz}) is not used to minimize the thermodynamic potential.} In Eq.~(\ref{Ansatz}), $\Delta$ serves as the Kondo condensate made of a light quark and a heavy quark. The value of $\Delta$ should be determined by a variational method on the free energy of the system. Despite the violation of $O(3)$ rotational invariance due to the magnetic field, we employ the ansatz~(\ref{Ansatz}) as an appropriate configuration of the Kondo effect since we make use of the linear response theory for a weak magnetic field. 

The nontrivial phase characterized by $\Delta\neq0$ is called the {\it Kondo phase}, whereas the trivial phase by $\Delta=0$ is called the {\it normal phase}.\footnote{Due to the ansatz~(\ref{Ansatz}), only positive-energy parts of light and heavy quarks are correlated to form the Kondo condensate, as is explicitly shown in Eq.~(\ref{Dispersions}). Such a condensate is reasonable since the heavy quark carries only positive-energy parts due to its heavy mass, and for the light quark, the Fermi sea is sufficiently formed where antiparticle contributions are rather suppressed. Therefore, the Kondo phase with the ansatz~(\ref{Ansatz}) is expected to be one of the thermodynamically preferred phases~\cite{Yasui:2017izi}.}

From the Lagrangian~(\ref{NJLStart2}) together with the ansatz~(\ref{Ansatz}), the inverse of Green's function of the fermions in the absence of a gauge field is read as
\begin{eqnarray}
i{\cal G}_0^{-1}(p_0,{\bm p})  =\left(
\begin{array}{ccc}
p_0+\mu & -{\bm p}\cdot{\bm \sigma} & \Delta^* \\
{\bm p}\cdot{\bm \sigma} & -p_0-\mu & -\Delta^*\hat{\bm p}\cdot{\bm \sigma} \\
\Delta & \Delta\hat{\bm p}\cdot{\bm \sigma} & p_0  \\ 
\end{array}
\right) \ .\label{G0Inverse}
 \end{eqnarray}
This Green's function is a $6\times6$ matrix since the light and heavy quarks are four-component and two-component spinors, respectively. The inverse of Eq.~(\ref{G0Inverse}) yields the Green's function which is of the form~\cite{Suenaga:2020oeu}
\begin{eqnarray}
\tilde{\cal G}_0(p_0,{\bm p}) = \left(
\begin{array}{cc}
\tilde{\cal G}_0^{\bar{\psi}\psi} (p_0,{\bm p}) & \tilde{\cal G}_0^{\bar{\psi}\Psi_v}  (p_0,{\bm p}) \\
\tilde{\cal G}_0^{{\Psi}_v^\dagger\psi} (p_0,{\bm p}) & \tilde{\cal G}_0^{{\Psi}_v^\dagger\Psi_v} (p_0,{\bm p}) \\  
\end{array}
\right) \ ,\label{GZeroAnother}
\end{eqnarray}
with the elements 
\begin{eqnarray}
\tilde{\cal G}_0^{\bar{\psi}\psi} (p_0,{\bm p})  &=& i\left[\frac{U_{+}({\bm p})}{p_0-E_{\bm p}^{+}}+ \frac{U_{-}({\bm p})}{p_0-E_{\bm p}^{-}}\right]\Lambda_{\rm p} \nonumber\\
&&  + i\frac{U_{\rm a}({\bm p})}{p_0-{E}_{\bm p}^{\rm a}}\Lambda_{\rm a} \, ,\nonumber\\
\tilde{\cal G}_0^{\bar{\psi}\Psi_v}(p_0,{\bm p}) &=& i\left[\frac{V^*_{+}({\bm p})}{p_0-E_{\bm p}^{+}}+ \frac{V^*_{-}({\bm p})}{p_0-E_{\bm p}^{-}}\right]\Lambda_{{\rm p}{\rm H}}\ , \nonumber\\
\tilde{\cal G}_0^{{\Psi}_v^\dagger\psi}(p_0,{\bm p}) &=& i\left[\frac{V_{+}({\bm p})}{p_0-E_{\bm p}^{+}}+ \frac{V_{-}({\bm p})}{p_0-E_{\bm p}^{-}}\right]\Lambda_{{\rm H}{\rm p}}\ , \nonumber\\
\tilde{\cal G}_0^{{\Psi}_v^\dagger\Psi_v}(p_0,{\bm p}) &=& i\left[\frac{W_{+}({\bm p})}{p_0-E_{\bm p}^{+}}+ \frac{W_{-}({\bm p})}{p_0-E_{\bm p}^{-}}\right]{\bm 1}\ . \label{GZeroElements}
\end{eqnarray}
The matrices $\Lambda_{\rm p}$ and $\Lambda_{\rm a}$ are the projection operators for positive- and negative-energy states of the light quark defined by 
\begin{eqnarray}
\Lambda_{\rm p} \equiv \frac{1+\hat{\bm p}\cdot{\bm \alpha}}{2}\gamma_0\ , \ \ \Lambda_{\rm a} \equiv \frac{1-\hat{\bm p}\cdot{\bm \alpha}}{2}\gamma_0\ ,
\end{eqnarray}
with ${\bm \alpha} = \gamma_0{\bm \gamma}$, respectively, and ${\bm 1}$ in Eq.~(\ref{GZeroElements}) is a $2\times2$ unit matrix. Similarly, $\Lambda_{\rm pH}$ and $\Lambda_{\rm Hp}$ are the operators mixing the positive-energy states of light and heavy quarks defined by
\begin{eqnarray}
\Lambda_{{\rm p}{\rm H}} \equiv \left(
\begin{array}{c}
1  \\
\hat{\bm p}\cdot{\bm \sigma} \\
\end{array}
\right)\ , \ \ \Lambda_{{\rm H}{\rm p}} \equiv \left(
\begin{array}{cc}
1  & -\hat{\bm p}\cdot{\bm \sigma} \\
\end{array}
\right)\ ,
\end{eqnarray}
respectively. The dispersion relations $E_{\bm p}^+$, $E_{\bm p}^-$, and $E_{\bm p}^{\rm a}$ are given by
\begin{eqnarray}
 E_{\bm p}^{+} &=& \frac{1}{2}\left(|{\bm p}|-\mu + \sqrt{(|{\bm p}|-\mu)^2+8|\Delta|^2}\right)\ , \nonumber\\
E_{\bm p}^{-} &=& \frac{1}{2}\left(|{\bm p}|-\mu - \sqrt{(|{\bm p}|-\mu)^2+8|\Delta|^2}\right)\ , \nonumber\\
E^{\rm a}_{\bm p} &=& -|{\bm p}|-\mu\ .  \label{Dispersions}
\end{eqnarray}
In the following analysis, the modes carrying the dispersions $E_{\bm p}^+$, $E_{\bm p}^-$, and $E_{\bm p}^{\rm a}$ are referred to as the Kondo quasiparticle (``q.p.''), the Kondo quasihole (``q.h.''), and the light antiparticle (``a.p.''), respectively. The factors $U({\bm p})$'s, $V({\bm p})$'s, and $W({\bm p})$'s in Eq.~(\ref{GZeroElements}) are 
\begin{eqnarray}
&&U_{+}({\bm p})= \frac{2 (|\Delta|^2+|{\bm p}|E_{\bm p}^{+})}{(E_{\bm p}^{-}-E_{\bm p}^{+})(E_{\bm p}^{\rm a}-E_{\bm p}^{+})} \ , \nonumber\\
&& U_{-}({\bm p}) = \frac{2 (|\Delta|^2+|{\bm p}|E_{\bm p}^{-})}{(E_{\bm p}^{+}-E_{\bm p}^{-})(E_{\bm p}^{\rm a}-E_{\bm p}^{-})} \ , \nonumber\\
&& U_{\rm a}({\bm p}) = 1\ , \nonumber\\
&& V_{+}({\bm p}) = \frac{\Delta}{E_{\bm p}^{-}-E_{\bm p}^{+}} \ , \ \ V_{-}({\bm p}) = \frac{\Delta}{E_{\bm p}^{+}-E_{\bm p}^{-}} \ , \nonumber\\
 && W_{+}({\bm p}) = \frac{E_{\bm p}^{+}-|{\bm p}|+\mu}{E_{\bm p}^{+}-E_{\bm p}^{-}} \ , \ \ W_{-}({\bm p}) = \frac{E_{\bm p}^{-}-|{\bm p}|+\mu}{E_{\bm p}^{-}-E_{\bm p}^{+}}\ . \nonumber\\ \label{WeightFactors}
\end{eqnarray}
The dispersion relations~(\ref{Dispersions}) and the factors~(\ref{WeightFactors}) clearly show that only the positive-energy component of light quarks couples to the Kondo condensate. It should be noted that $U_\pm({\bm p})$ and $W_\pm({\bm p})$ satisfy $U_+({\bm p})+U_-({\bm p})=1$ and $W_+({\bm p})+W_-({\bm p})=1$, respectively.

\begin{figure}[t]
\centering
\includegraphics*[scale=0.45]{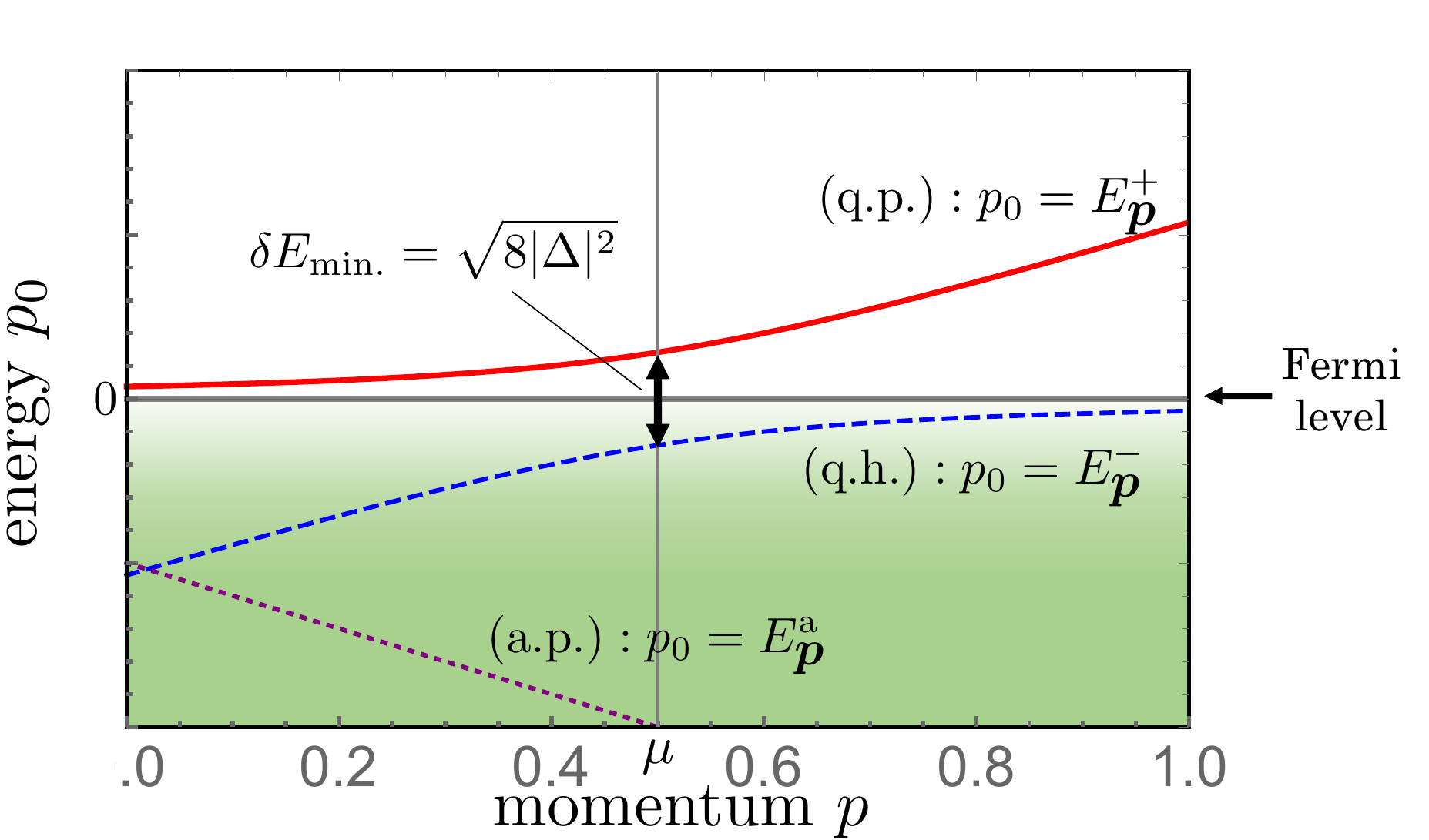}
\caption{A schematic picture of the dispersion relations of q.p., $p_0=E_{\bm p}^+$ (red solid line), q.h., $p_0=E_{\bm p}^-$ (blue dashed line), and a.p., $p_0=E_{\bm p}^{\rm a}$ (purple dotted line).}
\label{fig:Dispersions}
\end{figure}

A schematic picture of the dispersion relations of q.p. (red solid line), q.h. (blue dashed line), and a.p. (purple dotted line) is shown in Fig.~\ref{fig:Dispersions}. This figure shows that q.p. always lives above the Fermi level, while q.h. and a.p. are below the Fermi level. The energy gap between q.p. and q.h. at the identical momentum ${\bm p}$ satisfies 
\begin{eqnarray}
\delta E &\equiv& E_{\bm p}^+- E_{\bm p}^- \nonumber\\
&=& \sqrt{(|{\bm p}|-\mu)^2+8|\Delta|^2}  \nonumber\\
&\geq& \sqrt{8|\Delta|^2}\ ,
\end{eqnarray}
where the minimum value is given at $|{\bm p}|=\mu$. In other words, the threshold energy of pair creation or pair annihilation of q.p. $+$ q.h. at rest frame is found to be $\delta E_{\rm min.} = \sqrt{8|\Delta|^2}$ as indicated in Fig.~\ref{fig:Dispersions}.

From the Lagrangian~(\ref{NJLStart2}) with the ansatz~(\ref{Ansatz}) inserted, the thermodynamic potential in the absence of the gauge field is given by
\begin{eqnarray}
\Omega/V &=& -N_c\sum_{\zeta=+,-,{\rm a}}\int^{\Lambda}\frac{d^3p}{(2\pi)^3}\Bigg[{E}_{\bm p}^{\zeta}+\frac{2}{\beta}{\rm ln}\left(1+{\rm e}^{-\beta{E}^{\zeta}_{\bm p}}\right) \Bigg] \nonumber\\
&&  + \frac{2}{g}|\Delta|^2\ , \label{Thermodynamic}
\end{eqnarray}
with the three-dimensional volume and the inverse temperature, $V$ and $\beta=1/T$, respectively. The cutoff parameter $\Lambda$ is included to regularize the integral, and $N_c$ is the number of colors which is taken to be $N_c=3$. By taking a derivative of the thermodynamic potential~(\ref{Thermodynamic}) with respect to $\Delta$, the gap equation that the Kondo condensate satisfies is derived as
\begin{eqnarray}
2N_c\int^\Lambda\frac{d^3p}{(2\pi)^3}\frac{f_F(E_{\bm p}^+)-f_F(E_{\bm p}^-)}{E_{\bm p}^+-E_{\bm p}^-} + \frac{1}{g}= 0\ , \label{GapEquation}
\end{eqnarray}
with the Fermi-Dirac distribution function $f_F(\epsilon) = 1/({\rm e}^{\beta \epsilon}+1)$. At zero temperature, $f_F(E_{\bm p}^+)-f_F(E_{\bm p}^-) \to -1$, and hence an approximate solution of Eq.~(\ref{GapEquation}) at $T=0$ can be found as~\cite{Yasui:2016svc,Yasui:2017izi}
\begin{eqnarray}
\Delta \sim \Lambda {\rm exp}\left(-\frac{1}{N_c g\rho(\mu)}\right)\ , \label{DeltaBCS}
\end{eqnarray}
with the density of states of a massless quark at the Fermi level, $\rho(\mu) = \frac{\mu^2}{2\pi^2}$, under assumptions of $\Delta\ll \mu,\Lambda$ and $N_cg\rho(\mu)\ll 1$. The solution~(\ref{DeltaBCS}) shows that the Kondo condensate $\Delta$ significantly appears in cold dense quark matter. The integration in Eq.~(\ref{GapEquation}) is mostly governed by modes near the Fermi sphere for the noninteracting case $|{\bm p}|\sim\mu$. This fact allows us to estimate the temperature dependence of $\Delta$ in an easy way. Namely, the factor by distribution functions $f_F(E_{\bm p}^+)-f_F(E_{\bm p}^-)$ at the Fermi sphere turns into
\begin{eqnarray}
\Big[f_F(E_{\bm p}^+)-f_F(E_{\bm p}^-)\Big]_{|{\bm p}|\sim\mu} \sim 2f_F\big(\sqrt{2}|\Delta|\big) -1 \ .\label{TDepKondo}
\end{eqnarray}
Here, if $T \ll\sqrt{2}\Delta$, then thermal fluctuations are rather suppressed, leading to 
\begin{eqnarray}
\Big[f_F(E_{\bm p}^+)-f_F(E_{\bm p}^-)\Big]_{|{\bm p}|\sim\mu} \to  -1 \ ,\label{TDepKondo2}
\end{eqnarray}
which is the same value as that at $T=0$. In other words, because of the gapped nature of q.p. and q.h., at lower temperature, the magnitude of $\Delta$ does not change significantly from that at zero temperature. Numerically such a tendency was found in Ref.~\cite{Suzuki:2017gde}. 

We employ the gap equation~(\ref{GapEquation}) to determine the size of the Kondo condensate in the following analysis, since the magnetic field is treated perturbatively based on the linear response theory.

\section{Analytic evaluation}
\label{sec:Calculation}

In this section, we give analytic evaluation of the HQSP induced by the Kondo effect under a magnetic field with vertex corrections required by the $U(1)_{\rm EM}$ gauge symmetry.

\subsection{General properties}
\label{sec:Genaral}

The emergence of the HQSP is described by a thermodynamic expectation value of a spin polarization of the heavy quarks. For this purpose, we define
\begin{eqnarray}
\langle\tilde{S}_H^i(q_0,{\bm q}) \rangle_{\beta} &\equiv&  \int d^4x \langle\Psi^\dagger_v(t,{\bm x})S^i_h\Psi_v(t,{\bm x})\rangle_\beta {\rm e}^{iq\cdot x} \label{SpinStart}
\end{eqnarray}
with $S_h^i = \frac{\sigma^i}{2}$ being a spin operator of heavy quarks. Introducing six-component spinors $\Phi$ and $\bar{\Phi}$ containing the light- and heavy-quark fields, 
\begin{eqnarray}
\Phi \equiv \left(
\begin{array}{c}
\psi \\
\Psi_v \\
\end{array}
\right)\ , \ \ \bar{\Phi} \equiv \left(
\begin{array}{cc}
\bar{\psi} & \Psi_v^\dagger \\
\end{array}
\right)\ .
\end{eqnarray}
Equation~(\ref{SpinStart}) can be expressed in terms of $\Phi$ and $\bar{\Phi}$ as
\begin{eqnarray}
\langle\tilde{S}_H^i(q_0,{\bm q}) \rangle_{\beta} = \int d^4x \langle\bar{\Phi}(t,{\bm x})S_H^i\Phi(t,{\bm x})\rangle_\beta {\rm e}^{iq\cdot x} \ ,\label{STildeDef}
\end{eqnarray}
where $S_H^i$ is the $6\times6$ matrix spin operator of heavy quark
\begin{eqnarray}
S_H^i = \left(
\begin{array}{cc}
0 & 0 \\
0 & S_h^i \\
\end{array}
\right)\ .
\end{eqnarray}
One of the useful ways to calculate the expectation value~(\ref{STildeDef}) is the analytic continuation from the imaginary-time formalism~\cite{Kapusta:2006pm}. Namely, first we calculate the expectation value of $\langle\tilde{\cal S}^i_H (i\bar{\omega}_n,{\bm q})\rangle_\beta$ with $\bar{\omega}_n=2n\pi T$ the Matsubara frequency for bosons ($n$ is an integer), and next, we evaluate Eq.~(\ref{STildeDef}) by the following relation:
\begin{eqnarray}
\langle \tilde{\cal S}^i_H (q_0,{\bm q})\rangle_\beta = \langle \tilde{\cal S}^i_H (i\bar{\omega}_n,{\bm q})\rangle_\beta\Big|_{i\bar{\omega}_n \to q_0+i\eta}\ . \label{SAnalytic}
\end{eqnarray}
In this equation, $\eta$ is an infinitesimal positive number.

With the help of linear response theory, $\langle\tilde{\cal S}^i_H (i\bar{\omega}_n,{\bm q})\rangle_\beta$ at the leading order of the gauge field can be given as
\begin{eqnarray}
&& \langle \tilde{\cal S}^i_H (i\bar{\omega}_n,{\bm q})\rangle_\beta = -N_cT\sum_m\int^\Lambda\frac{d^3p}{(2\pi)^3} \nonumber\\
&& \times {\rm tr}\left[S_H^i \tilde{\cal G}_0(i\omega_m',{\bm p}_+)\Gamma_\mu\tilde{\cal G}_0(i\omega_m,{\bm p}_-)\right] \tilde{A}^\mu(i\bar{\omega}_n,{\bm q})\ , \label{SHImaginary}
\end{eqnarray}
in which we have defined $\tilde{A}^\mu(i\bar{\omega}_n,{\bm q})$ as the Fourier transformation of $A^\mu(t,{\bm x})$ within the imaginary-time formalism. Here, $\omega_m = (2m+1)\pi T$ is the Matsubara frequency for fermions ($m$ is an integer), and $i\omega_m' \equiv i\omega_m + i\bar{\omega}_n$, ${\bm p}_\pm\equiv {\bm p} \pm\frac{ {\bm q}}{2}$.\footnote{Here, we have defined the spatial part of loop momenta symmetrically as ${\bm p}_\pm\equiv {\bm p} \pm\frac{ {\bm q}}{2}$. This symmetric choice is useful in deriving the vertex corrections for $\Gamma_\mu$ in Sec.~\ref{sec:GaugeInv}.} Note that $\Gamma_\mu$ in Eq.~(\ref{SHImaginary}) is the vertex function responsible for the coupling between the fermions and the gauge field. Regarding the spin polarization~(\ref{SHImaginary}), the response function of the HQSP to a magnetic field $\tilde{\Pi}_H(i\bar{\omega}_n,{\bm q})$ is defined by
\begin{eqnarray}
 \langle \tilde{\cal S}^i_H (i\bar{\omega}_n,{\bm q})\rangle_\beta = e\tilde{B}^i(i\bar{\omega}_n,{\bm q}) \tilde{\Pi}_H(i\bar{\omega}_n,{\bm q})  \ , \label{PiHDefine}
\end{eqnarray}
where $\tilde{B}^i(i\bar{\omega}_n,{\bm q})$ is the magnetic field generated by the gauge field $\tilde{A}^\mu(i\bar{\omega}_n,{\bm q})$.

\subsection{Gauge invariance and vertex corrections}
\label{sec:GaugeInv}

As pointed out in our previous work~\cite{Suenaga:2020oeu}, naive adoption of the bare vertex read by the Lagrangian in Eq.~(\ref{NJLStart}) or Eq.~(\ref{NJLStart2}) leads to the violation of $U(1)_{\rm EM}$ gauge symmetry. One useful way to cure this problem is to correct the vertex $\Gamma^\mu$ in an {\it ad hoc} manner such that it recovers the $U(1)_{\rm EM}$ invariance. Here, we show the detailed procedure of this treatment in the real time.

The $U(1)_{\rm EM}$ gauge invariance of the Lagrangian~(\ref{NJLStart}) yields the Ward-Takahashi identity for the vertex ($\Gamma^\mu$) between the fermions and the gauge field as~\cite{Nambu:1960tm}
\begin{eqnarray}
q_\mu\Gamma^\mu &=&  i\tilde{\cal G}_0^{-1}(p_0^+,{\bm p}_+)Q -Q i\tilde{\cal G}_0^{-1}(p_0^-,{\bm p}_-) \ ,\label{WTIVertexQ}
\end{eqnarray}
with $p^\pm_0 = p_0\pm\frac{q_0}{2}$, where $\tilde{\cal G}^{-1}_0(p_0,{\bm p})$ has been defined in Eq.~(\ref{G0Inverse}), and $Q$ is the charge matrix
\begin{eqnarray}
Q \equiv \left(
\begin{array}{cc}
e_q & 0 \\
0 & e_Q \\
\end{array}
\right)\ . \label{ChargeMatrix}
\end{eqnarray}
Therefore, by denoting the vertex by
\begin{eqnarray}
\Gamma^\mu \equiv \left(
\begin{array}{cc}
\Gamma^\mu_{A\bar{\psi}\psi} & \Gamma^\mu_{A\bar{\psi}\Psi_v} \\
\Gamma^\mu_{A\Psi_v^\dagger\psi} & \Gamma^\mu_{A\Psi_v^\dagger\Psi_v} \\
\end{array}
\right)\ , \label{VertexMatrix}
\end{eqnarray}
the identity~(\ref{WTIVertexQ}) is explicitly given by
\begin{eqnarray}
&& q_\mu\Gamma^\mu_{A\bar{\psi}\psi} = e_q\Slash{q}\ ,  \label{WTIEach1} \\
&& q_\mu \Gamma^\mu_{A\bar{\psi}\Psi_v}=\Delta^* \left(
\begin{array}{c}
e_Q-e_q  \\
-e_Q\hat{\bm p}_+\cdot{\bm \sigma} + e_q\hat{\bm p}_-\cdot{\bm \sigma}   \\
\end{array}
\right)\ , \label{WTIEach2} \\
&&q_\mu\Gamma^\mu_{A\Psi_v^\dagger\psi} = \Delta\left(
\begin{array}{cc}
e_q-e_Q &   e_q \hat{\bm p}_+\cdot{\bm \sigma}-e_Q\hat{\bm p}_-\cdot{\bm \sigma}  \\
\end{array}
\right)\ , \label{WTIEach3} \\
&&q_\mu\Gamma^\mu_{A\Psi_v^\dagger\Psi_v} =e_Qq_0{\bm 1}\ . \label{WTIEach4}
\end{eqnarray}
Equations~(\ref{WTIEach1}) and~(\ref{WTIEach4}) show that we can safely employ the bare vertex as 
\begin{eqnarray}
\Gamma^\mu_{A\bar{\psi}\psi}  = e_q\gamma^\mu \ , \ \ \Gamma^\mu_{A\Psi_v^\dagger\Psi_v} = e_Q \delta^{\mu 0}\ , \label{GammaBareOK}
\end{eqnarray}
for the diagonal components of Eq.~(\ref{VertexMatrix}). On the other hand, the off-diagonal ones no longer vanish and have to be corrected to satisfy Eqs.~(\ref{WTIEach2}) and~(\ref{WTIEach3}).

In a later analysis, we study the HQSP for a small external momentum ${\bm q} \ll {\bm p}$. This restriction is reasonable since the loop integral in Eq.~(\ref{SHImaginary}) is dominated by modes around $|{\bm p}_-|\sim |{\bm p}_+| \sim \mu$ or those having sufficiently large density of states; the modes close to ${\bm p}\sim{\bm 0}$ do not contribute to the HQSP significantly. Hence, we expand the right-hand sides (rhs's) in Eqs.~(\ref{WTIEach2}) and~(\ref{WTIEach3}) up to ${\cal O}({\bm q}^1)$.\footnote{In order to treat this expansion symmetrically we have defined the spatial part of loop momenta as in Eq.~(\ref{SHImaginary}): ${\bm p}_\pm = {\bm p}\pm\frac{{\bm q}}{2}$.} In this approximation, the identities~(\ref{WTIEach2}) and~(\ref{WTIEach3}) are reduced to
\begin{eqnarray}
q_\mu \Gamma^\mu_{A\bar{\psi}\Psi_v} &\approx& \Delta^*(e_Q-e_q) \left(
\begin{array}{c}
{\bm 1} \\
\hat{\bm p}\cdot{\bm \sigma}   \\
\end{array}
\right)  \nonumber\\
&-& \Delta^* (e_q+e_Q) \left(
\begin{array}{c}
0  \\
\frac{{\bm q}\cdot{\bm \sigma}-(\hat{\bm p}\cdot{\bm \sigma})(\hat{\bm p}\cdot{\bm q})}{2|{\bm p}|}    \\
\end{array}
\right)\ , \label{WTIEach2_2} 
\end{eqnarray}
and
\begin{eqnarray}
q_\mu\Gamma^\mu_{A\Psi_v^\dagger\psi} &\approx&  \Delta (e_q-e_Q)\left(
\begin{array}{cc}
{\bm 1} &  \hat{\bm p}\cdot{\bm \sigma} \\
\end{array}
\right)  \nonumber\\
&+&  \Delta (e_q+e_Q)\left(
\begin{array}{cc}
0&  \frac{{\bm q}\cdot{\bm \sigma}-(\hat{\bm p}\cdot{\bm \sigma})(\hat{\bm p}\cdot{\bm q})}{2|{\bm p}|} \\
\end{array}
\right) \ , \label{WTIEach3_2} 
\end{eqnarray}
respectively. When we take a $q^\mu\to0$ limit in Eqs.~(\ref{WTIEach2_2}) and~(\ref{WTIEach3_2}), the second terms of the rhs's vanish while the first terms survive. The nonvanishing behavior of the first term implies that radiative contributions by a massless mode couple to the vertices. Namely, the first terms in Eqs.~(\ref{WTIEach2_2}) and~(\ref{WTIEach3_2}) are responsible for the Nambu-Goldstone (NG) mode contributions~\cite{Nambu:1960tm}. In fact, such contributions arise only when the Kondo condensate is electrically charged ($e_q\neq e_Q$) where the $U(1)_{\rm EM}$ symmetry is spontaneously broken.\footnote{The vertices stemming from the NG mode contributions are explicitly given in Eqs.~(\ref{NGApp1}) and~(\ref{NGApp2}).} On the other hand, the second terms in Eqs.~(\ref{WTIEach2_2}) and~(\ref{WTIEach3_2}) are important even when the condensate is electrically neutral ($e_q=e_Q$) since they are proportional to $e_q+e_Q$. Namely, regardless of the spontaneous breakdown of the $U(1)_{\rm EM}$ symmetry, the vertex corrections are necessary. Those contributions appear due to the nontrivial momentum dependence of the hedgehog ansatz in Eq.~(\ref{Ansatz}).

According to discussions in Appendix~\ref{sec:WTISol}, general solutions of Eqs.~(\ref{WTIEach2_2}) and~(\ref{WTIEach3_2}) relevant to the magnetic response are given by
\begin{eqnarray}
\Gamma^\mu_{A\bar{\psi}\Psi_v} &=& \frac{\Delta^*(e_q+e_Q)}{2|{\bm p}|^3}\delta^{\mu j} \left(
\begin{array}{c}
0  \\
({\bm p}_+\cdot{\bm p})\sigma^j-({\bm p}_+\cdot{\bm \sigma})p^j  \\
\end{array}
\right) \nonumber\\
&+& \delta^{\mu j}\hat{\Gamma}^-_{A\bar{\psi}\Psi_v} \left(
\begin{array}{c}
0  \\
({\bm p}\cdot{\bm q})\sigma^j-({\bm q}\cdot{\bm \sigma})p^j  \\
\end{array}
\right)  \ ,   \label{GammaCorrect1}
\end{eqnarray}
and
\begin{eqnarray}
\Gamma^\mu_{A\Psi_v^\dagger\psi} &=&  -\frac{\Delta (e_q+e_Q)}{2|{\bm p}|^3}\delta^{\mu j}\left(
\begin{array}{cc}
0 &  ({\bm p}_-\cdot{\bm p})\sigma^j-({\bm p}_-\cdot{\bm \sigma})p^j \\
\end{array}
\right)  \nonumber\\
&-&\delta^{\mu j}\hat{\Gamma}^+_{A\Psi_v^\dagger\psi}\left(
\begin{array}{cc}
0 & ({\bm p}\cdot{\bm q})\sigma^j-({\bm q}\cdot{\bm \sigma})p^j  \\
\end{array}
\right)   \ ,  \label{GammaCorrect2}
\end{eqnarray}
respectively. In these equations, $\hat{\Gamma}^-_{A\bar{\psi}\Psi_v}$ and $\hat{\Gamma}^+_{A\Psi_v^\dagger\psi}$ are arbitrary scalar functions of ${\bm p}$ and ${\bm q}$ which cannot be fixed (for more detail, see Appendix~\ref{sec:WTISol}). They appear because the identities~(\ref{WTIEach2_2}) and~(\ref{WTIEach3_2}) include the combinations of $q_\mu{\Gamma}^\mu_{A\bar{\psi}\Psi_v}$ and $q_\mu{\Gamma}^\mu_{A\Psi_v^\dagger\psi}$. In other words, $\hat{\Gamma}^-_{A\bar{\psi}\Psi_v}$ and $\hat{\Gamma}^+_{A\Psi_v^\dagger\psi}$ could be regarded as ``integration constants''. In the following analysis, we choose those values to be zero:
\begin{eqnarray}
\hat{\Gamma}^-_{A\bar{\psi}\Psi_v} = \hat{\Gamma}^+_{A\Psi_v^\dagger\psi} = 0\ . \label{ConstantZero}
\end{eqnarray}
This choice preserves the Hermiticity of the HQSP by the vertex corrections. Here, we note that the vertices proportional to $q^j$ have not been included in Eqs.~(\ref{GammaCorrect1}) and~(\ref{GammaCorrect2}) since such contributions vanish due to the transversality of the vertices as explained in Appendix~\ref{sec:NGVertex}. For the same reason, the NG mode contributions disappear because they are longitudinal.

In this subsection, we have shown that the $U(1)_{\rm EM}$ invariance requires us to employ the corrected vertices~(\ref{GammaCorrect1}) and~(\ref{GammaCorrect2}) in addition to the bare ones~(\ref{GammaBareOK}). Based on them, in Sec.~\ref{sec:Proceed}, we proceed with evaluation of the HQSP~(\ref{SHImaginary}).

\subsection{Evaluation of Eq.~(\ref{SHImaginary})}
\label{sec:Proceed}
Here, we proceed with analytical evaluation of the HQSP in Eq.~(\ref{SHImaginary}) with the corrected vertices obtained in Sec~\ref{sec:GaugeInv}.

Since the magnetic field enters only the spatial components of $A_\mu$, we take $A_0=0$. Thus using Eq.~(\ref{VertexMatrix}), Eq.~(\ref{SHImaginary}) is given by the following two parts:
\begin{eqnarray}
 \langle \tilde{\cal S}^i_H (i\bar{\omega}_n,{\bm q})\rangle_\beta  =  \langle \tilde{\cal S}^i_{0H} (i\bar{\omega}_n,{\bm q})\rangle_\beta  +  \langle \tilde{\cal S}^i_{\delta H} (i\bar{\omega}_n,{\bm q})\rangle_\beta \ . \nonumber\\
 \label{SHCalculation1}
\end{eqnarray}
Here $\langle \tilde{\cal S}^i_{0H} (i\bar{\omega}_n,{\bm q})\rangle_\beta$ stands for the HQSP obtained through the bare vertex in Eq.~(\ref{GammaBareOK}) as
\begin{widetext}
\begin{eqnarray}
 \langle \tilde{\cal S}^i_{0H} (i\bar{\omega}_n,{\bm q})\rangle_\beta = N_c\frac{A^j(i\bar{\omega}_n,{\bm q}) }{2}T\sum_m\int^\Lambda\frac{d^3p}{(2\pi)^3} {\rm tr}\Big[\sigma^i \tilde{\cal G}_0^{\Psi_v^\dagger\psi}(i\omega_m',{\bm p}_+)\Gamma^j_{A\bar{\psi}\psi}  \tilde{\cal G}_0^{\bar{\psi}\Psi_v}(i\omega_m,{\bm p}_-)  \Big] \ , \label{SHCalculationBare1}
\end{eqnarray}
and $\langle \tilde{\cal S}^i_{\delta H} (i\bar{\omega}_n,{\bm q})\rangle_\beta$ through the vertex corrections in Eqs.~(\ref{GammaCorrect1}) and~(\ref{GammaCorrect2}) with Eq.~(\ref{ConstantZero}) inserted as
\begin{eqnarray}
 \langle \tilde{\cal S}^i_{\delta H} (i\bar{\omega}_n,{\bm q})\rangle_\beta &=& N_c\frac{A^j(i\bar{\omega}_n,{\bm q}) }{2}T\sum_m\int^\Lambda\frac{d^3p}{(2\pi)^3}\nonumber\\
&\times&  {\rm tr}\Big[\sigma^i \tilde{\cal G}_0^{\Psi_v^\dagger \psi}(i\omega_m',{\bm p}_+) \Gamma^j_{A\bar{\psi}\Psi_v}\tilde{\cal G}_0^{\Psi_v^\dagger \Psi_v}(i\omega_m,{\bm p}_-) + \sigma^i \tilde{\cal G}_0^{\Psi_v^\dagger \Psi_v}(i\omega_m',{\bm p}_+)\Gamma^j_{A\Psi_v^\dagger\psi}\tilde{\cal G}_0^{\bar{\psi} \Psi_v}(i\omega_m,{\bm p}_-)\Big]\ .  \label{SHCalculationCorr1}
\end{eqnarray}
\end{widetext}
Correspondingly, the HQSP response function to the magnetic field given in Eq.~(\ref{PiHDefine}) is separated as
\begin{eqnarray}
\tilde{\Pi}_{H} (i\bar{\omega}_n,{\bm q}) = \tilde{\Pi}_{0H} (i\bar{\omega}_n,{\bm q}) + \tilde{\Pi}_{\delta H} (i\bar{\omega}_n,{\bm q})\ ,
\end{eqnarray}
with each response function defined by
\begin{eqnarray}
\langle \tilde{\cal S}^i_{0H} (i\bar{\omega}_n,{\bm q})\rangle_\beta &=& e \tilde{B}^i(i\bar{\omega}_n,{\bm q})   \tilde{\Pi}_{0H} (i\bar{\omega}_n,{\bm q})\ ,  \label{PiHZeroDefine}\\
\langle \tilde{\cal S}^i_{\delta H} (i\bar{\omega}_n,{\bm q})\rangle_\beta &=& e \tilde{B}^i(i\bar{\omega}_n,{\bm q})   \tilde{\Pi}_{\delta H} (i\bar{\omega}_n,{\bm q})\ . \label{PiHDeltaDefine}
\end{eqnarray}

In Eqs.~(\ref{SHCalculationBare1}) and~(\ref{SHCalculationCorr1}), the contribution proportional to $\tilde{\cal G}_0^{\Psi_v^\dagger \Psi_v}\Gamma^j_{A\Psi_v^\dagger\Psi_v}\tilde{\cal G}_0^{\Psi_v^\dagger \Psi_v}$ does not exist. This means that the magnetic field does not couple to the heavy quark directly; the HQSP in the absence of Kondo condensate vanishes. It is also easily understood by the lack of magnetic coupling between the heavy quark and the gauge field in Eq.~(\ref{NJLStart}). 

In the following calculation, we treat the bare-vertex part~(\ref{SHCalculationBare1}) in detail as a demonstration. From the Green's function in Eq.~(\ref{GZeroElements}) together with the bare vertex in Eq.~(\ref{GammaBareOK}), the calculation of Eq.~(\ref{SHCalculationBare1}) is further proceeded as
\begin{eqnarray}
  \langle \tilde{\cal S}^i_{0H} (i\bar{\omega}_n,{\bm q})\rangle_\beta 
&=&-e_qN_c\frac{A^j(i\bar{\omega}_n,{\bm q})}{2} \sum_{\zeta,\zeta'=\pm} \int^\Lambda\frac{d^3p}{(2\pi)^3} \nonumber\\
 &\times& {\cal I}_{0H}^{\zeta\zeta'}({\bm p}_+;{\bm p}_-){\rm tr}[\sigma^i\Lambda_{\rm Hp}({\bm p}_+)\gamma^j\Lambda_{\rm pH}({\bm p}_-)]   \ ,  \nonumber\\
 \label{SHCalculationBare2}
\end{eqnarray} 
with 
\begin{eqnarray}
{\cal I}_{0H}^{\zeta\zeta'}({\bm p}_+;{\bm p}_-) &\equiv&  \frac{V_\zeta({\bm p}_+)V^*_{\zeta'}({\bm p}_-)}{i\bar{\omega}_n-E_{{\bm p}_+}^{\zeta}+E_{{\bm p}_-}^{\zeta'}} \nonumber\\
&\times&\left[{f}_F(E_{{\bm p}_-}^{\zeta'})-{f}_F(E_{{\bm p}_+}^{\zeta})\right]\ . 
\label{Size}
\end{eqnarray}
In getting Eq.~(\ref{SHCalculationBare2}), we have made use of the Matsubara summation formula 
\begin{eqnarray}
T\sum_m \frac{1}{(i\omega'_m-\epsilon_{{\bm p}_+})(i\omega_m-\epsilon_{{\bm p}_-})} = \frac{f_F(\epsilon_{{\bm p}_-})-f_F(\epsilon_{{\bm p}_+})}{i\bar{\omega}_n-\epsilon_{{\bm p}_+}+\epsilon_{{\bm p}_-}}\, .\nonumber\\ \label{Matsubara}
\end{eqnarray}
The constituent~(\ref{Size}) includes information on the quantum and thermal fluctuations of each process upon the Kondo effect, which is not specific to the response to a magnetic field. Namely, the constituent~(\ref{Size}) always appears when we evaluate such fluctuations at one loop regardless of the detail of interactions. On the other hand, the kinetic factor ${\rm tr}[\sigma^i\Lambda_{\rm Hp}({\bm p}_+)\gamma^j\Lambda_{\rm pH}({\bm p}_-)] $ in Eq.~(\ref{SHCalculationBare2}) reflects information on the couplings among the spins of the light quark, heavy quark, and gauge field, contributing to the HQSP.

\begin{figure*}[htbp]
\centering
\includegraphics*[scale=0.55]{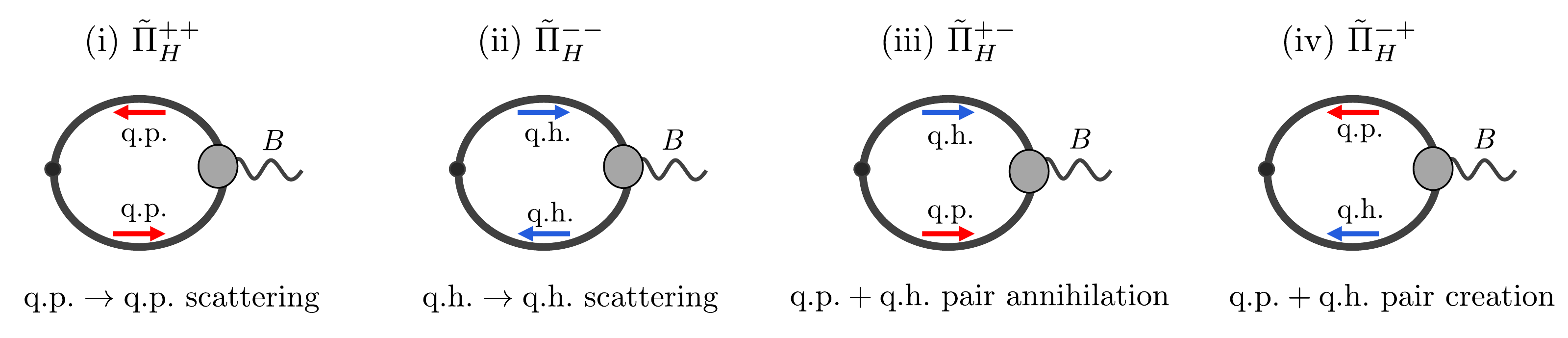}
\caption{The Feynman diagrams contributing to the HQSP. Blobs represent the corrected vertices required by $U(1)_{\rm EM}$ gauge invariance. The diagrams (i) and (ii) correspond to the intraband processes while (iii) and (iv) the interband ones.}
\label{fig:Diagrams}
\end{figure*}

By making use of a trace formula
\begin{eqnarray}
{\rm tr}[\sigma^i\Lambda_{\rm Hp}({\bm p}_+)\gamma^j\Lambda_{\rm pH}({\bm p}_-)] = 2i\epsilon^{ijk}(\hat{p}_-^k-\hat{p}_+^k)\ ,
\end{eqnarray}
Eq.~(\ref{SHCalculationBare2}) turns into
\begin{eqnarray}
 \langle \tilde{\cal S}^i_{0H} (i\bar{\omega}_n,{\bm q})\rangle_\beta &=& -ie_qN_c\epsilon^{ijk}A^j(i\bar{\omega}_n,{\bm q}) \sum_{\zeta,\zeta'=\pm} \int^\Lambda\frac{d^3p}{(2\pi)^3} \nonumber\\
 &\times& {\cal I}_{0H}^{\zeta\zeta'}({\bm p}_+;{\bm p}_-)  (\hat{p}_-^k-\hat{p}_+^k) \nonumber\\
&=& e\tilde{B}^i(i\bar{\omega}_n,{\bm q})\sum_{\zeta,\zeta'=\pm}  \frac{\hat{e}_qN_c}{|{\bm q}|^2} \int^\Lambda\frac{d^3p}{(2\pi)^3}  \nonumber\\
 &\times& {\cal I}_{0H}^{\zeta\zeta'}({\bm p}_+;{\bm p}_-) (\hat{\bm p}_-\cdot{\bm q}-\hat{\bm p}_+\cdot{\bm q}) 
 \ . 
\label{SHCalculationBare3}
\end{eqnarray}
In obtaining the second equality, we have used $\tilde{B}^i(i\bar{\omega}_n,{\bm q}) = i\epsilon^{ijk}q^j\tilde{A}^k(i\bar{\omega}_n,{\bm q})$ and defined $\hat{e}_q$ via $e_q=\hat{e}_qe$. Therefore, comparing Eqs.~(\ref{PiHZeroDefine}) and~(\ref{SHCalculationBare3}), the response function  from the bare-vertex part $\tilde{\Pi}_{0H}(i\bar{\omega}_n,{\bm q})$ is evaluated as
\begin{eqnarray}
\tilde{\Pi}_{0H}(i\bar{\omega}_n,{\bm q}) &\equiv& \sum_{\zeta,\zeta'=\pm} \tilde{\Pi}_{0H}^{\zeta\zeta'}(i\bar{\omega}_n,{\bm q}) \ , 
\end{eqnarray}
with
\begin{eqnarray}
\tilde{\Pi}_{0H}^{\zeta\zeta'}(i\bar{\omega}_n,{\bm q}) &=& 
 \frac{\hat{e}_qN_c}{|{\bm q}|^2} \int^\Lambda\frac{d^3p}{(2\pi)^3}  \nonumber\\
 &\times& {\cal I}_{0H}^{\zeta\zeta'}({\bm p}_+;{\bm p}_-) (\hat{\bm p}_-\cdot{\bm q}-\hat{\bm p}_+\cdot{\bm q}) \ .  \label{PiH1}
\end{eqnarray}

In a similar way, the response function from the corrected-vertex part $\tilde{\Pi}_{\delta H}(i\bar{\omega}_n,{\bm q})$ in Eq.~(\ref{PiHDeltaDefine}) reads
\begin{eqnarray}
\tilde{\Pi}_{\delta H}(i\bar{\omega}_n,{\bm q}) &\equiv& \sum_{\zeta,\zeta'=\pm} \tilde{\Pi}_{\delta H}^{\zeta\zeta'}(i\bar{\omega}_n,{\bm q}) \ , 
\end{eqnarray}
with 
\begin{eqnarray}
\tilde{\Pi}_{\delta H}^{\zeta\zeta'}(i\bar{\omega}_n,{\bm q}) &=&
 \frac{(\hat{e}_q+\hat{e}_Q)N_c}{2|{\bm q}|^2}\int^\Lambda\frac{d^3p}{(2\pi)^3} \nonumber\\
&\times&\Bigg[\Delta^*{\cal I}_{\delta H+}^{\zeta\zeta'}({\bm p}_+;{\bm p}_-)\frac{({\bm p}_+\cdot{\bm p})(\hat{\bm p}_+\cdot{\bm q})}{|{\bm p}|^3}  \nonumber\\
&& -\Delta {\cal I}_{\delta H-}^{\zeta\zeta'}({\bm p}_+;{\bm p}_-)\frac{({\bm p}_-\cdot{\bm p})(\hat{\bm p}_-\cdot{\bm q})}{|{\bm p}|^3} \Bigg] \ ,  \nonumber\\
\label{PiH2}
\end{eqnarray}
where $\hat{e}_Q$ is defined via $e_Q=\hat{e}_Qe$. In Eq.~(\ref{PiH2}) we have defined the constituent 
\begin{eqnarray}
{\cal I}_{\delta H+}^{\zeta\zeta'}({\bm p}_+;{\bm p}_-) &\equiv& \frac{V_\zeta({\bm p}_+)W_{\zeta'}({\bm p}_-)}{i\bar{\omega}_n-E_{{\bm p}_+}^{\zeta}+E_{{\bm p}_-}^{\zeta'}} \nonumber\\
&\times&\left[{f}_F(E_{{\bm p}_-}^{\zeta'})-{f}_F(E_{{\bm p}_+}^{\zeta})\right]\ , \nonumber\\
{\cal I}_{\delta H-}^{\zeta\zeta'}({\bm p}_+;{\bm p}_-) &\equiv& \frac{W_\zeta({\bm p}_+)V^*_{\zeta'}({\bm p}_-)}{i\bar{\omega}_n-E_{{\bm p}_+}^{\zeta}+E_{{\bm p}_-}^{\zeta'}} \nonumber\\
&\times&\left[{f}_F(E_{{\bm p}_-}^{\zeta'})-{f}_F(E_{{\bm p}_+}^{\zeta})\right]\ , 
\label{SizeDelta}
\end{eqnarray}
that accounts for only quantum and thermal fluctuations stemming from the vertex corrections as an analogue of Eq.~(\ref{Size}). As explained in Sec.~\ref{sec:GaugeInv}, the vertex corrections are necessary although the NG mode does not couple to a magnetic field, because of the momentum dependence of the Kondo condensate. Indeed, Eq.~(\ref{PiH2}) accounts for such corrections to the HQSP.

The Feynman diagrams for $\tilde{\Pi}_H^{++}$, $\tilde{\Pi}_H^{--}$, $\tilde{\Pi}_H^{+-}$, and $\tilde{\Pi}_H^{-+}$ are depicted in Fig.~\ref{fig:Diagrams}. These contributions correspond to (i) ${\rm q.p.}\to {\rm q.p.}$ scattering, (ii) ${\rm q.h.}\to {\rm q.h.}$ scattering, (iii) ${\rm q.p.}+{\rm q.h.}$ pair annihilation, and (iv) ${\rm q.p.}+{\rm q.h.}$ pair creation, respectively. The diagrams (i) and (ii) are often referred to as the {\it intraband} processes, while the remaining (iii) and (iv) are as the {\it interband} processes.\footnote{The intra-band process means that exclusively either q.p. or q.h. participates in the loops. On the other hand, the inter-band process means that both q.p. and q.h. participate.} In this figure blobs represent the corrected vertices.

The HQSP $\langle\tilde{S}_H^i(q_0,{\bm q})\rangle_\beta$ in the real time can be obtained via the analytic continuation in Eq.~(\ref{SAnalytic}). In the same manner, the response function $\tilde{\Pi}_H(q_0,{\bm q})$ in the real time can be evaluated. In our present paper, we investigate the HQSP response functions with vertex corrections 
\begin{eqnarray}
\tilde{\Pi}_H(q_0,{\bm q}) = \tilde{\Pi}_{0H}(q_0,{\bm q}) + \tilde{\Pi}_{\delta H}(q_0,{\bm q})\ ,
\end{eqnarray}
and without them $\tilde{\Pi}_{0H}(q_0,{\bm q})$, to examine the importance of gauge invariance clearly. In what follows, we take a phase of $\Delta$ such that $\Delta$ is always real without loss of generality.

\section{The HQSP in the dynamical and static limits}
\label{sec:TwoLimits}

In our present paper, we investigate the HQSP response function for vanishing spatial momentum $\tilde{\Pi}_{0H}(q_0,{\bm 0})$ and $\tilde{\Pi}_{H}(q_0,{\bm 0})$ (timelike) and for vanishing frequency $\tilde{\Pi}_{0H}(0,{\bm q})$ and $\tilde{\Pi}_{H}(0,{\bm q})$ (spacelike) toward understanding of the HQSP in the two distinct momentum regions in the clearest way. Physically, the former (latter) describes the HQSP whose time dependence is faster (slower) than the equilibration of the spatial part of the system. In particular, it is useful to examine the response function in the {\it dynamical limit},
\begin{eqnarray}
\tilde{\Pi}_{0H}^{\rm dyn} &\equiv& \lim_{q_0\to 0}\tilde{\Pi}_{0H}(q_0,{\bm 0})\ , \nonumber\\
\tilde{\Pi}_{H}^{\rm dyn} &\equiv& \lim_{q_0\to0}\tilde{\Pi}_{H}(q_0,{\bm 0})\ ,\label{PiDynamical}
\end{eqnarray}
and in the {\it static limit}
\begin{eqnarray}
\tilde{\Pi}_{0H}^{\rm sta} &\equiv& \lim_{{\bm q}\to{\bm 0}}\tilde{\Pi}_{0H}(0,{\bm q})\ , \nonumber\\
\tilde{\Pi}_{H}^{\rm sta} &\equiv& \lim_{{\bm q}\to{\bm 0}}\tilde{\Pi}_{H}(0,{\bm q})\ ,\label{PiStatic}
\end{eqnarray}
in detail to see differences in the two momentum regimes~\cite{PhysRev.135.A1505}. Before moving on to numerical computations, in this section, we analytically study the dynamical and static response functions in Eqs.~(\ref{PiDynamical}) and~(\ref{PiStatic}).

\subsection{The bare-vertex parts $\tilde{\Pi}_{0H}$}
\label{sec:BareVertex}

Here, we evaluate analytically the HQSP response functions $\tilde{\Pi}_{0H}(q_0,{\bm q})$ from the bare-vertex parts. The $q_0$ (or $i\bar{\omega}_n$ in the imaginary time) dependence of the response functions is rather trivial as can be seen from Eq.~(\ref{Size}), and hence we expand $\tilde{\Pi}_{0H}(q_0,{\bm q})$ with respect to a small momentum ${\bm q}$.

From Eq.~(\ref{Size}), ${\cal I}_{0H}^{\zeta\zeta'}({\bm p}_+;{\bm p}_-)$ with small ${\bm q}$ for the intraband processes described by $\zeta=\zeta'$ can be evaluated as
\begin{eqnarray}
{\cal I}_{0H}^{\zeta\zeta}({\bm p}_+;{\bm p}_-) &\approx& -\frac{V_\zeta({\bm p})V_{\zeta}({\bm p})}{i\bar{\omega}_n-\frac{\partial E_{{\bm p}}^{\zeta}}{\partial|{\bm p}|}(\hat{\bm p}\cdot{\bm q})}\frac{\partial{f}_F(E_{{\bm p}}^{\zeta})}{\partial |{\bm p}|}(\hat{\bm p}\cdot{\bm q})\ , \nonumber\\
\end{eqnarray}
while for the interband ones described by $\zeta\neq\zeta'$ as
\begin{eqnarray}
{\cal I}_{0H}^{\zeta\zeta'}({\bm p}_+;{\bm p}_-) \approx  \frac{V_\zeta({\bm p})V_{\zeta'}({\bm p})}{i\bar{\omega}_n-E_{{\bm p}}^{\zeta}+E_{\bm p}^{\zeta'}}\left[{f}_F(E_{\bm p}^{\zeta'})-{f}_F(E_{{\bm p}}^{\zeta})\right]\ . \nonumber\\
\end{eqnarray}
Besides, the common kinetic factor $\hat{\bm p}_-\cdot{\bm q}-\hat{\bm p}_+\cdot{\bm q}$ is expanded as
\begin{eqnarray}
\hat{\bm p}_-\cdot{\bm q}-\hat{\bm p}_+\cdot{\bm q} = \frac{(\hat{\bm p}\cdot{\bm q})^2-|{\bm q}|^2}{|{\bm p}|} + {\cal O}({\bm q}^3)\ . \label{KineticBare}
\end{eqnarray}
Hence, the response functions with a small momentum ${\bm q}$ are of the forms
\begin{eqnarray}
\tilde{\Pi}_{0H}^{\zeta\zeta}(i\bar{\omega}_n,{\bm q}) &\approx&  \hat{e}_q
 N_c \int^\Lambda\frac{d^3p}{(2\pi)^3} \frac{V_\zeta({\bm p})V_{\zeta}({\bm p})}{i\bar{\omega}_n-\frac{\partial E_{{\bm p}}^{\zeta}}{\partial|{\bm p}|}(\hat{\bm p}\cdot{\bm q})} \nonumber\\
&\times&\frac{1-(\hat{\bm p}\cdot\hat{\bm q})^2}{|{\bm p}|} \frac{\partial{f}_F(E_{{\bm p}}^{\zeta})}{\partial |{\bm p}|}(\hat{\bm p}\cdot{\bm q}) \label{Pi0Intra}
\end{eqnarray}
for the intraband processes, while
\begin{eqnarray}
\tilde{\Pi}_{0H}^{\zeta\zeta'}(i\bar{\omega}_n,{\bm q}) &\approx&  -\hat{e}_q
 N_c \int^\Lambda\frac{d^3p}{(2\pi)^3}  \frac{V_\zeta({\bm p})V_{\zeta'}({\bm p})}{i\bar{\omega}_n-E_{{\bm p}}^{\zeta}+E_{\bm p}^{\zeta'}} \nonumber\\
 &\times&\frac{1-(\hat{\bm p}\cdot\hat{\bm q})^2}{|{\bm p}|} \left[{f}_F(E_{\bm p}^{\zeta'})-{f}_F(E_{{\bm p}}^{\zeta})\right] \label{Pi0Inter}
\end{eqnarray}
for the interband processes.

Equation~(\ref{Pi0Intra}) shows that the HQSP response functions in the real time for the intraband processes lead to distinct results in the dynamical and static limits,
\begin{eqnarray}
\lim_{q_0\to0}\tilde{\Pi}^{\zeta\zeta}_{0H}(q_0,{\bm 0}) &=& 0 \ , \label{IntraD}
\end{eqnarray}
and
\begin{eqnarray}
\lim_{{\bm q}\to{\bm 0}}\tilde{\Pi}^{\zeta\zeta}_{0H}(0,{\bm q}) &=& -\hat{e}_q
 N_c \int_0^\Lambda\frac{d|{\bm p}|}{3\pi^2}\, |{\bm p}|  \nonumber\\
&\times& V_\zeta({\bm p})V_{\zeta}({\bm p})\frac{\partial{f}_F(E_{{\bm p}}^{\zeta})}{\partial E_{\bm p}^\zeta}\ . \label{IntraS}
\end{eqnarray}
Equation~(\ref{IntraD}) indicates that $\tilde{\Pi}^{\zeta\zeta}_{0H}$ in the dynamical limit is always zero. In the static limit, at zero temperature $\partial f_F(E_{\bm p}^\zeta)/\partial E_{\bm p}^\zeta$ in Eq.~(\ref{IntraS}) becomes zero since the density of states of q.p. or q.h. vanishes at the Fermi level, and hence 
\begin{eqnarray}
\lim_{{\bm q}\to{\bm 0}}\tilde{\Pi}^{\zeta\zeta}_{0H}(0,{\bm q}) \overset{T=0}{=} 0 . \label{IntraST0}
\end{eqnarray}
At finite temperature, the factor $\partial f_F(E_{\bm p}^\zeta)/\partial E_{\bm p}^\zeta$ begins to collect contributions around the Fermi level by thermal fluctuations. In the present Kondo system, due to the flatness of the dispersion relation of q.p. or q.h. as shown in Fig.~\ref{fig:Dispersions}, the thermal fluctuations can pick up modes slightly far from the Fermi sphere but having large density of states. As a result, $\tilde{\Pi}^{\zeta\zeta}_{0H}$ in the static limit at finite temperature becomes nonzero.

On the other hand, from Eq.~(\ref{Pi0Inter}), we can find that the HQSP response functions for the interband processes in the two limits yield the identical result,
\begin{eqnarray}
\lim_{{\bm q}\to{\bm 0}}\tilde{\Pi}^{\zeta\zeta'}_{0H}(0,{\bm q}) &=& \lim_{q_0\to0}\tilde{\Pi}^{\zeta\zeta'}_{0H}(q_0,{\bm 0})  \nonumber\\
&=&  -\hat{e}_q
 N_c \int_0^\Lambda\frac{d|{\bm p}|}{3\pi^2}\, |{\bm p}| \nonumber\\
 &\times&  \frac{V_\zeta({\bm p})V_{\zeta'}({\bm p})}{E_{\bm p}^{\zeta'}-E_{{\bm p}}^{\zeta}} \left[{f}_F(E_{\bm p}^{\zeta'})-{f}_F(E_{{\bm p}}^{\zeta})\right] \ . \nonumber\\ \label{InterSD}
\end{eqnarray}
The integrand for the interband processes~(\ref{InterSD}) is dominated by contributions near the Fermi sphere $|{\bm p}| \sim\mu$ due to the factor ${f}_F(E_{\bm p}^{\zeta'})-{f}_F(E_{{\bm p}}^{\zeta})$, which is in contrast to the intraband ones. Besides, because of this factor Eq.~(\ref{InterSD}) is less temperature dependent similarly to the Kondo condensate as explained around Eq.~(\ref{TDepKondo}).

\begin{figure*}[t]
  \begin{center}
    \begin{tabular}{cc}

      \begin{minipage}[c]{0.3\hsize}
       \centering
       \hspace*{-0.2cm} 
         \includegraphics*[scale=0.48]{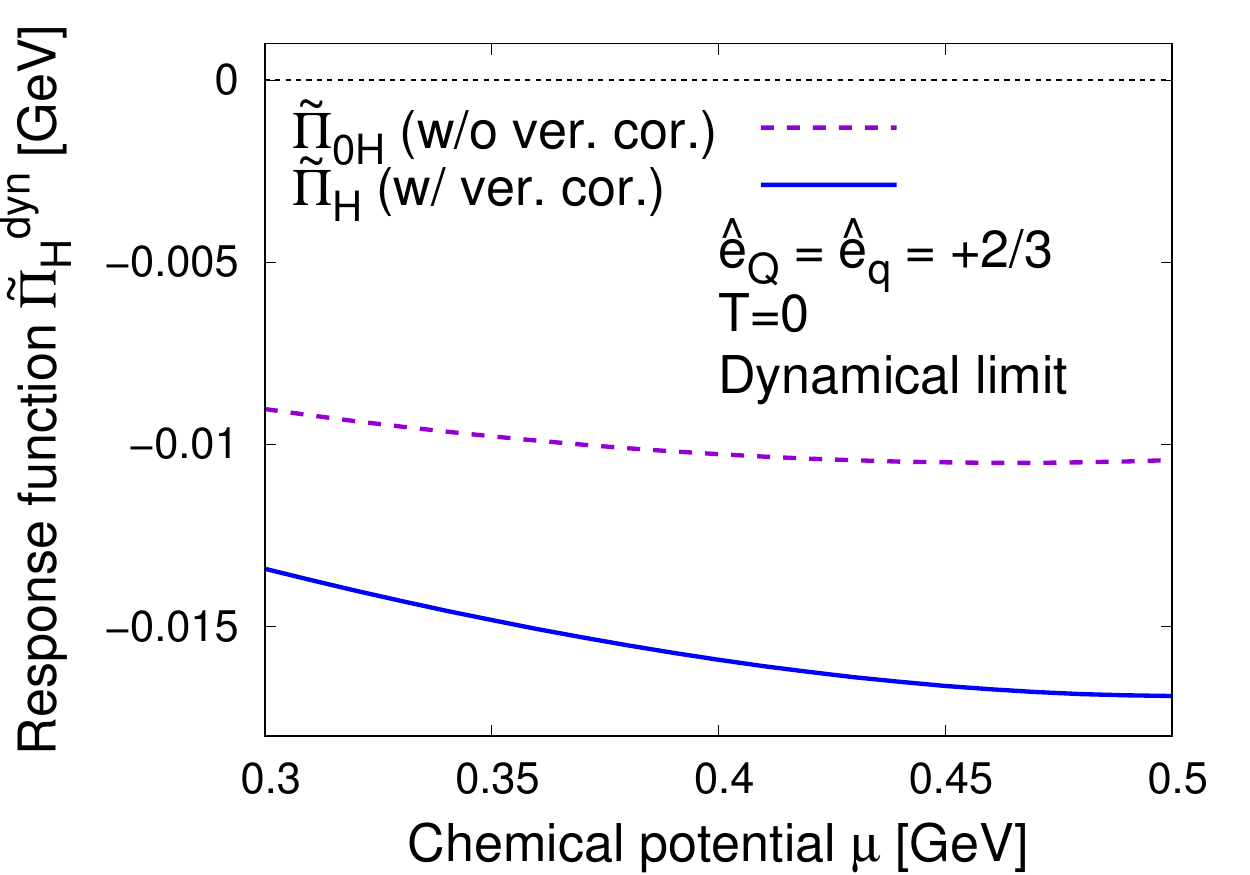}\\
         \end{minipage}

      \begin{minipage}[c]{0.4\hsize}
       \centering
        \hspace*{-0.3cm} 
          \includegraphics*[scale=0.48]{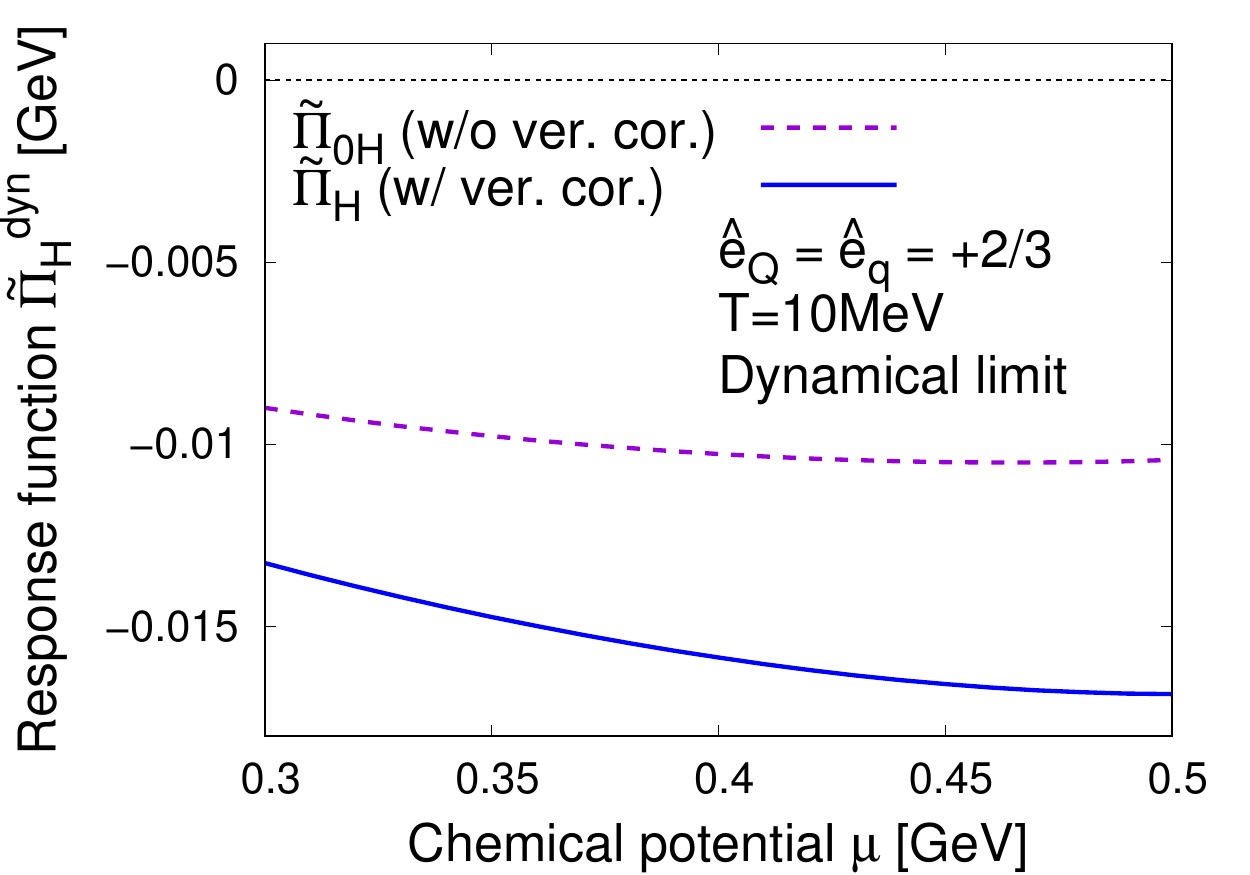}\\
      \end{minipage}
      
      \begin{minipage}[c]{0.4\hsize}
       \centering
        \hspace*{-2.8cm} 
          \includegraphics*[scale=0.48]{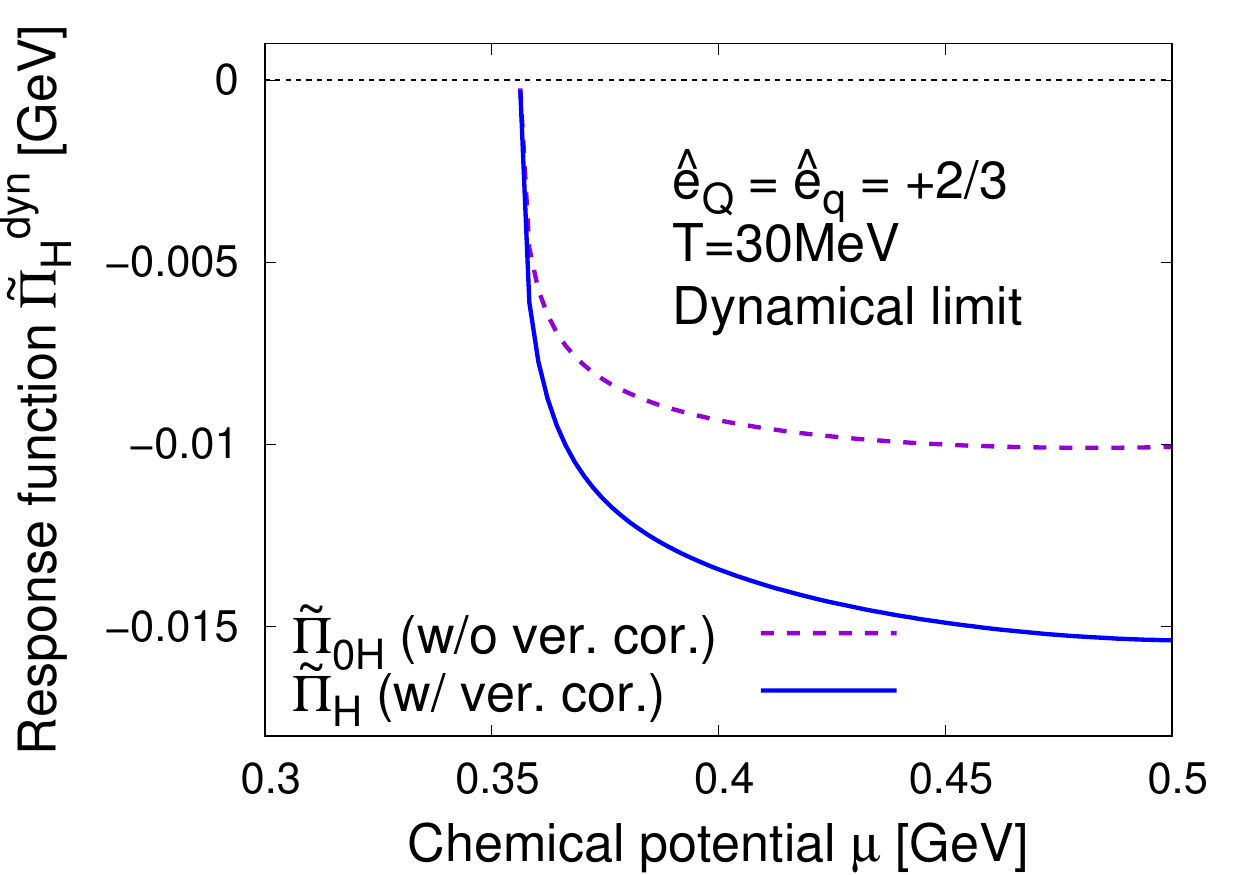}\\
      \end{minipage}

      \end{tabular}
\caption{The $\mu$ dependence of the HQSP response function in the dynamical limit. In this figure, $\tilde{\Pi}_{0H}^{\rm dyn}$ (dashed purple line) and $\tilde{\Pi}_{H}^{\rm dyn}$ (solid blue line) with $\hat{e}_Q=\hat{e}_q=+\frac{2}{3}$ are plotted.}
\label{fig:PiDyn1}
  \end{center}
\end{figure*}

\begin{figure*}[t]
  \begin{center}
    \begin{tabular}{cc}

      \begin{minipage}[c]{0.3\hsize}
       \centering
       \hspace*{-0.2cm} 
         \includegraphics*[scale=0.48]{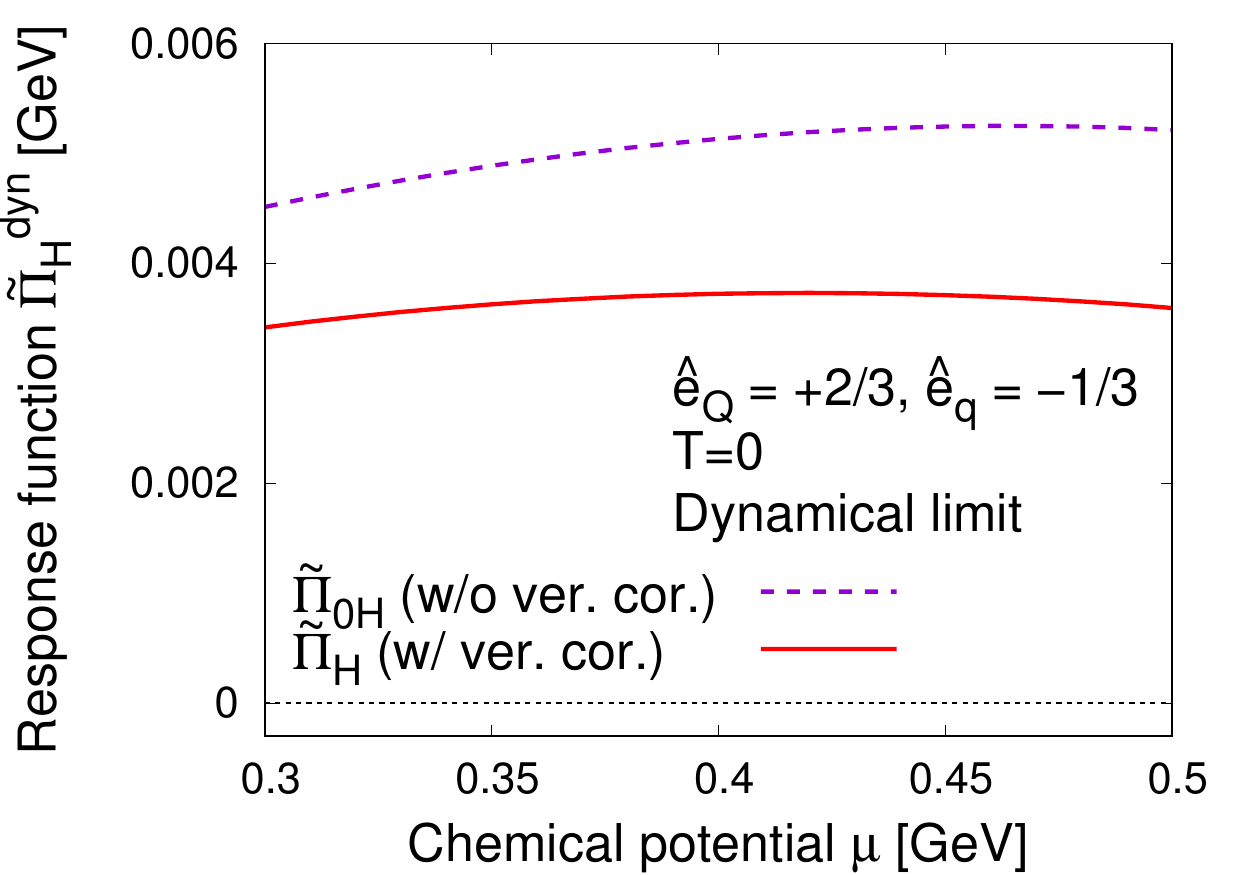}\\
         \end{minipage}

      \begin{minipage}[c]{0.4\hsize}
       \centering
        \hspace*{-0.3cm} 
          \includegraphics*[scale=0.48]{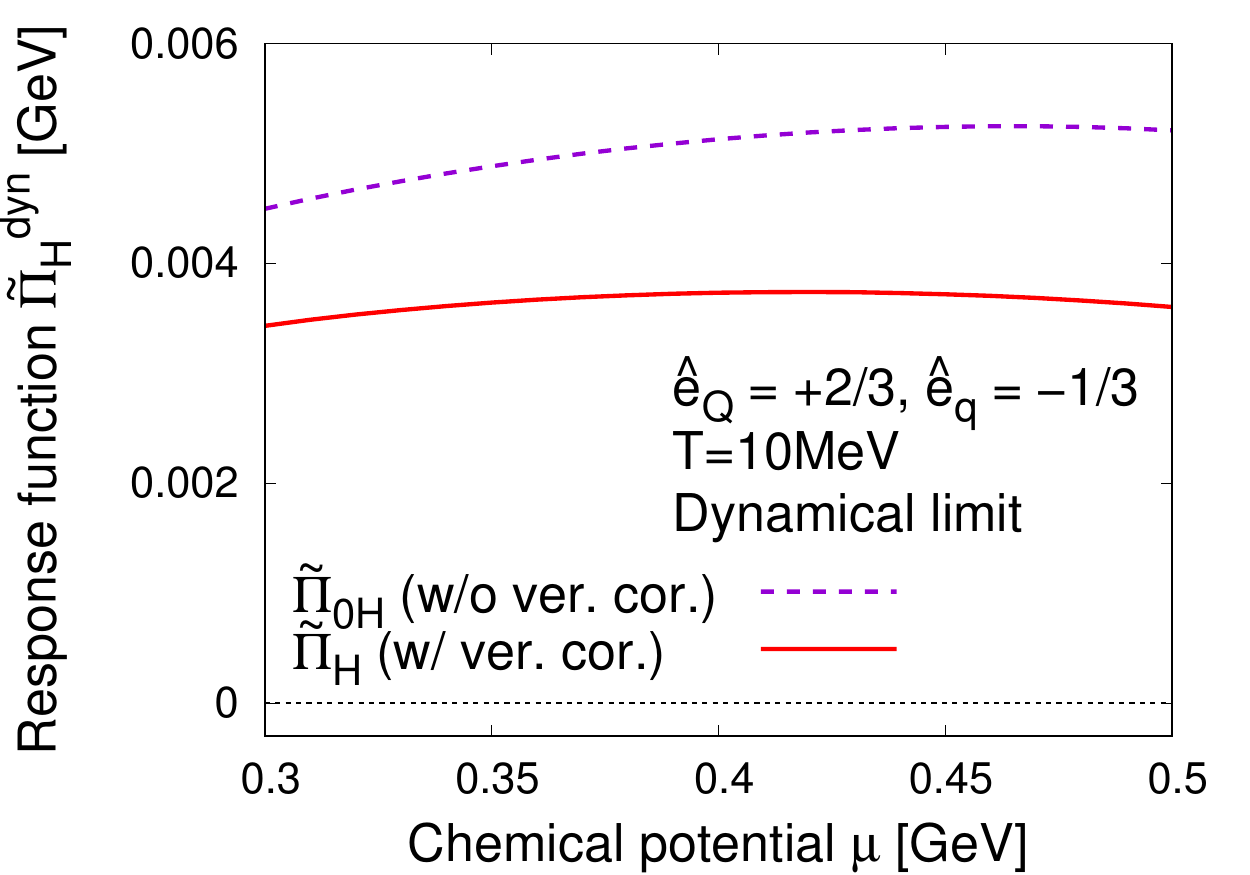}\\
      \end{minipage}
      
      \begin{minipage}[c]{0.4\hsize}
       \centering
        \hspace*{-2.8cm} 
          \includegraphics*[scale=0.48]{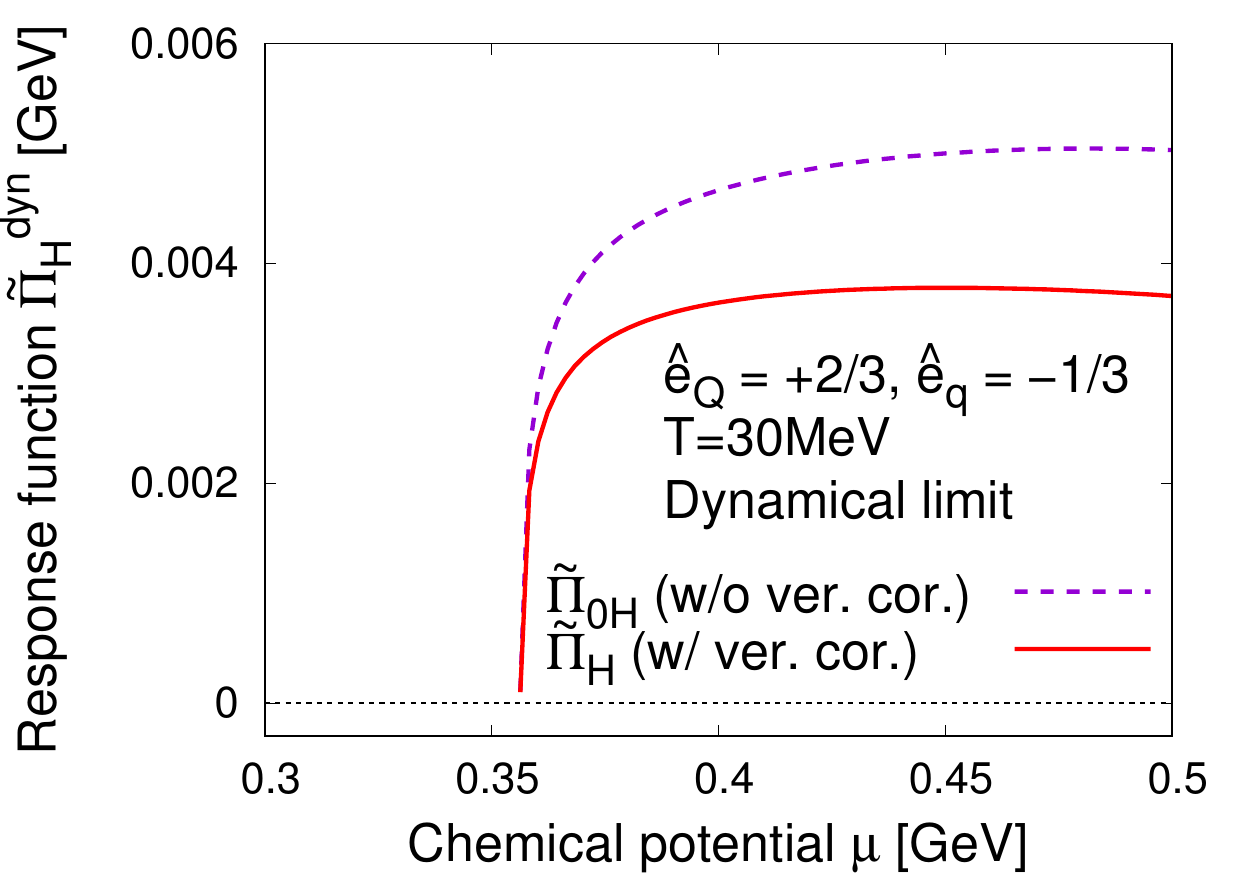}\\
      \end{minipage}

      \end{tabular}
\caption{The $\mu$ dependence of the HQSP response function in the dynamical limit. In this figure, $\tilde{\Pi}_{0H}^{\rm dyn}$ (dashed purple line) and $\tilde{\Pi}_{H}^{\rm dyn}$ (solid red line) with $\hat{e}_Q=+\frac{2}{3},\hat{e}_q=-\frac{1}{3}$ are plotted.}
\label{fig:PiDyn2}
  \end{center}
\end{figure*}

\subsection{The corrected-vertex parts $\tilde{\Pi}_{\delta H}$}
\label{sec:CorrectedVertex}

The dynamical and static limits for the corrected-vertex parts $\tilde{\Pi}_{\delta H}(q_0,{\bm q})$ can be evaluated in a similar way to Sec.~\ref{sec:BareVertex}. Expressions for $\tilde{\Pi}_{\delta H}(q_0,{\bm q})$ with a small momentum ${\bm q}$ and resultant dynamical and static response functions are so lengthy that we show them in Appendix~\ref{sec:AppCorrectedVertex}. Here we only give important comments on them.

The significant difference from the bare-vertex parts is that the kinetic factor originating from the spin couplings between the fermions is now proportional to $({\bm p}_\pm\cdot{\bm p})(\hat{\bm p}_\pm\cdot{\bm q})$ which is of ${\cal O}({\bm q}^1)$, whereas that for the bare-vertex parts in Eq.~(\ref{KineticBare}) is of ${\cal O}({\bm q}^2)$. Complexity of the evaluation of $\tilde{\Pi}_{\delta H}(q_0,{\bm q})$ with small ${\bm q}$ essentially stems from this difference, but qualitative properties are similar to those for the bare-vertex parts. In fact, as seen from Eqs.~(\ref{IntraDeltaDApp}) and~(\ref{IntraDeltaSApp}), for the intraband processes the response function in the dynamical limit vanishes and the result in the static limit is proportional to the derivative $\partial f_F(E_{\bm p}^\zeta)/\partial E_{\bm p}^\zeta$.

For the interband processes, as seen from Eq.~(\ref{InterDeltaSApp}) both the dynamical and static limits for the corrected-vertex parts yield the identical response similarly to the bare-vertex parts,
\begin{eqnarray}
\lim_{q_0\to{0}}\tilde{\Pi}^{\zeta\zeta'}_{\delta H}(q_0,{\bm 0}) = \lim_{{\bm q}\to{\bm 0}}\tilde{\Pi}^{\zeta\zeta'}_{\delta H}(0,{\bm q})\ .
\end{eqnarray}
Unlike the bare-vertex parts in Eq.~(\ref{InterSD}), the corrected-vertex parts for the interband processes include contributions proportional to $\partial f_F(E_{\bm p}^{\zeta^{(\prime)}})/\partial|{\bm p}|$ in addition to those proportional to ${f}_F(E_{\bm p}^{\zeta'})-{f}_F(E_{{\bm p}}^{\zeta})$. Thus, the corrected-vertex parts in the dynamical limit can be rather temperature dependent compared to the bare-vertex ones.

Before closing this section we summarize important properties of the HQSP response function $\tilde{\Pi}_H$ in the static and dynamical limits. At zero temperature, $\tilde{\Pi}_H$ in the static and dynamical limits coincide since the density of states at the Fermi level are absent. On the other hand, at finite temperature, the intraband processes are allowed due to the thermal excitations, and the resultant $\tilde{\Pi}_H$ in the static and dynamical limits can differ.

\section{Numerical results}
\label{sec:Results}

In Sec.~\ref{sec:Proceed}, we have presented analytic evaluation of the HQSP response function induced by the Kondo effect under a magnetic field. In this section, based on it, we show the numerical results of the response functions.

First, in Sec.~\ref{sec:DynamicalPiH}, we show the resultant HQSP response function in the timelike regime. Next, in Sec.~\ref{sec:StaticPiH}, we show the result in the spacetime regime. For all computations, the value of the Kondo condensate $\Delta$ at a given $\mu$ and $T$ is determined by solving the gap equation~(\ref{GapEquation}), with $g=9.47$ GeV$^{-2}$ and $\Lambda=0.65$ GeV~\cite{Yasui:2016svc,Yasui:2017izi}. We note that in our present paper we assume $c$ quark as impurities, and hence, $\hat{e}_Q=+\frac{2}{3}$.
Throughout this paper, we show the calculated values of the response functions in the unit GeV for clarity of discussion.
The response of 1 GeV corresponds to the HQSP of the density $7.70\times 10^{30}$ m$^{-3}$ induced under a magnetic field of 1T.

\begin{figure*}[t]
  \begin{center}
    \begin{tabular}{cc}

      \begin{minipage}[c]{0.3\hsize}
       \centering
       \hspace*{-2cm} 
         \includegraphics*[scale=0.65]{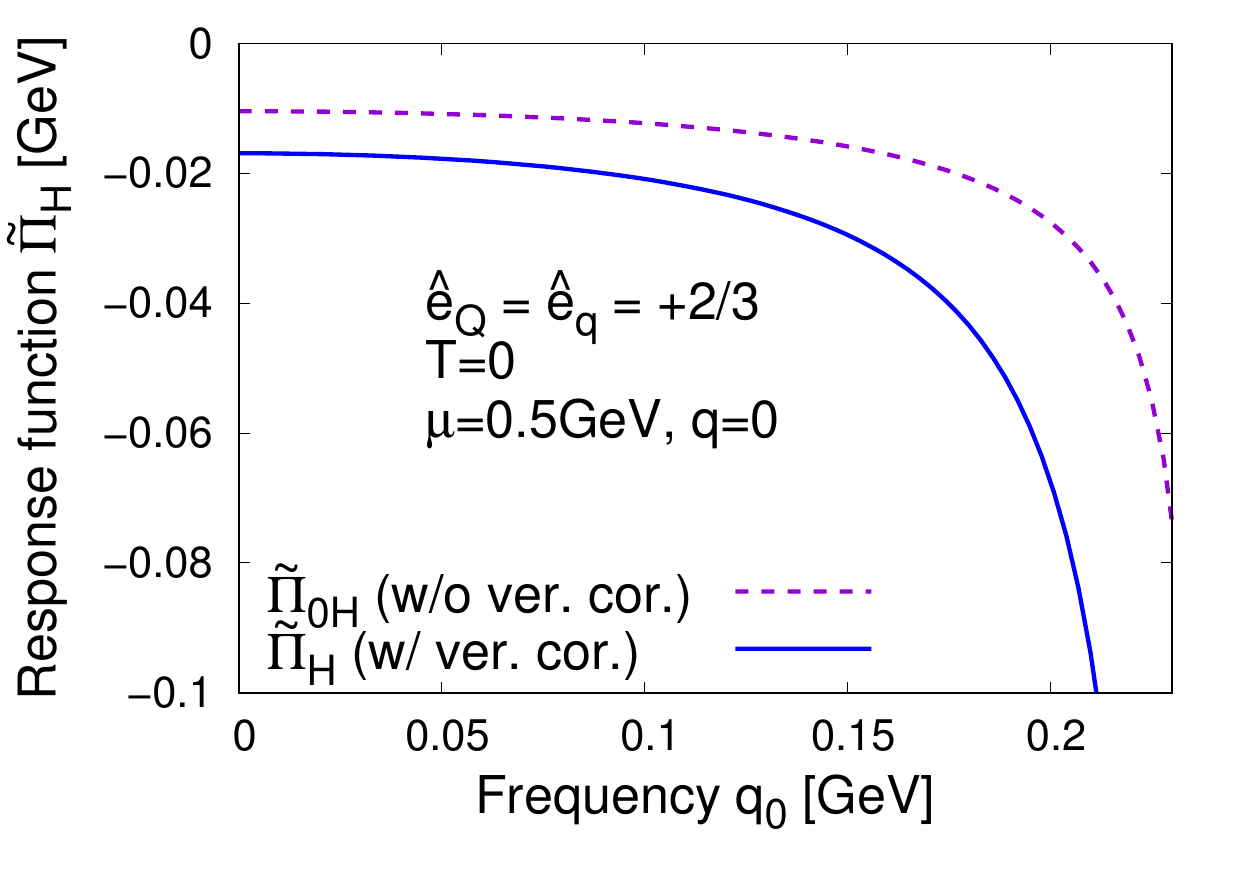}\\
         \end{minipage}

      \begin{minipage}[c]{0.4\hsize}
       \centering
        \hspace*{1cm} 
          \includegraphics*[scale=0.65]{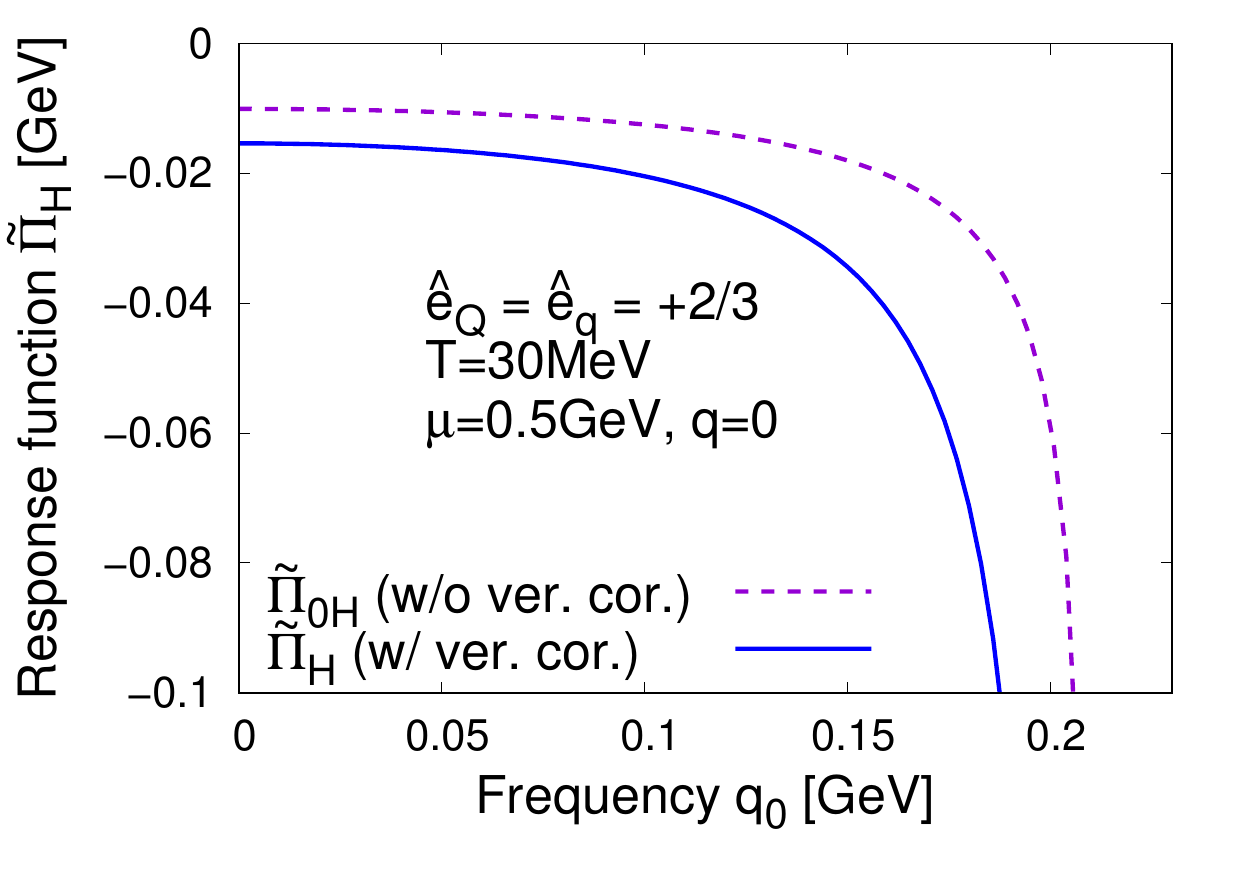}\\
      \end{minipage}

      \end{tabular}
\caption{The $q_0$ dependence of the HQSP response function for vanishing ${\bm q}$ at $\mu=0.5$ GeV. In this figure, $\tilde{\Pi}_{0H}$ (dashed purple line) and $\tilde{\Pi}_{H}(q_0,{\bm 0})$ (solid blue line) with $\hat{e}_Q=\hat{e}_q=+\frac{2}{3}$ are plotted.}
\label{fig:PiQ01}
  \end{center}
\end{figure*}

\begin{figure*}[t]
  \begin{center}
    \begin{tabular}{cc}

      \begin{minipage}[c]{0.3\hsize}
       \centering
       \hspace*{-2cm} 
         \includegraphics*[scale=0.65]{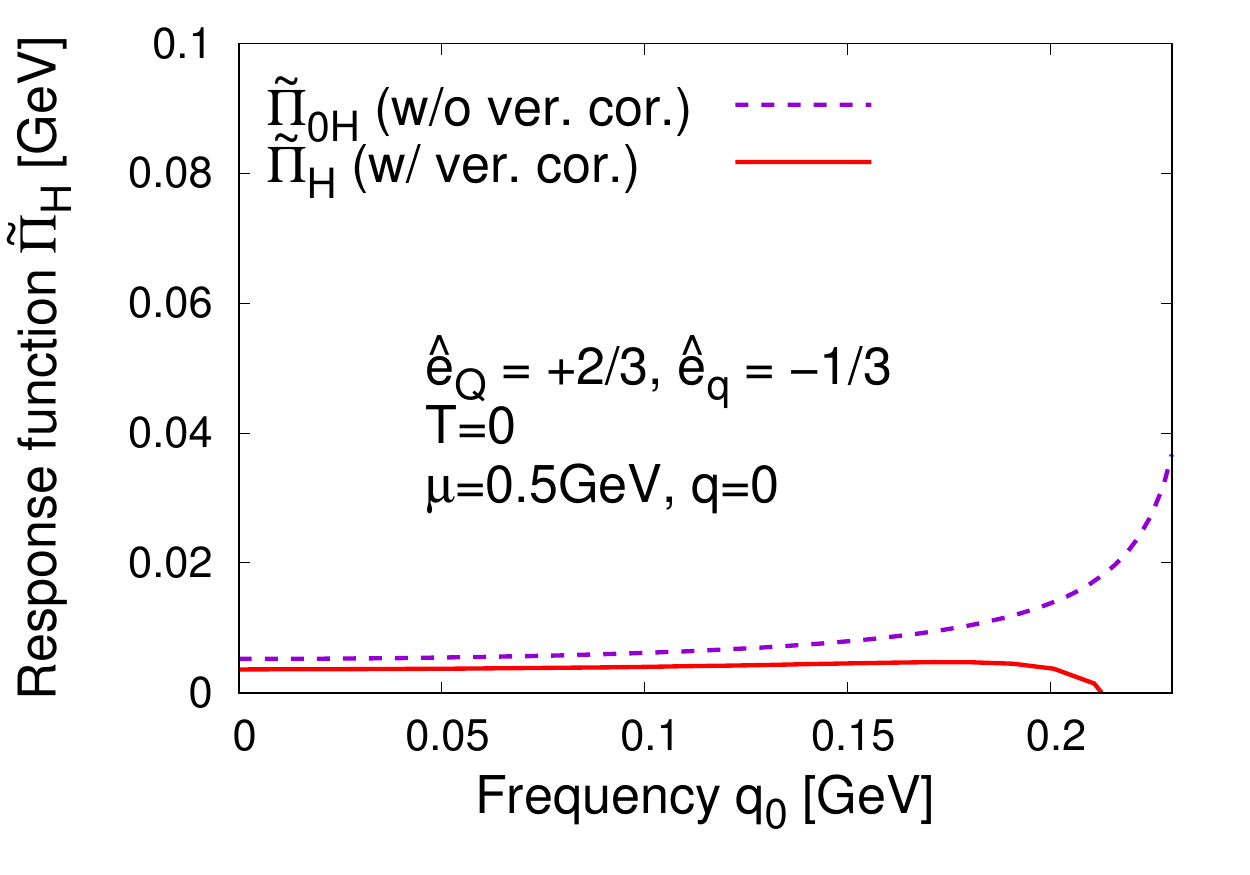}\\
         \end{minipage}

      \begin{minipage}[c]{0.4\hsize}
       \centering
        \hspace*{1cm} 
          \includegraphics*[scale=0.65]{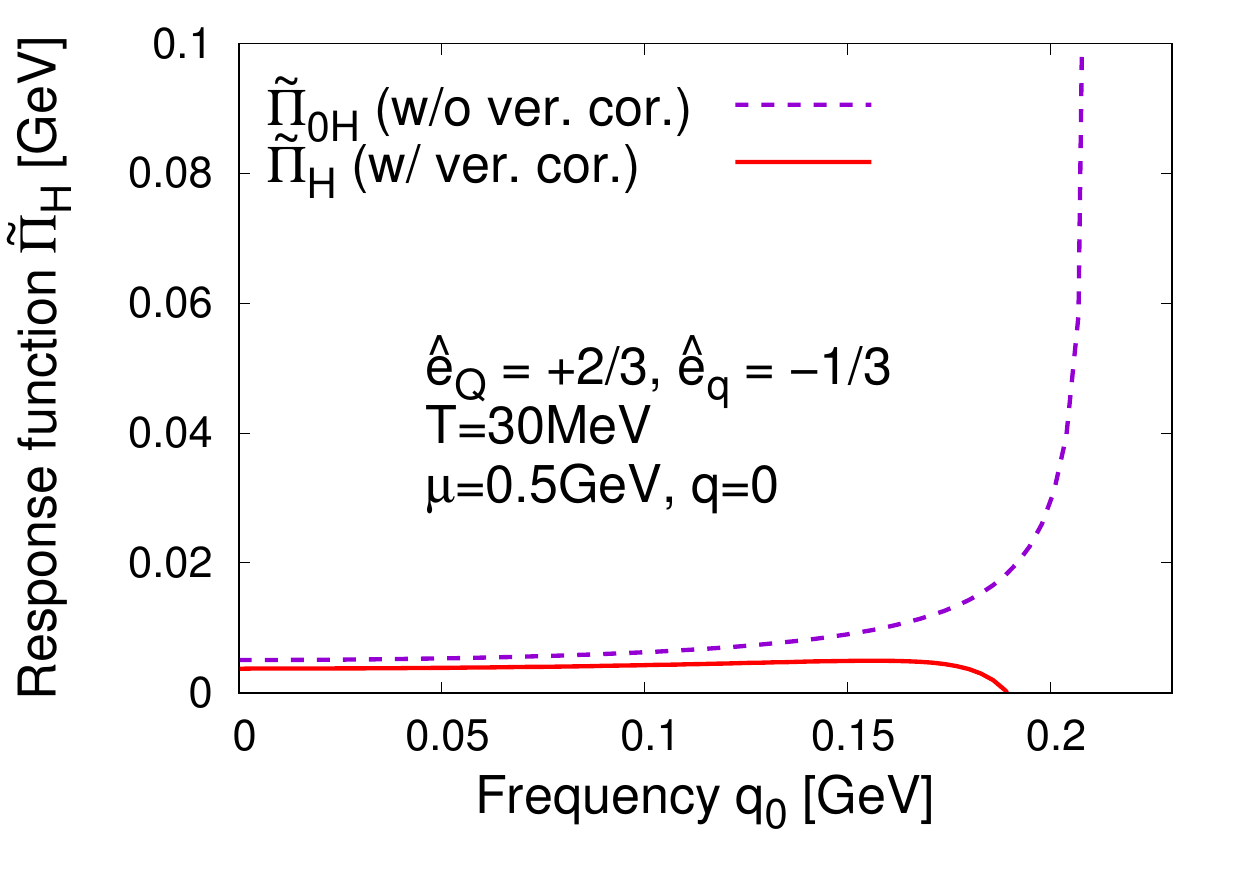}\\
      \end{minipage}

      \end{tabular}
\caption{The $q_0$ dependence of the HQSP response function for vanishing ${\bm q}$ at $\mu=0.5$ GeV. In this figure, $\tilde{\Pi}_{0H}$ (dashed purple line) and $\tilde{\Pi}_{H}(q_0,{\bm 0})$ (solid red line) with $\hat{e}_Q=+\frac{2}{3}, \hat{e}_q=-\frac{1}{3}$ are plotted.}
\label{fig:PiQ02}
  \end{center}
\end{figure*}

\subsection{The HQSP response function in the timelike regime}
\label{sec:DynamicalPiH}

Here, we show the $\mu$ dependence of the HQSP response function in the timelike regime. 

First, we examine the response function in the dynamical limit defined in Eq.~(\ref{PiDynamical}). In this limit the contributions from intraband processes, or the diagrams (i) and (ii) in Fig.~\ref{fig:Diagrams}, vanish.
This reduction allows us to interpret the numerical results in an easier way. Figure~\ref{fig:PiDyn1} represents the results at $T=0$, $T=0.01$ GeV, and $T=0.03$ GeV where the electric charge of light quarks is $\hat{e}_q = +\frac{2}{3}$, meaning that quark matter is filled by $u$ quarks. In this figure, we show $\tilde{\Pi}_{0H}^{\rm dyn}$ (dashed purple line) and $\tilde{\Pi}_{H}^{\rm dyn}$ (solid blue line).  Similarly, Fig.~\ref{fig:PiDyn2} represents the results of $\tilde{\Pi}_{0H}^{\rm dyn}$ (dashed purple line) and $\tilde{\Pi}_{H}^{\rm dyn}$ (solid red) with $\hat{e}_q=-\frac{1}{3}$, where $d$ quarks compose matter.

The figures show that the HQSP is significantly driven by the Kondo effect under a magnetic field.\footnote{In order to clarify significance of the HQSP driven by the Kondo effect, we compare the results with those from the Zeeman interaction in Sec.~\ref{sec:Discussions}.} Besides, the vertex corrections lead to an enhancement of the HQSP for $\hat{e}_q=\hat{e}_Q$ as seen from Fig.~\ref{fig:PiDyn1}, while they suppress the HQSP for $\hat{e}_q\neq\hat{e}_Q$ from Fig.~\ref{fig:PiDyn2}. Interestingly, the magnitude of the HQSP response function does not depend on the chemical potential significantly while this tendency cannot be seen easily by analytic evaluation. We note that the nonanalyticity at $\mu\approx 0.36$ GeV for $T=0.03$ GeV corresponds to the phase transition from the normal phase ($\Delta=0$) to the Kondo phase ($\Delta\neq0$). Thus, the HQSP vanishes in the normal phase.

In the figures, we have shown the spin polarization of $c$ quarks regarded as impurities in $u$ quark matter and in $d$ quark matter separately. When quark matter is composed of both $u$ and $d$ quarks, we expect that the Kondo condensates made of $c$ and $u$ quarks are more favored than those of $c$ and $d$ quarks, because the condensates become electrically neutral in this choice. In this case, the HQSP response function is mostly given by Fig.~\ref{fig:PiDyn1}. Even when both the electrically neutral and charged Kondo condensates equally contribute, the resultant HQSP response function is given by the averaged value of Figs.~\ref{fig:PiDyn1} and~\ref{fig:PiDyn2}, yielding a finite negative response.

\begin{figure*}[t]
  \begin{center}
    \begin{tabular}{cc}

      \begin{minipage}[c]{0.3\hsize}
       \centering
       \hspace*{-0.2cm} 
         \includegraphics*[scale=0.48]{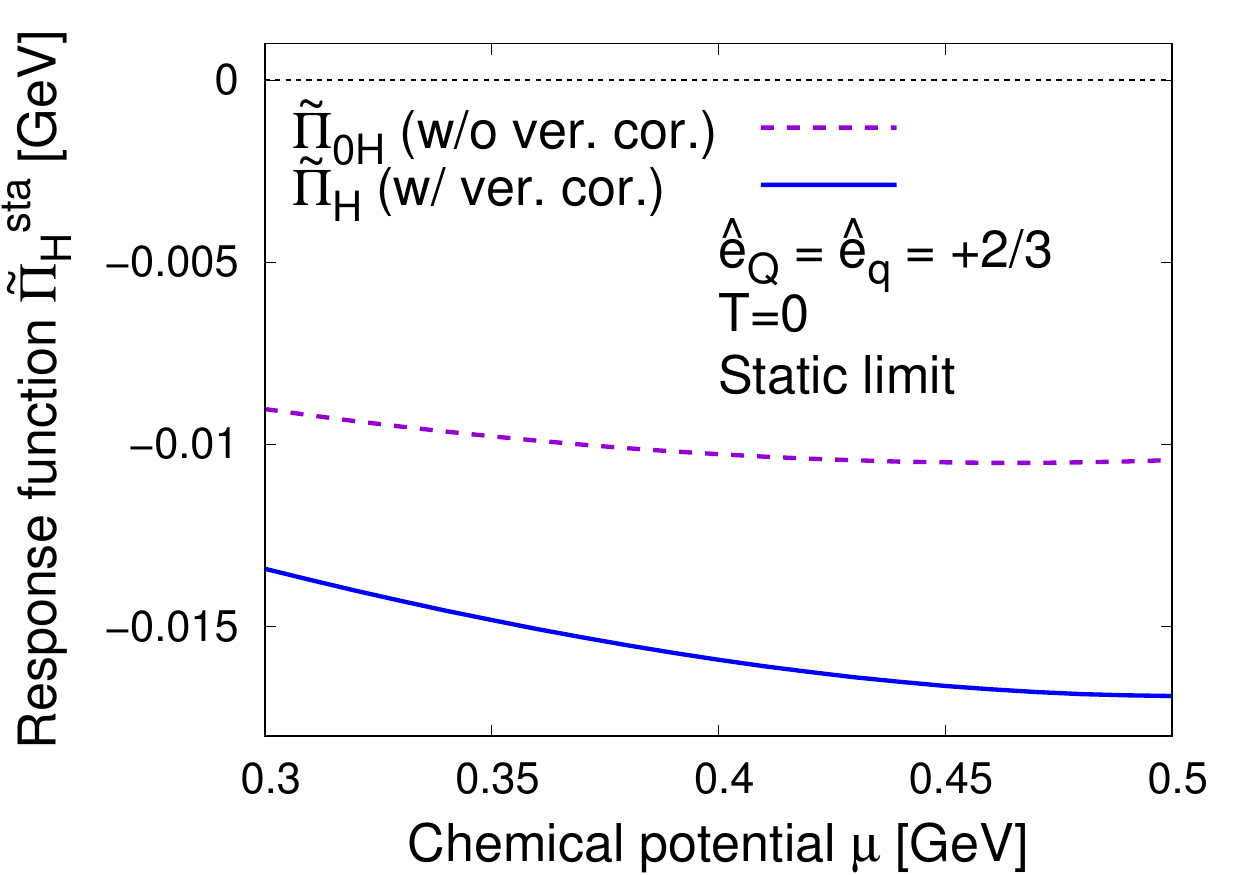}\\
         \end{minipage}

      \begin{minipage}[c]{0.4\hsize}
       \centering
        \hspace*{-0.3cm} 
          \includegraphics*[scale=0.48]{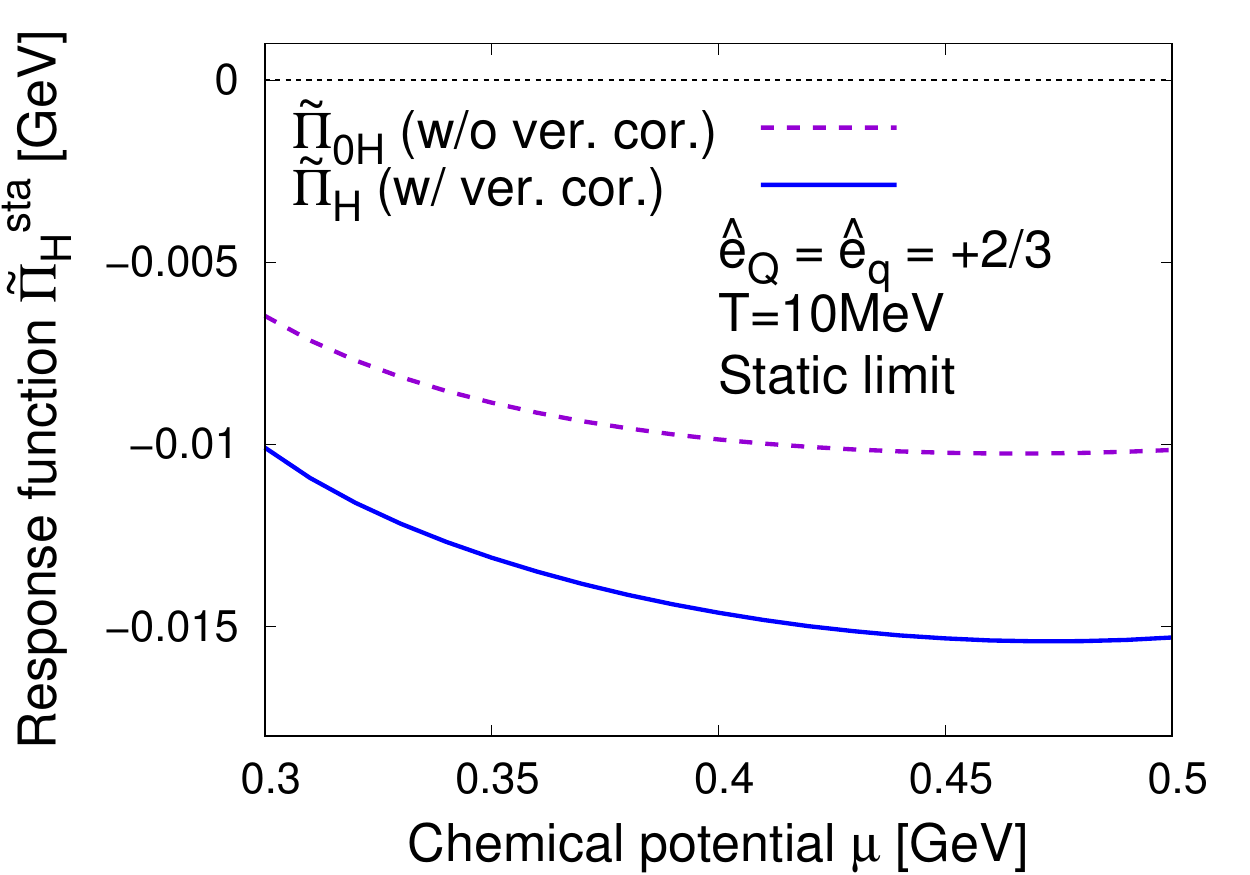}\\
      \end{minipage}
      
      \begin{minipage}[c]{0.4\hsize}
       \centering
        \hspace*{-2.8cm} 
          \includegraphics*[scale=0.48]{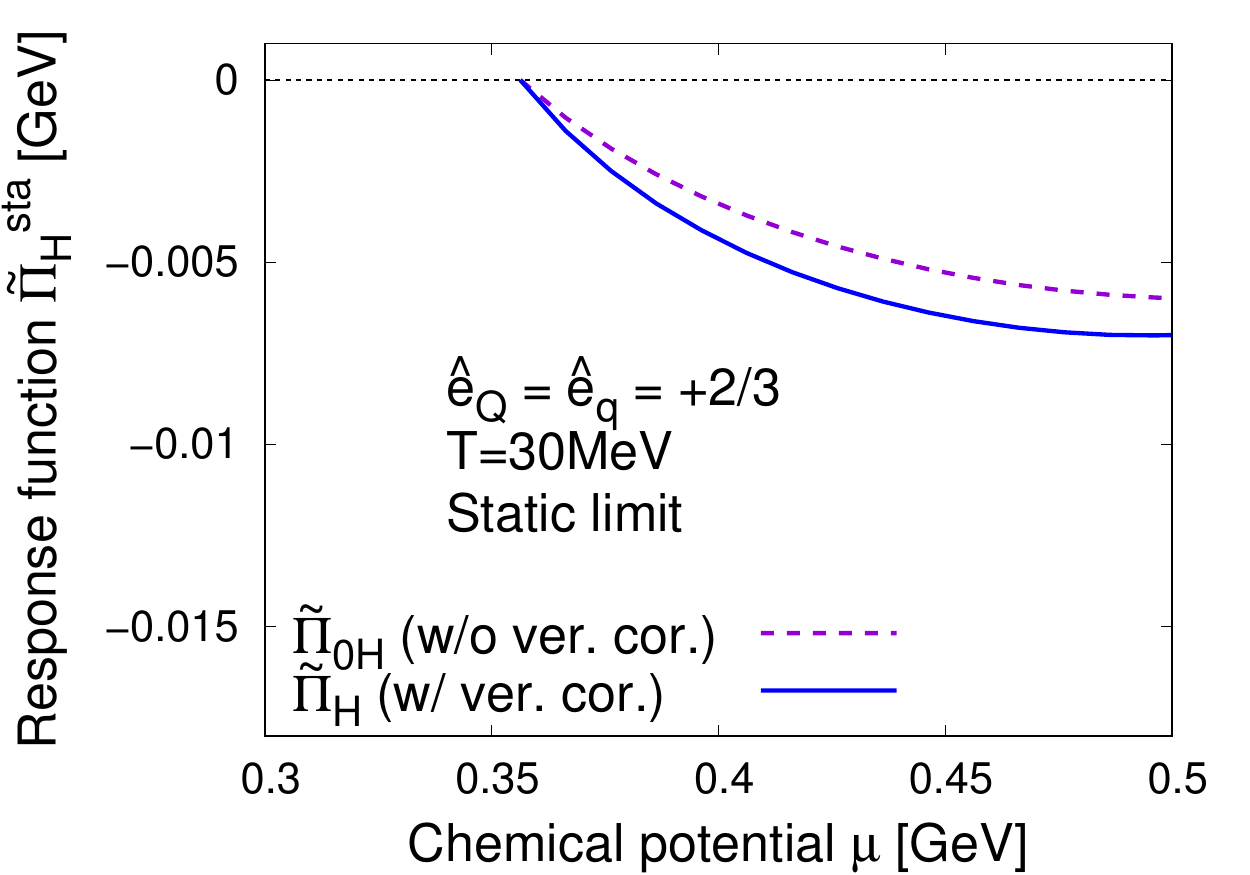}\\
      \end{minipage}

      \end{tabular}
\caption{The $\mu$ dependence of the HQSP response function in the static limit. In this figure, $\tilde{\Pi}_{0H}^{\rm sta}$ (dashed purple line) and $\tilde{\Pi}_{H}^{\rm sta}$ (solid blue line) with $\hat{e}_Q=\hat{e}_q=+\frac{2}{3}$ are plotted.}
\label{fig:PiSta1}
  \end{center}
\end{figure*}

\begin{figure*}[t]
  \begin{center}
    \begin{tabular}{cc}

      \begin{minipage}[c]{0.3\hsize}
       \centering
       \hspace*{-0.2cm} 
         \includegraphics*[scale=0.48]{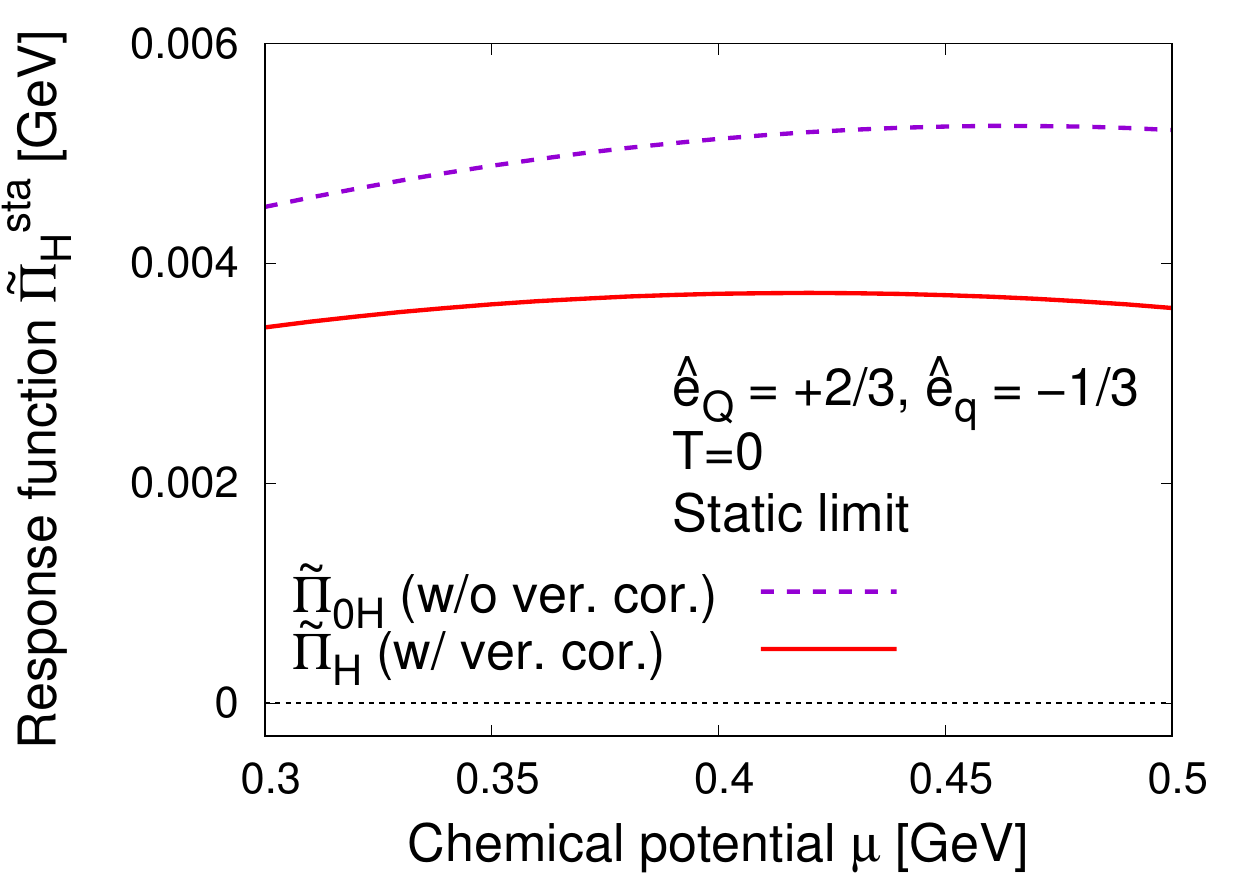}\\
         \end{minipage}

      \begin{minipage}[c]{0.4\hsize}
       \centering
        \hspace*{-0.3cm} 
          \includegraphics*[scale=0.48]{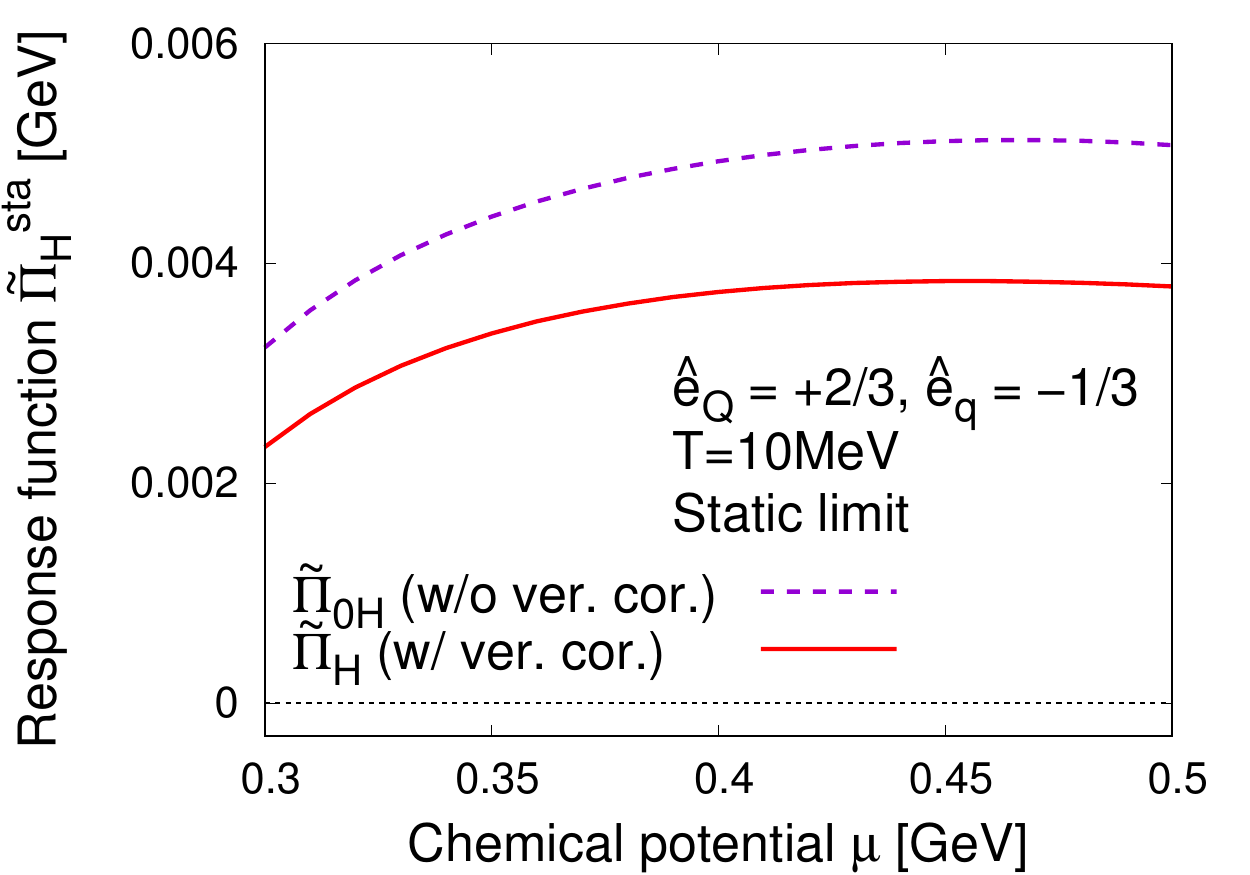}\\
      \end{minipage}
      
      \begin{minipage}[c]{0.4\hsize}
       \centering
        \hspace*{-2.8cm} 
          \includegraphics*[scale=0.48]{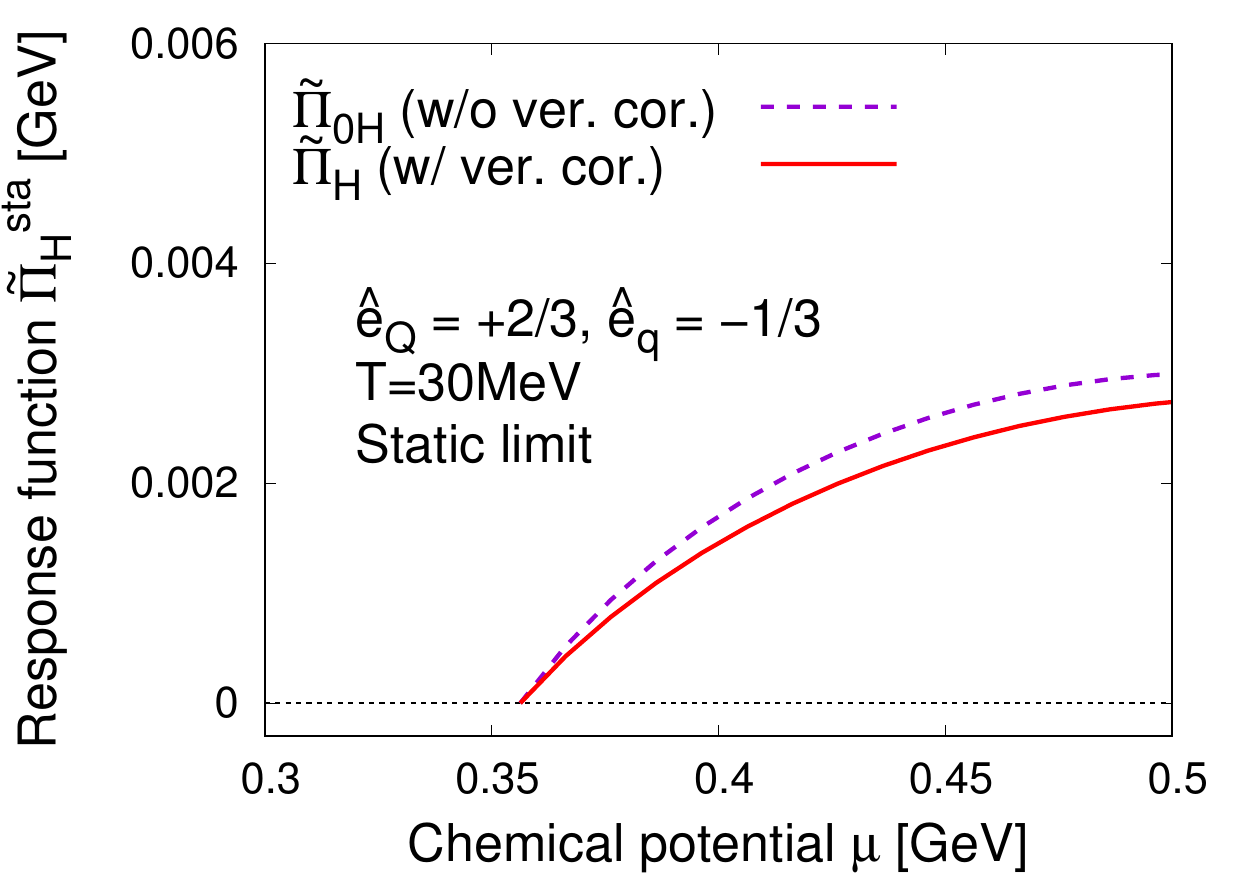}\\
      \end{minipage}

      \end{tabular}
\caption{The $\mu$ dependence of the HQSP response function in the static limit. In this figure, $\tilde{\Pi}_{0H}^{\rm sta}$ (dashed purple line) and $\tilde{\Pi}_{H}^{\rm sta}$ (solid red lline) with $\hat{e}_Q=+\frac{2}{3},\hat{e}_q=-\frac{1}{3}$ are plotted.}
\label{fig:PiSta2}
  \end{center}
\end{figure*}

\begin{figure*}[t]
  \begin{center}
    \begin{tabular}{cc}

      \begin{minipage}[c]{0.3\hsize}
       \centering
       \hspace*{-2cm} 
         \includegraphics*[scale=0.65]{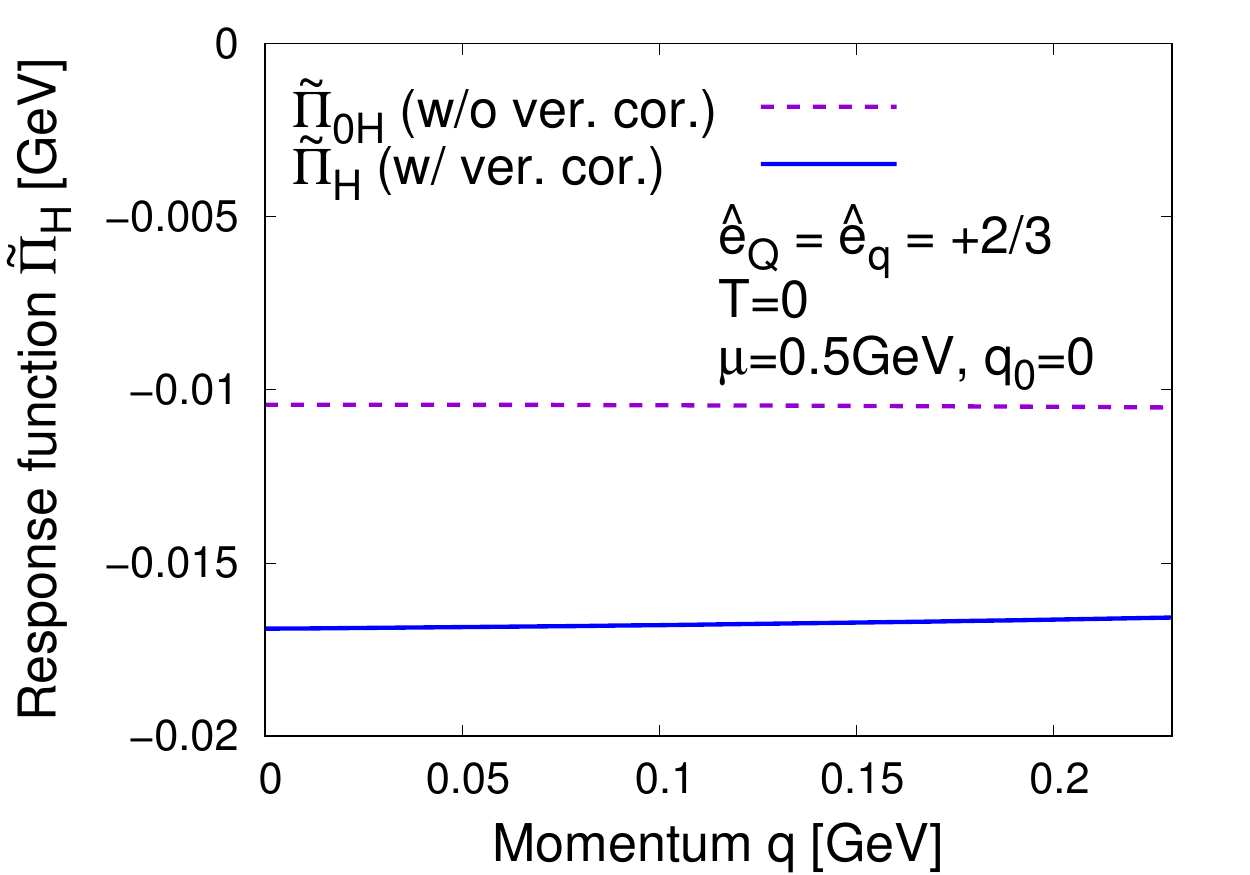}\\
         \end{minipage}

      \begin{minipage}[c]{0.4\hsize}
       \centering
        \hspace*{1cm} 
          \includegraphics*[scale=0.65]{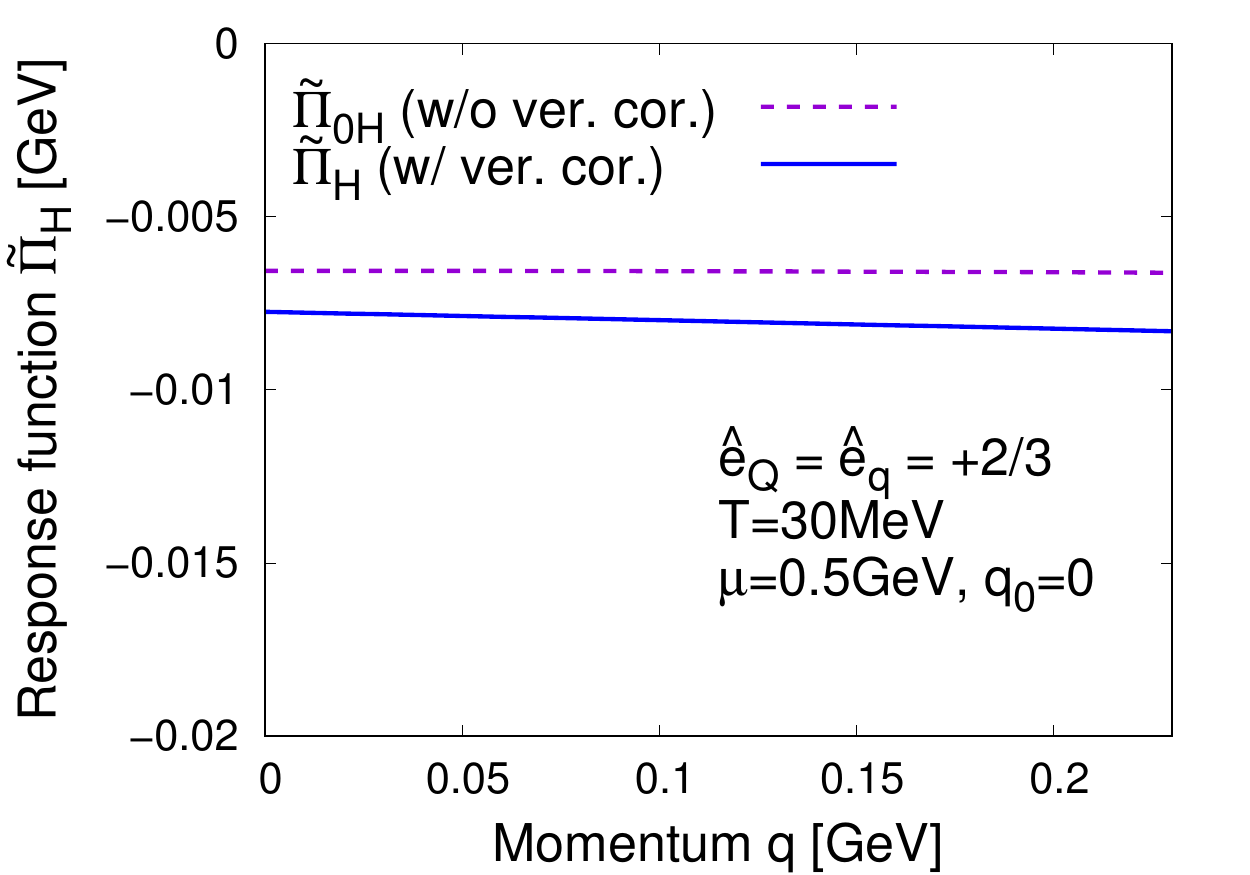}\\
      \end{minipage}

      \end{tabular}
\caption{The $q$ dependence of the HQSP response function for vanishing $q_0$ at $\mu=0.5$ GeV ($q\equiv|{\bm q}|$). In this figure, $\tilde{\Pi}_{0H}$ (dashed purple line) and $\tilde{\Pi}_{H}(q_0,{\bm 0})$ (solid blue line) with $\hat{e}_Q=\hat{e}_q=+\frac{2}{3}$ are plotted.}
\label{fig:PiQ1}
  \end{center}
\end{figure*}

\begin{figure*}[t]
  \begin{center}
    \begin{tabular}{cc}

      \begin{minipage}[c]{0.3\hsize}
       \centering
       \hspace*{-2cm} 
         \includegraphics*[scale=0.65]{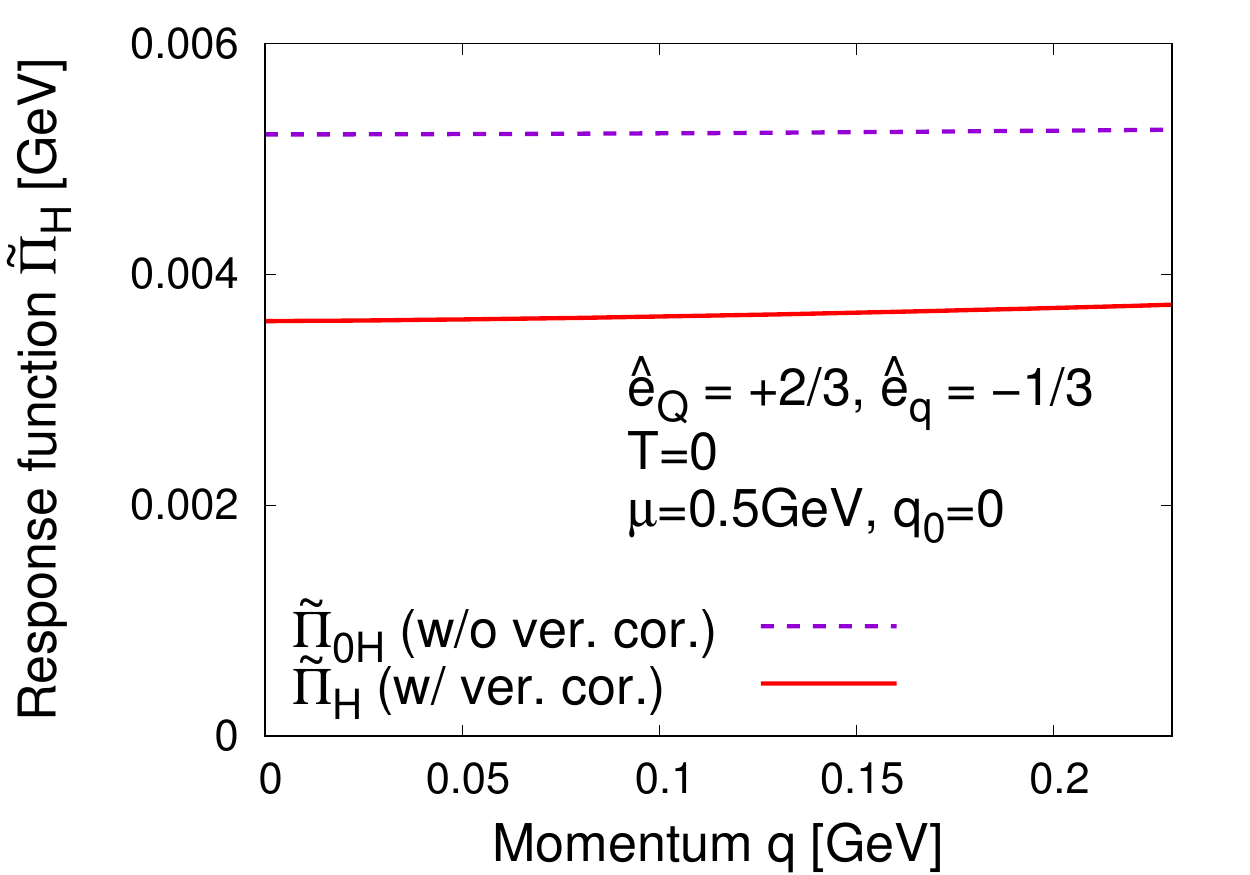}\\
         \end{minipage}

      \begin{minipage}[c]{0.4\hsize}
       \centering
        \hspace*{1cm} 
          \includegraphics*[scale=0.65]{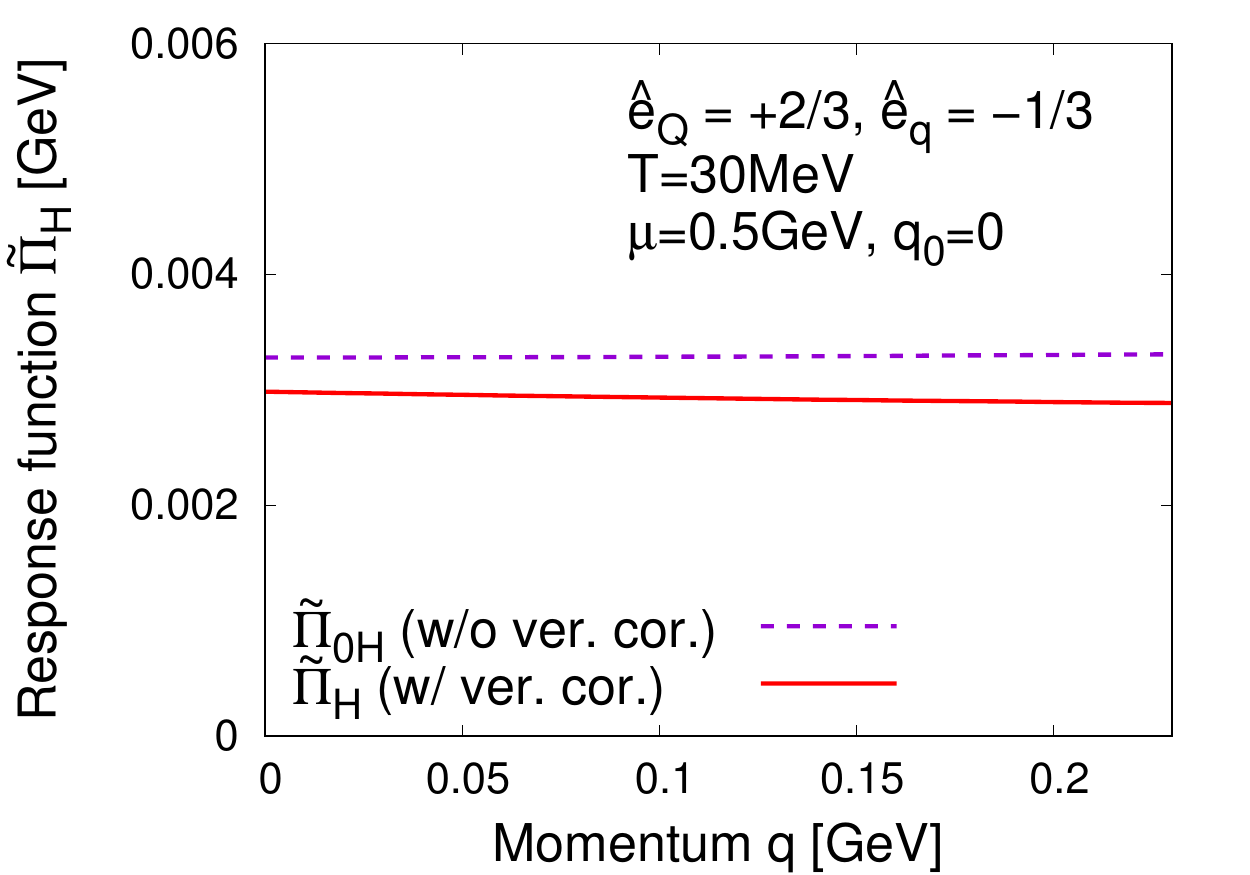}\\
      \end{minipage}

      \end{tabular}
\caption{The $q$ dependence of the HQSP response function for vanishing $q_0$ at $\mu=0.5$ GeV ($q\equiv|{\bm q}|$). In this figure, $\tilde{\Pi}_{0H}$ (dashed purple line) and $\tilde{\Pi}_{H}(q_0,{\bm 0})$ (solid red line) with $\hat{e}_Q=+\frac{2}{3}, \hat{e}_q=-\frac{1}{3}$ are plotted.}
\label{fig:PiQ2}
  \end{center}
\end{figure*}

Next, we show numerical results of $q_0$ dependence of the response function for vanishing ${\bm q}$ at $\mu=0.5$ GeV. Figure~\ref{fig:PiQ01} represents the results of $\tilde{\Pi}_{0H}(q_0,{\bm 0})$ (dashed purple line) and $\tilde{\Pi}_{H}(q_0,{\bm 0})$ (solid blue line) for $\hat{e}_q=+\frac{2}{3}$ at $T=0$ and $T=0.03$ GeV. Similarly, in Fig.~\ref{fig:PiQ02}, we show $\tilde{\Pi}_{0H}(q_0,{\bm 0})$ (dashed purple line) and $\tilde{\Pi}_{H}(q_0,{\bm 0})$ (solid red line) for $\hat{e}_q=-\frac{1}{3}$. Note that we plot the result for only $q_0>0$, since the response function is symmetric with respect to an inversion of $q_0\to -q_0$ as $\tilde{\Pi}_{H}(q_0,{\bm 0}) = \tilde{\Pi}_{H}(-q_0,{\bm 0})$, due to time-reversal symmetry of the Kondo phase.

In what follows, we discuss the divergent behaviors near $q_0\sim 0.2$ GeV in Figs.~\ref{fig:PiQ01} and \ref{fig:PiQ02}, where the difference between $\tilde{\Pi}_{0}(q_0,{\bm 0})$ and $\tilde{\Pi}_{H}(q_0,{\bm 0})$ becomes more significant. The reason is as follows.
The threshold energy for the pair annihilation and pair creation of q.p. and q.h. [the diagrams (iii) and (iv) in Fig.~\ref{fig:Diagrams}] is given by $\delta E_{\rm min} = \sqrt{8\Delta^2}$, as indicated in Fig.~\ref{fig:Dispersions}. Thus, when $q_0$ becomes larger than $\delta E_{\rm min} = \sqrt{8\Delta^2}$, the imaginary part corresponding to the above physical processes appears. The value of threshold energy is $\delta E_{\rm min}\approx0.239$ GeV for $T=0$ and $\delta E_{\rm min}\approx0.210$ GeV for $T=0.03$ GeV, at $\mu=0.5$ GeV. Above this threshold, we encounter a divergence that cannot be regulated by the UV cutoff $\Lambda$, when we include the vertex corrections as well. This divergence stems from the interband processes of corrected-vertex parts of the form 
\begin{eqnarray}
&& \tilde{\Pi}_{\delta H}^{\zeta\zeta'}(q_0,{\bm 0}) \nonumber\\
 &=& \Delta\frac{(\hat{e}_q+\hat{e}_Q)N_c}{2}\int\frac{d^3p}{(2\pi)^3}\frac{1}{(q_0+i\eta- E_{\bm p}^{\zeta}+E_{\bm p}^{\zeta'})^2}\Big(\cdots\Big) \nonumber\\
&& + ({\rm other\ terms})
 \label{Q0DepDiv}
\end{eqnarray}
[see Eq.~(\ref{InterDeltaApp}) for the explicit expression]. Namely, the divergence arises from the factor $1/(q_0+i\eta- E_{\bm p}^{\zeta}+E_{\bm p}^{\zeta'})^2$ and cannot be removed by, e.g., making use of the principal value integral. For this reason, in Figs.~\ref{fig:PiQ01} and~\ref{fig:PiQ02}, we have shown the results where $q_0$ is smaller than the threshold.

In more realistic situations, the problematic factor $1/(q_0+i\eta- E_{\bm p}^{\zeta}+E_{\bm p}^{\zeta'})^2$ is replaced by $1/(q_0+i\tau_R^{-1}-E_{\bm p}^\zeta+E_{\bm p}^{\zeta'})^2$ due to a finite relaxation time $\tau_R$. In this case, the divergence will be smeared. In field-theoretical treatments, the relaxation time $\tau_R$ can be evaluated by a self-energy of the Green's function of fermions beyond the perturbative calculation. Namely, for a feasible treatment in the higher frequency regime, we need to employ a nonperturbative method to determine the self-energy, such as the Dyson-Schwinger equations coupling with the Ward-Takahashi identity~(\ref{WTIVertexQ}). Although the behavior above the threshold is problematic in our present treatment, we can see a significant enhancement of the HQSP as the frequency $q_0$ approaches the threshold from below, which is a universal behavior of response functions.

\subsection{The HQSP response function in the spacelike regime}
\label{sec:StaticPiH}

In this subsection, we show the $\mu$ dependence of the response function in the spacelike regime.

First, we examine the response function in the static limit defined in Eq.~(\ref{PiDynamical}). In this limit, the intraband processes in addition to the interband ones, namely, all of the diagrams in Fig.~\ref{fig:Diagrams} contribute. Figure~\ref{fig:PiSta1} represents the results of $\tilde{\Pi}_{0H}^{\rm sta}$ (dashed purple line) and $\tilde{\Pi}_{H}^{\rm sta}$ (solid blue line) for $\hat{e}_q=+\frac{2}{3}$ at $T=0$, $T=0.01$ GeV and $T=0.03$ GeV. Similarly, in Fig.~\ref{fig:PiSta2} we show $\tilde{\Pi}_{0H}^{\rm sta}$ (dashed purple line) and $\tilde{\Pi}_{H}^{\rm sta}$ (solid red line) for $\hat{e}_q=-\frac{1}{3}$. 

As in the dynamical limit, the vertex corrections enhance the HQSP response function for $\hat{e}_q=\hat{e}_Q$, while they suppress the HQSP for $\hat{e}_q\neq\hat{e}_Q$. At finite temperature, the HQSP in the static limit is suppressed compared to that in the dynamical limit for any electric charges. This suppression shows that the intraband processes driven by the mechanism explained in Sec.~\ref{sec:TwoLimits} are considerably large. It is worth noting that such intraband contributions at finite temperature were also found in the CSE with the Kondo effect~\cite{Suenaga:2020oeu}. In addition, we can see that the vertex corrections becomes relatively insignificant at $T=0.03$ GeV.

Here, we summarize differences between the HQSP with vertex corrections in the dynamical limit (Figs.~\ref{fig:PiDyn1} and~\ref{fig:PiDyn2}) and that in the static limit (Figs.~\ref{fig:PiSta1} and~\ref{fig:PiSta2}). At zero temperature, the $\tilde{\Pi}_H^{\rm dyn}$ and $\tilde{\Pi}_H^{\rm sta}$ coincide, while at finite temperature they differ, which is consistent with the properties found analytically in Sec.~\ref{sec:TwoLimits}. Besides, numerically, we have found that the magnitude of $\tilde{\Pi}_H^{\rm dyn}$ is always larger than that of $\tilde{\Pi}_H^{\rm sta}$ for $\hat{e}_Q=\hat{e}_q=+\frac{2}{3}$. On the other hand, for $\hat{e}_Q=+\frac{2}{3}$ and $\hat{e}_q=-\frac{1}{3}$, the difference of magnitudes depends on temperature.

Next, we show numerical results of $|{\bm q}|$ dependence of the response function for vanishing $q_0$ at $\mu=0.5$ GeV. Because of the Gauss's law for the magnetic field, the momentum ${\bm q}$ is transverse to the field: $q^i\tilde{B}^i(0,{\bm q})=0$. Figure~\ref{fig:PiQ1} represents the results of $\tilde{\Pi}_{0H}(0,{\bm q})$ (dashed purple line) and $\tilde{\Pi}_{H}(0,{\bm q})$ (solid blue line) for $\hat{e}_q=+\frac{2}{3}$ at $T=0$ and $T=0.03$ GeV. Similarly, in Fig.~\ref{fig:PiQ2} we show $\tilde{\Pi}_{0H}(0,{\bm q})$ (dashed purple line) and $\tilde{\Pi}_{H}(0,{\bm q})$ (solid red line) for $\hat{e}_q=-\frac{1}{3}$. It should be noted that we have plotted the results for the momentum up to $q=0.3$ GeV, while we assumed that $q$ is sufficiently small compared to $\mu$.

Unlike the frequency dependence of the HQSP response function studied in Sec.~\ref{sec:DynamicalPiH}, Figs.~\ref{fig:PiQ1} and~\ref{fig:PiQ2} show that the magnitude of $\tilde{\Pi}_H(0,{\bm q})$ does not change significantly. This is because there is no notable kinetic effects such as the existence of thresholds in the spacelike region. In particular, for the small $|{\bm q}|$ regime such a stable behavior is understood well. Namely, since the loop integrals in Eqs.~(\ref{PiH1}) and~(\ref{PiH2}) are dominated by modes at $|{\bm p}_+|\sim |{\bm p}_-| \sim \mu$ or those having sufficiently large density of states, the small $q$ does not change the HQSP significantly.

\section{The HQSP from the Zeeman interaction}
\label{sec:Discussions}

Up to this point we have investigated the HQSP induced by the Kondo effect under a magnetic field which arises at ${\cal O}(1/m^0_Q)$ within the HQET. When we go beyond this order, the ordinary Zeeman interaction (ZI) at ${\cal O}(1/m_Q^1)$ is expected to become another source of the HQSP. Hence, in this section we discuss the HQSP induced by the ZI in the absence of the Kondo effect and compare its magnitude with that driven through the Kondo effect of ${\cal O}(1/m_Q^0)$. In the following analysis, we employ the grand canonical picture such that the chemical potential of heavy quarks measured within the nonrelativistic framework is always zero as done for the Kondo effect in this paper.

In the Lagrangian~(\ref{NJLStart}), we have described the heavy quark within the leading order of the HQET, where only terms of ${\cal O}(1/m_Q^0)$ have been taken into account. When we include contributions of ${\cal O}(1/m_Q^1)$ as well, the Lagrangian for the heavy quark can be given by
\begin{eqnarray}
{\cal L}_{\rm ZI} = \Psi_v^\dagger\left(i\partial_0+ \frac{{\bm \nabla}^2}{2m_Q} + \frac{e_Q}{m_Q}{\bm S}_ h\cdot{\bm B} \right)\Psi_v  \ . \label{LNRQFT_2}
\end{eqnarray}
Here, we have left only the kinetic and ZI terms with the spin operator for the heavy quark ${\bm S}_h =\frac{{\bm \sigma}}{2}$. As done in Sec.~\ref{sec:Calculation}, from Eq.~(\ref{LNRQFT_2}), the HQSP by the ZI under the weak magnetic field within the linear response theory is evaluated as
\begin{eqnarray}
\langle \tilde{\cal S}^{{\rm ZI},i}_H(i\bar{\omega}_n,{\bm q})\rangle &=& N_c\frac{e_Q}{m_Q}\tilde{B}^j(i\bar{\omega}_n,{\bm q})T\sum_m\int\frac{d^3p}{(2\pi)^3} \nonumber\\
&\times& {\rm tr}\left[S_h^iG_{\rm NR}(i\omega_m',{\bm p}')S_h^jG_{\rm NR}(i\omega_m,{\bm p})\right] \ ,\nonumber\\ \label{SNonRel}
\end{eqnarray}
where
\begin{eqnarray}
G_{\rm NR} (i\omega_m,{\bm p}) = \frac{i}{i\omega_m-E_{\bm p}^{\rm NR}}
\end{eqnarray}
is the Green's function for the heavy quark in the nonrelativistic framework with $E_{\bm p}^{\rm NR} = |{\bm p}|^2/(2m_Q)$, and ${\bm p}'={\bm p}+{\bm q}$. Performing the Matsubara summation with Eq.~(\ref{Matsubara}), the HQSP~(\ref{SNonRel}) with a small ${\bm q}$ reads
\begin{eqnarray}
\langle \tilde{S}^{{\rm ZI},i}_H(i\bar{\omega}_n,{\bm q})\rangle &\approx&
N_c\frac{e_Q\tilde{B}^i(i\bar{\omega}_n,{\bm q})}{2m_Q}\int\frac{d^3p}{(2\pi)^3} \nonumber\\
&\times&\frac{1}{i\bar{\omega}_n-\frac{\partial E_{\bm p}^{\rm NR}}{\partial|{\bm p}|}(\hat{\bm p}\cdot{\bm q})}\frac{\partial f_F(E_{\bm p}^{\rm NR})}{\partial|{\bm p}|}(\hat{\bm p}\cdot{\bm q})\ . \nonumber\\
\end{eqnarray}
Thus, defining the response function $\tilde{\Pi}_H^{\rm ZI}(i\bar{\omega}_n,{\bm q})$ via 
\begin{eqnarray}
\langle \tilde{S}^{{\rm ZI},i}_H(i\bar{\omega}_n,{\bm q})\rangle = e \tilde{B}^i(i\bar{\omega}_n,{\bm q})\tilde{\Pi}_H^{\rm ZI}(i\bar{\omega}_n,{\bm q})\ ,
\end{eqnarray}
$\tilde{\Pi}_H^{\rm ZI}(i\bar{\omega}_n,{\bm q})$ in the dynamical and static limits in the real time are evaluated as
\begin{eqnarray}
\lim_{q_0\to0}\tilde{\Pi}_H^{\rm ZI}(q_0,{\bm 0}) = 0\ ,  \label{ZIDyn}
\end{eqnarray}
and
\begin{eqnarray}
\lim_{{\bm q}\to{\bm 0}}\tilde{\Pi}_H^{\rm ZI}(0,{\bm q}) &=& -N_c\frac{\hat{e}_Q}{2m_Q}\int\frac{d^3p}{(2\pi)^3}\frac{\partial f_F(E_{\bm p}^{\rm NR})}{\partial E_{\bm p}^{\rm NR}} \nonumber\\
&=&N_c\frac{\hat{e}_QB(2\pi m_QT)^{1/2}}{8\pi^2}(1-\sqrt{2})\zeta(1/2) \ , \nonumber\\ \label{ZISta}
\end{eqnarray}
respectively. In Eq.~(\ref{ZISta}), $\zeta(1/2)=-1.46035\cdots$ is the Riemann zeta function $\zeta(n)$ for $n=1/2$.

\begin{figure}[t]
\centering
\includegraphics*[scale=0.65]{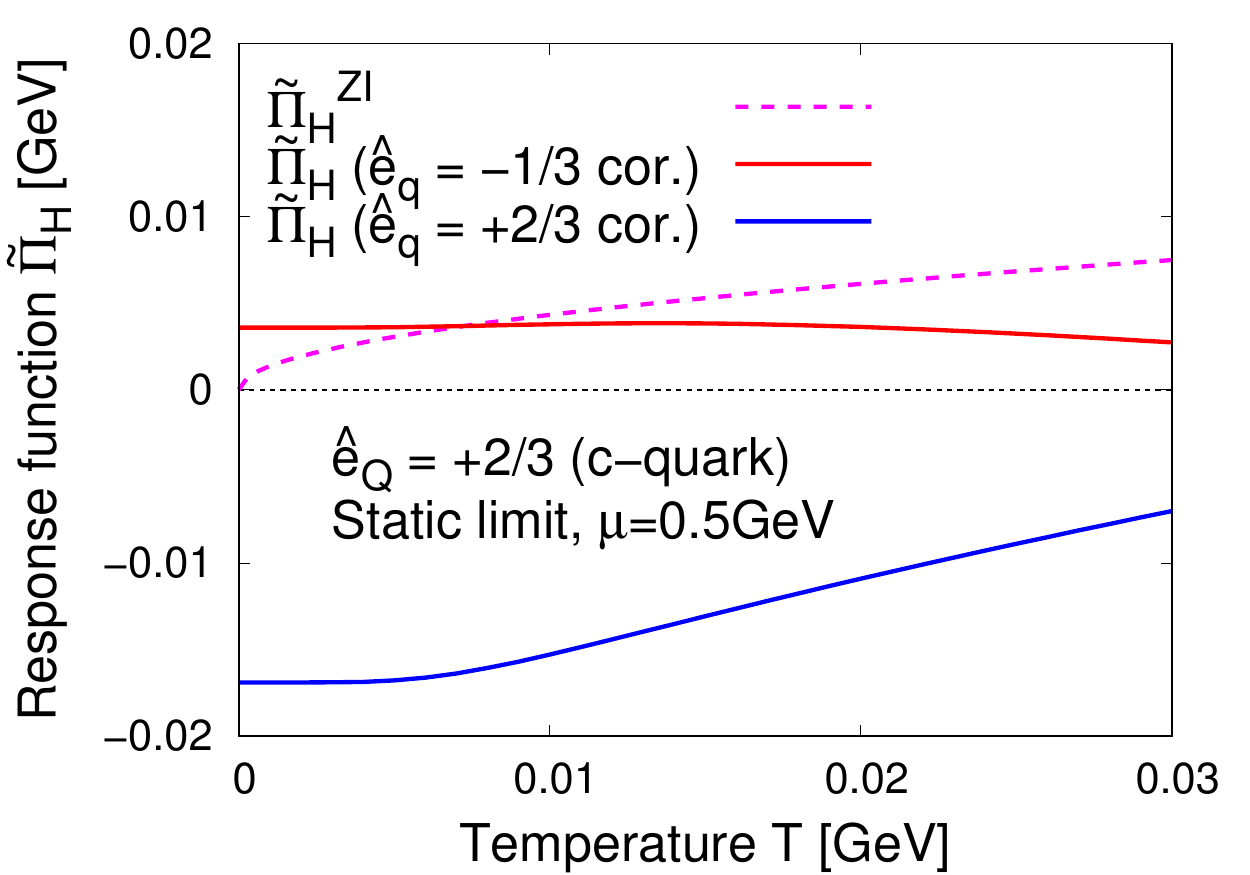}
\caption{The HQSP response function induced by the Kondo effect and that by the ZI as functions of $T$. For details see the text.}
\label{fig:ZeemanPiH}
\end{figure}

Equation~(\ref{ZIDyn}) shows that $\tilde{\Pi}_H^{\rm ZI}$ in the dynamical limit vanishes, because interband processes for the HQSP driven by the ZI are absent. In other words, the emergence of HQSP in the dynamical limit can be regarded as a peculiar phenomenon induced by the Kondo effect under magnetic field. From Eq.~(\ref{ZISta}) we can see that $\tilde{\Pi}_H^{\rm ZI}$ in the static limit is always positive for finite $T$. The positive $\tilde{\Pi}_H^{\rm ZI}$ is understood as follows: from the ZI term in Eq.~(\ref{LNRQFT_2}), the corresponding ZI Hamiltonian is always negative when the heavy-quark spin and the magnetic field are parallel for $e_Q=+\frac{2}{3}$. Thus, the positive spin polarization is thermodynamically favored more than the negative one, resulting in the positive HQSP. Besides, $\tilde{\Pi}_H^{\rm ZI}$ at $T=0$ is zero, since in the three-dimensional space $\tilde{\Pi}_H^{\rm ZI}$ is of ${\cal O}(B^{3/2})$ and can be neglected within the linear response theory.

In Fig.~\ref{fig:ZeemanPiH}, we compare the $T$ dependence of the HQSP response function induced by the Kondo effect and that by the ZI in the static limit. The solid blue line corresponds to the result by the Kondo effect ($\tilde{\Pi}_H$) with $\hat{e}_Q=\hat{e}_q=+\frac{2}{3}$, while the solid red one with $\hat{e}_Q=+\frac{2}{3}, \hat{e}_q=-\frac{1}{3}$, at $\mu=0.5$ GeV (recall that this $\mu$ is the chemical potential for light quarks). The dotted pink line shows the result by the ZI ($\tilde{\Pi}_H^{\rm ZI}$) for the $c$ quark with $m_Q=1.27$ GeV. Figure~\ref{fig:ZeemanPiH} shows that the HQSP is clearly dominated by the contribution from the Kondo effect at lower temperature particularly for $\hat{e}_Q=\hat{e}_q=+\frac{2}{3}$ even in the static limit. 

\section{Conclusions}
\label{sec:Conclusions}

In this paper, we have proposed a new mechanism of the HQSP in quark matter induced by the Kondo effect under a magnetic field. By employing the NJL type model, we have indeed shown that the HQSP is driven through the Kondo condensate from light quarks coupling with the magnetic field, although a magnetic coupling of the heavy quarks themselves is absent in the heavy-quark limit. In particular, we have demonstrated the emergence of HQSP in the distinct momentum regimes: the timelike and spacelike momentum regions. Physically, the former (latter) describes the HQSP whose time dependence is faster (slower) than the equilibration of the spatial part of the system. The effects of vertex corrections required by the $U(1)_{\rm EM}$ gauge symmetry have been also examined. 

Our analysis shows that the response function of the HQSP is significantly driven in both the momentum regimes. Also, we have found that the vertex corrections enhance the resultant HQSP compared to that with the bare vertices, when the Kondo condensate is electrically neutral. In addition, the timelike HQSP is significantly enhanced as the frequency of the magnetic field approaches the threshold of pair creation (annihilation) of the quasiparticle and quasihole induced by the Kondo effect, whereas the spacelike one does not vary largely with the momentum. 

We have also discussed the HQSP induced by the Zeeman interaction of heavy quarks, as corrections from violation of the heavy-quark limit. As a result, we have found that the HQSP induced by the Zeeman interaction  in the dynamical limit (zero frequency and momentum limits from the timelike regime) always vanishes and that in the static limit (similar limit from the spacelike regime) at lower temperature is largely suppressed. Therefore, emergence of the HQSP particularly in such regimes can be a useful signal of the Kondo effect.

Experimentally, the peripheral and low-energy heavy-ion collisions (HICs) are expected to become a testing ground to investigate the HQSP induced by the Kondo effect. In fact, in such HICs heavy quarks as impurities are produced by hard processes mediated by gluons, together with a magnetic field and a sufficient quark chemical potential. In this case, the HQSP is converted into the spin polarization of heavy hadrons which can be observables~\cite{Araki:2020rok}.

The HQSP together with the emergence of the Kondo condensate can be examined in detail in future lattice simulations of QCD (or QCD-like theories). The HQSP is measured by computing the operator $\langle \Psi^\dagger(t,{\bm x}) \frac{{\bm \sigma}}{2}\Psi(t,{\bm x})\rangle$, where  $\Psi(t,{\bm x})$ and ${\bm \sigma}$ are the heavy-quark field and the Pauli matrix parallel to a magnetic field ${\bm B}$, respectively. For the lattice simulations, one has to introduce, in addition to the usual QCD, the three additional backgrounds: (i) a nonzero chemical potential for light quarks, (ii) a magnetic field, and (iii) heavy impurities.
For (i), although Monte Carlo simulations for systems with a nonzero quark chemical potential are usually useless due to the sign problem, one may utilize other {\it sign-problem-free systems}, such as two-color QCD, isospin chemical potential $\mu_I$, and chiral chemical potential $\mu_5$.\footnote{See, e.g., Refs.~\cite{Nakamura:1984uz,Hands:1999md,Kogut:2001na,Muroya:2002ry,Muroya:2003qs,Hands:2006ve,Hands:2010gd,Cotter:2012mb}, for earlier works on lattice simulations of two-color dense QCD, Refs.~\cite{Kogut:2002tm,Kogut:2002zg,Kogut:2004zg,Brandt:2017oyy} for $\mu_I$, and Refs.~\cite{Yamamoto:2011gk,Yamamoto:2011ks,Braguta:2015zta,Braguta:2015owi} for $\mu_5$.
See Refs.~\cite{Suenaga:2019jqu,Kanazawa:2020xje} for model studies of the Kondo effect with $\mu_5$.}
Such simulations have been devoted to elucidating the properties of dense QCD in a background magnetic field~\cite{Endrodi:2014lja,Puhr:2016kzp,Buividovich:2020gnl,Buividovich:2021fsa}, and model studies to intuitively understand the simulations have been also done~\cite{Duarte:2015ppa,Cao:2015xja,Adhikari:2015wva,Adhikari:2018fwm,Mao:2020xvc,Canfora:2020uwf,Suenaga:2021bjz}.
For example, our present analysis can be easily applied to the two-color system by changing the number of colors from $N_c=3$ to $N_c=2$.
Namely, our proposal in this paper is testable in two-color QCD simulations without major changes. In particular, the static three-dimensional momentum dependence of the HQSP response function within the linear response regime can be directly measured by the lattice simulations.
For (iii), we note that inclusion of heavy quarks as impurities does not spoil the sign-problem-free advantage unless we introduce a ``chemical potential'' for the heavy quarks as done in this paper.
We expect that a better understanding of the role of heavy impurities and their properties under external fields in quark matter will be promoted by such simulations in the future.

\section*{ACKNOWLEDGEMENTS}

D. S. wishes to thank Keio University and Japan Atomic Energy Agency (JAEA) for their hospitalities during his stay there. Y. A. is supported by the Leading Initiative for Excellent Young Researchers (LEADER) and the Japan Society for the Promotion of Science KAKENHI (Grant No. 20H01830). K. S. is supported by JSPS KAKENHI (Grants No. JP17K14277 and JP20K14476). S. Y. is supported by JSPS KAKENHI (Grant No. JP17K05435), by the Ministry of Education, Culture, Sports, Science (MEXT)-Supported Program for the Strategic Research Foundation at Private Universities Topological Science (Grant No. S1511006), and by the Interdisciplinary Theoretical and Mathematical Sciences Program
(iTHEMS) at RIKEN.

\appendix

\section{TRANSVERSALITY OF THE ELECTROMAGNETIC VERTEX}
\label{sec:NGVertex}
Here, we give a detailed explanation for the transversality of the electromagnetic vertex $\Gamma^\mu$.

In our present study, the gauge field only includes its spatial parts $\tilde{A}^i(q_0,{\bm q})$ since the magnetic field is generated by
\begin{eqnarray}
\tilde{B}^i(q_0,{\bm q}) = i\epsilon^{ijk}q^j\tilde{A}^k(q_0,{\bm q})\ . \label{BDefA}
\end{eqnarray}
In general the (three-dimensional) gauge field $\tilde{A}^i(q_0,{\bm q})$ can be decomposed into longitudinal and transverse parts as
\begin{eqnarray}
\tilde{A}^i(q_0,{\bm q}) &=& P_L^{ij} \tilde{A}^j(q_0,{\bm q}) + P_T^{ij}\tilde{A}^j(q_0,{\bm q}) \nonumber\\
&\equiv&  \tilde{A}_L^i(q_0,{\bm q}) +\tilde{A}_T^i(q_0,{\bm q})\ , \label{ADec}
\end{eqnarray}
with $\tilde{A}_L^i = P^{ij}_L \tilde{A}^j$, $\tilde{A}_T^i = P^{ij}_T \tilde{A}^j$, where longitudinal and transverse projection operators are defined by
\begin{eqnarray}
P_L^{ij} = \frac{q^i q^j}{|{\bm q}|^2} \ , \ \ P_T^{ij} = \delta^{ij}-\frac{q^i q^j}{|{\bm q}|^2}\ .
\end{eqnarray}
Here, we can show that the $\tilde{A}_L^i$ and $\tilde{A}_T^i$ satisfy
\begin{eqnarray}
\epsilon^{ijk}q^j\tilde{A}_L^k(q_0,{\bm q}) &=& 0 \ , \nonumber\\
\epsilon^{ijk}q^j\tilde{A}_T^k(q_0,{\bm q}) &=& \epsilon^{ijk}q^j\tilde{A}^k(q_0,{\bm q}) \ ,
\end{eqnarray}
with $\epsilon^{ijk}q^j q^k= 0$. Namely, the relation~(\ref{BDefA}) can be rewritten to
\begin{eqnarray}
\tilde{B}^i(q_0,{\bm q})= i\epsilon^{ijk}q^j\tilde{A}_T^k(q_0,{\bm q})\ , \label{BDefA2}
\end{eqnarray}
showing that the magnetic field is generated solely by the transverse part of the gauge field. In other words, in our present study we can replace the gauge field with its transverse part,
\begin{eqnarray}
\tilde{A}^i(q_0,{\bm q}) \to \tilde{A}^i_T(q_0,{\bm q})\ . \label{AReplace}
\end{eqnarray}

Equation~(\ref{AReplace}) shows that, when the electromagnetic vertex $\Gamma^\mu$ is proportional to $q^\mu$, the coupling turns into

\begin{eqnarray}
 \Gamma^\mu\tilde{A}_\mu(q_0,{\bm q}) \propto  q^i\tilde{A}_T^i(q_0,{\bm q}) = 0\ , \label{TransGamma}
\end{eqnarray}
due to the transverse nature of $\tilde{A}^i_T$: $q^i\tilde{A}^i_T(q_0,{\bm q}) = 0$. Therefore, in our present paper, longitudinal parts of the vertex vanish and only transverse parts survive.

\section{SOLUTION OF THE IDENTITIES IN EQS.~(\ref{WTIEach2_2}) AND~(\ref{WTIEach3_2}) }
\label{sec:WTISol}

Here, we show solutions of the identities~(\ref{WTIEach2_2}) and~(\ref{WTIEach3_2}). For convenience, we decompose the vertices $\Gamma^\mu_{A\bar{\psi}\Psi_v}$ and $\Gamma^\mu_{A\Psi_v^\dagger\psi}$ as
\begin{eqnarray}
\Gamma^\mu_{A\bar{\psi}\Psi_v} &=&  \Gamma^{{\rm NG}, \mu}_{A\bar{\psi}\Psi_v} + \hat{\Gamma}^\mu_{A\bar{\psi}\Psi_v}\ , \nonumber\\
\Gamma^\mu_{A\Psi_v^\dagger\psi} &=& \Gamma^{{\rm NG},\mu}_{A\Psi_v^\dagger\psi} + \hat{\Gamma}^\mu_{A\Psi_v^\dagger\psi} \ ,
\end{eqnarray}
respectively. In these equations, $ \Gamma^{{\rm NG}, \mu}_{A\bar{\psi}\Psi_v}$ and $\Gamma^{{\rm NG},\mu}_{A\Psi_v^\dagger\psi}$ are the vertices stemming from the NG mode contributions while $\hat{\Gamma}^\mu_{A\bar{\psi}\Psi_v}$ and $\hat{\Gamma}^\mu_{A\Psi_v^\dagger\psi}$ are from non-NG mode contributions. Namely, from Eqs.~(\ref{WTIEach2_2}) and~(\ref{WTIEach3_2}), those vertices satisfy
\begin{eqnarray}
q_\mu\Gamma^{{\rm NG}, \mu}_{A\bar{\psi}\Psi_v} &=& \Delta^*(e_Q-e_q) \left(
\begin{array}{c}
{\bm 1} \\
\hat{\bm p}\cdot{\bm \sigma}   \\
\end{array}
\right)  \ , \nonumber\\
q_\mu \Gamma^{{\rm NG},\mu}_{A\Psi_v^\dagger\psi} &=&  \Delta (e_q-e_Q)\left(
\begin{array}{cc}
{\bm 1} &  \hat{\bm p}\cdot{\bm \sigma} \\
\end{array}
\right)  \ , \label{WTINGApp}
\end{eqnarray}
and
\begin{eqnarray}
q_\mu \hat{\Gamma}^\mu_{A\bar{\psi}\Psi_v} &=& -\Delta^* (e_q+e_Q) \left(
\begin{array}{c}
0  \\
\frac{{\bm q}\cdot{\bm \sigma}-(\hat{\bm p}\cdot{\bm \sigma})(\hat{\bm p}\cdot{\bm q})}{2|{\bm p}|}    \\
\end{array}
\right)\ , \nonumber\\
q_\mu\hat{\Gamma}^\mu_{A\Psi_v^\dagger\psi} &=&   \Delta (e_q+e_Q)\left(
\begin{array}{cc}
0&  \frac{{\bm q}\cdot{\bm \sigma}-(\hat{\bm p}\cdot{\bm \sigma})(\hat{\bm p}\cdot{\bm q})}{2|{\bm p}|} \\
\end{array}
\right) \ . \label{WTIEachApp} 
\end{eqnarray}

First, from the identity~(\ref{WTINGApp}), the corrected vertices from the NG mode contributions are easily found as
\begin{eqnarray}
\Gamma^{{\rm NG}, \mu}_{A\bar{\psi}\Psi_v}&=& \Delta^*(e_Q-e_q) \nonumber\\
&& \times \frac{\delta^{\mu 0}+v_{\rm NG}|{\bm q}|^{\alpha-2}q^j\delta^{\mu j}}{q_0-v_{\rm NG}|{\bm q}|^\alpha} \left(
\begin{array}{c}
{\bm 1}\\
\hat{\bm p}\cdot{\bm \sigma}\\
\end{array}
\right) \ , \label{NGApp1}
\end{eqnarray}
and
\begin{eqnarray}
 \Gamma^{{\rm NG},\mu}_{A\Psi_v^\dagger\psi} &=& \Delta(e_q-e_Q)\nonumber\\
&& \times  \frac{\delta^{\mu 0}+v_{\rm NG}|{\bm q}|^{\alpha-2}q^j\delta^{\mu j}}{q_0-v_{\rm NG}|{\bm q}|^\alpha}\left(
\begin{array}{cc}
{\bm 1}& \hat{\bm p}\cdot{\bm \sigma} \\
\end{array}
\right)\ , \nonumber\\
\label{NGApp2}
\end{eqnarray}
where $v_{\rm NG}$ is a constant related to a velocity of the NG mode, and $\alpha$ is an integer. Equations~(\ref{NGApp1}) and~(\ref{NGApp2}) show that the spatial parts of the vertices from NG mode contributions are proportional to $q^j$ which is longitudinal. Hence, due to the transversality of the coupling derived in Eq.~(\ref{TransGamma}), the contributions from the NG modes disappear.

Next, we solve the identity for the vertices from the non-NG mode contributions. From Eq.~(\ref{WTIEachApp}), only the lower component of $\hat{\Gamma}^\mu_{A\bar{\psi}\Psi_v}$ and the right component of $\hat{\Gamma}^\mu_{A\Psi_v^\dagger\psi}$ become nonzero. Thus, the solution of Eq.~(\ref{WTIEachApp}) is of the form
\begin{eqnarray}
\hat{\Gamma}^\mu_{A\bar{\psi}\Psi_v} = \delta^{\mu j}\left(
\begin{array}{c}
0  \\
\hat{\Gamma}^{{\rm sub},j}_{A\bar{\psi}\Psi_v}   \\
\end{array}
\right)\ , \nonumber\\
 \hat{\Gamma}^\mu_{A\Psi_v^\dagger\psi}  = \delta^{\mu j} \left(
\begin{array}{cc}
0& \hat{\Gamma}^{{\rm sub}, j}_{A\Psi_v^\dagger\psi} \\
\end{array}
\right) \ . \label{GammaSub}
\end{eqnarray}
In Eq.~(\ref{GammaSub}), we have left only spatial components of the vertices, since the time component vanishes as seen from Eq.~(\ref{WTIEachApp}) by inserting ${\bm q}={\bm 0}$. In our present paper, our goal is to study the HQSP under a magnetic field. Namely, we need to get a tensor structure $\epsilon^{ijk}q^j\tilde{A}^k(q_0,{\bm q})$ in evaluating the formula in Eq.~(\ref{SHImaginary}). For this reason $\hat{\Gamma}^{{\rm sub},j}_{A\bar{\psi}\Psi_v} $ and $ \hat{\Gamma}^{{\rm sub}, j}_{A\Psi_v^\dagger\psi} $ must include only the terms proportional to one Pauli matrix. By taking this consideration into account, general forms of $\hat{\Gamma}^{{\rm sub},j}_{A\bar{\psi}\Psi_v} $ and $ \hat{\Gamma}^{{\rm sub}, j}_{A\Psi_v^\dagger\psi} $ are given by
\begin{eqnarray}
\hat{\Gamma}^{{\rm sub},j}_{A\bar{\psi}\Psi_v} &=& \hat{\Gamma}^{\sigma}_{A\bar{\psi}\Psi_v} \sigma^j +  \hat{\Gamma}^{+}_{A\bar{\psi}\Psi_v} ({\bm p}_+\cdot{\bm \sigma}) p^j \nonumber\\
&& + \hat{\Gamma}^{-}_{A\bar{\psi}\Psi_v} ({\bm p}_-\cdot{\bm \sigma})p^j   \, ,\nonumber\\
\hat{\Gamma}^{{\rm sub},j}_{A\Psi_v^\dagger\psi} &=& \hat{\Gamma}^{\sigma}_{A\Psi_v^\dagger\psi} \sigma^j +  \hat{\Gamma}^{+}_{A\Psi_v^\dagger\psi} ({\bm p}_+\cdot{\bm \sigma}) p^j  \nonumber\\
&& + \hat{\Gamma}^{-}_{A\Psi_v^\dagger\psi} ({\bm p}_-\cdot{\bm \sigma})p^j \, , \label{GammaSubDec}
\end{eqnarray}
respectively. In this equation, $\hat{\Gamma}^{s}_{A\bar{\psi}\Psi_v}$ and $\hat{\Gamma}^{s}_{A\Psi_v^\dagger\psi} $ ($s=\sigma,+,-$) are arbitrary scalar functions of ${\bm p}$ and ${\bm q}$. It should be noted that we have not included terms proportional to $q^j$ in Eq.~(\ref{GammaSubDec}) because such terms will disappear due to the transversality as in Eq.~(\ref{TransGamma}). Inserting Eqs.~(\ref{GammaSubDec}) and~(\ref{GammaSub}) into Eq.~(\ref{WTIEachApp}) and after some algebraic calculation, $\hat{\Gamma}^{s}_{A\bar{\psi}\Psi_v}$ and $\hat{\Gamma}^{s}_{A\Psi_v^\dagger\psi} $ are found to be 
\begin{eqnarray}
\hat{\Gamma}^{\sigma}_{A\bar{\psi}\Psi_v} &=& \Delta^*(e_q+e_Q)\frac{({\bm p}\cdot{\bm p}_+)}{2|{\bm p}|^3} + ({\bm p}\cdot{\bm q}) \hat{\Gamma}^{-}_{A\bar{\psi}\Psi_v} \ ,\nonumber\\
\hat{\Gamma}^{+}_{A\bar{\psi}\Psi_v} &=& -\Delta^*(e_q+e_Q)\frac{1}{2|{\bm p}|^3}- \hat{\Gamma}^{-}_{A\bar{\psi}\Psi_v} \ ,\label{GammaRed1}
 \end{eqnarray}
and
 \begin{eqnarray}
 \hat{\Gamma}^{\sigma}_{A\Psi_v^\dagger\psi} &=& -\Delta(e_q+e_Q)\frac{({\bm p}\cdot{\bm p}_-)}{2|{\bm p}|^3}-({\bm p}\cdot{\bm q}) \hat{\Gamma}^{+}_{A\Psi_v^\dagger\psi} \ , \nonumber\\
 \hat{\Gamma}^{-}_{A\Psi_v^\dagger\psi} &=& \Delta(e_q+e_Q)\frac{1}{2|{\bm p}|^3}- \hat{\Gamma}^{+}_{A\Psi_v^\dagger\psi} \ , \label{GammaRed2}
 \end{eqnarray}
respectively. As found in Eqs.~(\ref{GammaRed1}) and~(\ref{GammaRed2}), it seems that all $\hat{\Gamma}^{s}_{A\bar{\psi}\Psi_v}$ and $\hat{\Gamma}^{s}_{A\Psi_v^\dagger\psi} $ cannot be fixed uniquely, which means that the integration constant remains as a free parameter.

To summarize the above calculation, the corrected vertices satisfying the identities~(\ref{WTIEach2_2}) and~(\ref{WTIEach3_2}) which survive under a magnetic field are 
\begin{eqnarray}
\Gamma^\mu_{A\bar{\psi}\Psi_v} &=& \frac{\Delta^*(e_q+e_Q)}{2|{\bm p}|^3}\delta^{\mu j} \left(
\begin{array}{c}
0  \\
({\bm p}_+\cdot{\bm p})\sigma^j-({\bm p}_+\cdot{\bm \sigma})p^j  \\
\end{array}
\right) \nonumber\\
&+& \delta^{\mu j}\hat{\Gamma}^-_{A\bar{\psi}\Psi_v} \left(
\begin{array}{c}
0  \\
({\bm p}\cdot{\bm q})\sigma^j-({\bm q}\cdot{\bm \sigma})p^j  \\
\end{array}
\right)  \ ,   \label{GammaCorrectApp1}
\end{eqnarray}
and
\begin{eqnarray}
\Gamma^\mu_{A\Psi_v^\dagger\psi} &=&  -\frac{\Delta (e_q+e_Q)}{2|{\bm p}|^3}\delta^{\mu j}\left(
\begin{array}{cc}
0 &  ({\bm p}_-\cdot{\bm p})\sigma^j-({\bm p}_-\cdot{\bm \sigma})p^j \\
\end{array}
\right)  \nonumber\\
&-&\delta^{\mu j}\hat{\Gamma}^+_{A\Psi_v^\dagger\psi}\left(
\begin{array}{cc}
0 & ({\bm p}\cdot{\bm q})\sigma^j-({\bm q}\cdot{\bm \sigma})p^j  \\
\end{array}
\right)   \ .  \label{GammaCorrectApp2}
\end{eqnarray}

\section{THE STATIC AND DYNAMICAL LIMITS FOR CORRECTED-VERTEX PARTS $\tilde{\Pi}_{\delta H}$}
\label{sec:AppCorrectedVertex}

Here, we show detailed evaluation of the HQSP response functions in the dynamical and static limits from the corrected-vertex parts $\tilde{\Pi}_{\delta H}(q_0,{\bm q})$ in Eq.~(\ref{PiH2}).

As done in Sec.~\ref{sec:BareVertex} for $\tilde{\Pi}_{0 H}(q_0,{\bm q})$, we expand $\tilde{\Pi}_{\delta H}(q_0,{\bm q})$ in terms of a small momentum ${\bm q}$. For $\tilde{\Pi}_{\delta H}(q_0,{\bm q})$, the kinetic factor originating from the spin couplings between the fermions is proportional to $({\bm p}_\pm\cdot{\bm p})(\hat{\bm p}_\pm\cdot{\bm q})$. This factor is expanded as
\begin{eqnarray}
({\bm p}_+\cdot{\bm p})(\hat{\bm p}_+\cdot{\bm q}) &=& |{\bm p}|({\bm p}\cdot{\bm q})\pm\frac{1}{2}|{\bm p}||{\bm q}|^2 + {\cal O}({\bm q}^3)\ ,
\end{eqnarray}
which is of ${\cal O}({\bm q}^1)$, unlike that for the bare-vertex parts in Eq.~(\ref{KineticBare}) of ${\cal O}({\bm q}^2)$. Hence we may need to expand the numerators and denominators of ${\cal I}_{\delta H\pm}^{\zeta\zeta'}({\bm p}_+;{\bm p}_-)$ in Eq.~(\ref{SizeDelta}) up to of ${\cal O}({\bm q}^2)$.

Keeping in mind the above remarks, after a lengthy but straightforward calculation, $\tilde{\Pi}_{\delta H}^{\zeta\zeta'}(i\bar{\omega}_n,{\bm q})$ for the intraband and interband processes with a small momentum ${\bm q}$ are reduced to
\begin{widetext}
\begin{eqnarray}
\Pi_{\delta H}^{\zeta\zeta}(q) \approx -\Delta\frac{(\hat{e}_q+\hat{e}_Q)N_c}{2}\int\frac{d^3p}{(2\pi)^3} \frac{1}{|{\bm p}|^3}\frac{V_\zeta({\bm p})W_\zeta({\bm p})|{\bm p}|+\left[\frac{\partial V_\zeta({\bm p})}{\partial|{\bm p}|}W_\zeta({\bm p})-V_\zeta({\bm p})\frac{\partial W_\zeta({\bm p})}{\partial|{\bm p}|}\right]({\bm p}\cdot\hat{\bm q})^2}{q_0-\frac{\partial E_{\bm p}^\zeta}{\partial|{\bm p}|}(\hat{\bm p}\cdot{\bm q})}\frac{\partial{f}_F(E_{\bm p}^\zeta)}{\partial|{\bm p}|}(\hat{\bm p}\cdot{\bm q}) \ , \nonumber\\ \label{IntraDeltaApp}
\end{eqnarray} 
and
\begin{eqnarray}
\Pi_{\delta H}^{\zeta\zeta'}(q) &\approx& \Delta\frac{(\hat{e}_q+\hat{e}_Q)N_c}{2}\int\frac{d^3p}{(2\pi)^3}\frac{1}{2|{\bm p}|^3} \frac{1}{q_0-E_{\bm p}^\zeta+E_{\bm p}^{\zeta'}}  \Bigg\{\Big({f}_F(E_{\bm p}^{\zeta'})-{f}_F(E_{\bm p}^\zeta)\Big)\Bigg[ \Big(V_\zeta({\bm p}) W_{\zeta'}({\bm p})+W_\zeta ({\bm p})V_{\zeta'}({\bm p})\Big)|{\bm p}|  \nonumber\\
&+& \left(\frac{\partial V_\zeta({\bm p})}{\partial|{\bm p}|} W_{\zeta'}({\bm p})+W_\zeta ({\bm p})\frac{\partial V_{\zeta'}({\bm p})}{\partial|{\bm p}|}-\frac{\partial W_\zeta({\bm p})}{\partial|{\bm p}|} V_{\zeta'}({\bm p})-V_\zeta({\bm p}) \frac{\partial W_{\zeta'}({\bm p})}{\partial|{\bm p}|}\right)({\bm p}\cdot\hat{\bm q})^2\Bigg] -\Bigg[ \left(\frac{\partial{f}_F(E_{\bm p}^{\zeta'})}{\partial|{\bm p}|}+\frac{\partial{f}_F(E_{\bm p}^\zeta)}{\partial|{\bm p}|}\right)  \nonumber\\
&-& \frac{1}{q_0-E_{\bm p}^\zeta+E_{\bm p}^{\zeta'}}\left( \frac{\partial E_{\bm p}^\zeta}{\partial|{\bm p}|} +\frac{\partial E_{\bm p}^{\zeta'}}{\partial|{\bm p}|}\right) \Big({f}_F(E_{\bm p}^{\zeta'})-{f}_F(E_{\bm p}^\zeta)\Big)\Bigg]\Big(V_\zeta({\bm p}) W_{\zeta'}({\bm p})-W_\zeta({\bm p}) V_{\zeta'}({\bm p})\Big)({\bm p}\cdot\hat{\bm q})^2\Bigg\}  \ , \label{InterDeltaApp}
\end{eqnarray}
\end{widetext}
respectively. It should be noted that unfamiliar contributions including $1/(q_0-E_{\bm p}^\zeta+E_{\bm p}^{\zeta'})^2$ appear in Eq.~(\ref{InterDeltaApp}). They cause divergences for higher $q_0$ which cannot be removed by the UV cutoff $\Lambda$. Taking low momentum limits properly in Eq.~(\ref{IntraDeltaApp}), the response functions from the corrected-vertex parts for the intraband processes in the dynamical and static limits read 
\begin{eqnarray}
\lim_{q_0\to0}\tilde{\Pi}^{\zeta\zeta}_{\delta H}(q_0,{\bm 0}) &=& 0 \ , \label{IntraDeltaDApp}
\end{eqnarray}
and
\begin{widetext}
\begin{eqnarray}
\lim_{{\bm q}\to{\bm 0}}\tilde{\Pi}^{\zeta\zeta}_{\delta H}(0,{\bm q}) &=& \Delta\frac{(\hat{e}_q+\hat{e}_Q)N_c}{2}\int_0^\Lambda\frac{d|{\bm p}|}{2\pi^2}\Bigg[V_\zeta({\bm p})W_\zeta({\bm p})+\left(\frac{\partial V_\zeta({\bm p})}{\partial|{\bm p}|}W_\zeta({\bm p})-V_\zeta({\bm p})\frac{\partial W_\zeta({\bm p})}{\partial|{\bm p}|}\right)\frac{|{\bm p}|}{3}\Bigg]\frac{\partial{f}_F(E_{\bm p}^\zeta)}{\partial E_{\bm p}^\zeta} \ , \label{IntraDeltaSApp}
\end{eqnarray}
respectively. The vanishing result in the dynamical limit and the contributions proportional to $\partial f_F(E_{\bm p}^\zeta)/\partial E_{\bm p}^\zeta$ in the static limit are the same as those from the bare-vertex parts. Similarly, from Eq.~(\ref{InterDeltaApp}) the response function for the interband processes in the dynamical and static limits are evaluated as
\begin{eqnarray}
 && \lim_{{\bm q}\to{\bm 0}}\tilde{\Pi}^{\zeta\zeta'}_{0H}(0,{\bm q}) = \lim_{q_0\to0}\tilde{\Pi}^{\zeta\zeta'}_{0H}(q_0,{\bm 0})  \nonumber\\ 
 &=& \Delta\frac{(\hat{e}_q+\hat{e}_Q)N_c}{2}\int_0^\Lambda\frac{d|{\bm p}|}{4\pi^2}\frac{1}{E_{\bm p}^{\zeta'}-E_{\bm p}^\zeta}  \Bigg\{\Big({f}_F(E_{\bm p}^{\zeta'})-{f}_F(E_{\bm p}^\zeta)\Big)\Bigg[ \Big(V_\zeta({\bm p}) W_{\zeta'}({\bm p})+W_\zeta ({\bm p})V_{\zeta'}({\bm p})\Big) \nonumber\\
&& + \left(\frac{\partial V_\zeta({\bm p})}{\partial|{\bm p}|} W_{\zeta'}({\bm p})+W_\zeta({\bm p}) \frac{\partial V_{\zeta'}({\bm p})}{\partial|{\bm p}|}-\frac{\partial W_\zeta({\bm p})}{\partial|{\bm p}|} V_{\zeta'}({\bm p}) -V_\zeta({\bm p}) \frac{\partial W_{\zeta'}({\bm p})}{\partial|{\bm p}|}\right)\frac{|{\bm p}|}{3}\Bigg] -\Bigg[ \left(\frac{\partial{f}_F(E_{\bm p}^{\zeta'})}{\partial|{\bm p}|}+\frac{\partial{f}_F(E_{\bm p}^\zeta)}{\partial|{\bm p}|}\right)  \nonumber\\
&& -\frac{1}{E_{\bm p}^{\zeta'}-E_{\bm p}^\zeta}\left( \frac{\partial E_{\bm p}^\zeta}{\partial|{\bm p}|} +\frac{\partial E_{\bm p}^{\zeta'}}{\partial|{\bm p}|}\right) \Big({f}_F(E_{\bm p}^{\zeta'})-{f}_F(E_{\bm p}^\zeta)\Big)\Bigg]\Big(V_\zeta({\bm p}) W_{\zeta'} ({\bm p}) -W_\zeta ({\bm p}) V_{\zeta'}({\bm p})\Big)\frac{|{\bm p}|}{3}\Bigg\}  \ . \label{InterDeltaSApp}
\end{eqnarray}
\end{widetext}
Namely, similarly to the bare-vertex parts, the interband processes yield the identical results in the two limits.

\bibliography{reference}

\end{document}